\documentclass[a4paper,11pt]{article}
\pdfoutput=1  
\usepackage{jheppub} 
\usepackage[T1]{fontenc}
\usepackage{amsmath} 
\usepackage{multirow}
\usepackage{slashed}
\usepackage[usenames,dvipsnames]{color}
\usepackage[colorinlistoftodos]{todonotes}

\title{Distinguishing between Dirac and Majorana neutrinos at FASER}
\author[a]{ShivaSankar K.A.,}
\author[b]{Alakabha Datta,}
\author[c]{Danny Marfatia}

\affiliation[a]{Department of Physics, Hokkaido University, Sapporo 060-0810, Japan}
\affiliation[b]{Department of Physics and Astronomy, 108 Lewis Hall, University of Mississippi, Oxford, MS 38677-1848, USA.}
\affiliation[c]{Department of Physics \& Astronomy, University of Hawaii at Manoa, 2505 Correa Rd., Honolulu, HI 96822, USA
}

\emailAdd{a-shiva@particle.sci.hokudai.ac.jp}
\emailAdd{datta@phy.olemiss.edu}
\emailAdd{dmarf8@hawaii.edu}

\abstract{
Some of the simplest models for the origin of neutrino mass involve right-handed neutrinos (RHNs), which could be either Dirac or Majorana particles — a distinction that has profound implications for lepton number conservation and the fundamental nature of neutrinos. We investigate the potential of the FASER experiment to distinguish between these two possibilities using signatures predicted by the Standard Model Neutrino Effective Field Theory (SMNEFT), where RHNs interact with Standard Model particles through higher-dimensional operators. We focus on RHNs produced via $B$, $D$, $K$, and $\pi$ meson decays at the Large Hadron Collider and their subsequent three-body decays within the FASER detector. The kinematic and angular distributions of the decay products in the RHN rest frame differ significantly for Dirac and Majorana RHNs, and these differences manifest as distinct spatial distributions of electron-positron pairs at FASER. Using Monte Carlo simulations and a $\chi^2$ analysis, we demonstrate that these spatial observables provide a robust experimental probe for determining the Dirac or Majorana nature of RHNs. For select production and decay operator combinations and RHN masses around 0.1~GeV, FASER can achieve discrimination at the $3\sigma$ level.}

\begin{document}
\maketitle
\flushbottom

\section{Introduction}
Neutrino oscillation experiments have firmly established that neutrinos have tiny but non-zero masses, which is one of the strongest hints for physics beyond the Standard Model (SM). Although many models have been proposed to explain the origin of neutrino mass, no direct experimental evidence has yet pointed to even a class of models. One of the simplest and most compelling solutions is to extend the SM by introducing SM singlet fermions like right handed neutrinos (RHNs) 
which can naturally generate small neutrino masses through the seesaw mechanism, while also explaining the baryon asymmetry of our Universe via leptogenesis. 
Since RHNs are fermions with zero electric charge, a very interesting question is whether they are Dirac or Majorana particles, a distinction that has profound implications for fundamental physics. If such RHNs are Majorana particles, they allow for lepton number violation, which could be tested in experiments. However, if they are Dirac particles, lepton number is conserved, leading to different experimental signatures. 

Rather than focusing on specific ultraviolet models with RHNs, a powerful and systematic approach is to employ a model independent effective field theory (EFT). The effects of heavy new physics at a scale $\Lambda$ are encoded in higher-dimensional operators constructed from SM fields and, in the present case, RHNs. The Standard Model Effective Field Theory (SMEFT), which includes only SM fields, has been widely used to study new physics effects involving only SM fields~\cite{Buchmuller:1985jz, Grzadkowski:2010es, Henning:2014wua, Brivio:2017vri}. 

For a RHN with a mass well below the electroweak scale, 
one can extend the SMEFT framework to include RHNs explicitly, leading to the so-called Standard Model Neutrino Effective Field Theory (SMNEFT)~\cite{delAguila:2008ir, Aparici:2009fh, Bhattacharya:2015vja, Liao:2016qyd, Bischer:2019ttk}. SMNEFT systematically captures all possible interactions between RHNs and SM particles that are consistent with SM gauge symmetries. This includes a set of dimension-six operators, such as the four-fermion contact terms. 
These operators provide production and decay channels for RHNs, leading to observable signatures at current and future experiments. 

In this work, we explore a novel strategy to distinguish Dirac from Majorana RHNs by exploiting the kinematic and angular distributions of their decay products. Specifically, we focus on RHNs produced in the decays of $B$, $D$, $K$ and $\pi$ mesons at the Large Hadron Collider (LHC), which subsequently decay into electron-positron pairs and neutrinos. The angular distributions of the decay products in the RHN rest frame differ significantly between the Dirac and Majorana scenarios. When boosted to the laboratory frame, these differences manifest as distinct spatial distributions in the detector. To quantify the statistical significance of these differences, we simulate RHN production and decay, reconstruct the spatial distributions of dileptons at FASER, and employ a $\chi^2$ test to compare the Dirac and Majorana hypotheses. Our results indicate that the spatial separation of decay products could provide a powerful experimental handle for probing the nature of RHNs.

This paper is organized as follows. In Section~\ref{Section2}, we establish the theoretical foundation of our analysis, including the structure of SMNEFT, the mechanisms of RHN production and decay, and the distinctive angular distributions that arise in each scenario. Section~\ref{Section3} details our simulation methodology and detector configuration. In Section~\ref{Section4}, we present our results, featuring the spatial distributions and statistical analysis that quantifies the discriminating power between Dirac and Majorana RHNs. We also discuss the implications of our findings and evaluate the experimental feasibility. In Section~\ref{Section5}, we summarize our findings.

\section{Theoretical framework}
\label{Section2}

In this section, we establish the theoretical foundation for our analysis of RHNs within the SMNEFT framework. We begin by outlining the theory behind SMNEFT, then detail the production mechanisms of RHNs through meson decays, and finally examine their subsequent decay processes. 

\subsection{Standard Model Neutrino Effective Field Theory (SMNEFT)}
\label{SM1}

SMEFT provides a systematic framework for parameterizing new physics effects through higher-dimensional operators constructed from SM fields. However, when considering RHNs with masses potentially below the electroweak scale, it becomes necessary to extend this framework by explicitly including the RHN field as a dynamical degree of freedom. This extension, known as the SMNEFT, allows for a model-independent description of RHN interactions. In SMNEFT, we introduce a gauge-singlet fermion field $N$ that represents the RHN. This field can be either a Dirac or Majorana fermion, enabling us to investigate both scenarios within a unified theoretical framework. The SMNEFT Lagrangian at dimension-six, preserving baryon and lepton number conservation ($\Delta B = \Delta L = 0$), is given by
\begin{equation}
\mathcal{L}^{(6)}_{\text{SMNEFT}} = \mathcal{L}_{\rm{SM}} + i \bar{N}\slashed{\partial}N + \sum_{i} \mathcal{C}_i\mathcal{O}_i\,,
\end{equation}
where $\mathcal{L}_{\rm{SM}}$ is the SM Lagrangian, $\mathcal{C}_i$ are dimensionless Wilson coefficients (WCs) that encode the strength of new physics effects at a scale $\Lambda$, and $\mathcal{O}_i$ represent the dimension-six operators involving the RHN field $N$ and SM fields. These operators are constructed to respect the SM gauge symmetries and provide all possible interactions between RHNs and SM particles at the dimension-six level. It is important to clarify how we treat Dirac versus Majorana RHNs in our analysis. In the Dirac case, where lepton number is conserved, $N$ and $\bar{N}$ are distinct. For most of our analysis, we present results only for the $\bar{N}$ state, unless there are explicit physical differences between $N$ and $\bar{N}$ (such as in angular distributions of decay products). In the Majorana case, this distinction vanishes. For consistency and clarity throughout our analysis, we uniformly refer to the RHN as $\bar{N}$ in both scenarios, but account for the appropriate physics in each case.

In the SMNEFT framework, we need to carefully distinguish between Dirac and Majorana scenarios. For a Dirac RHN, the field $N$ represents a four-component spinor with distinct right-chiral ($N_R$) and left-chiral ($N_L$) components, where the SMNEFT operators couple only to $N_R$. In contrast, for a Majorana RHN, the particle is its own antiparticle, and the right-chiral field $N_R$ in the SMNEFT completely describes the physical particle.  In either case, we remain agnostic about the specific mass mechanism. Our focus is on the phenomenological implications arising from RHN interactions in the SMNEFT framework.


\subsection{RHN production}
  
The dominant production channels at the LHC are meson decays, particularly two-body and three-body processes involving $B$, $D$, $K$, and $\pi$ mesons produced in the forward region. These decays are mediated by higher-dimensional operators in SMNEFT, and their rates depend critically on both the operator structure and the RHN mass. Understanding these production mechanisms is essential for establishing the event rates and kinematics that will ultimately enable discrimination between Dirac and Majorana RHNs. The general effective Lagrangian responsible for RHN production through meson decays can be written as
\begin{equation}\label{eq:lagrangian_smneft}
    \mathcal{L}_{\text{prod}} = 2\sqrt{2}G_F \sum_{X,\alpha} \mathcal{C}^X_{\alpha R} \mathcal{O}^X_{\alpha R} + \text{h.c.}\,,
\end{equation}
where $G_F$ is the Fermi constant, and the sum runs over Lorentz structures $X = \{S, V, T\}$ (scalar, vector, tensor) and chiralities $\alpha = \{L, R\}$. The CKM matrix element $V_{ud}$ between an up type and a down type quark relevant to all the decays has been absorbed into the Wilson coefficients for simplicity. The corresponding operators are explicitly given by
\begin{align}
    \mathcal{O}^S_{\alpha R} &= (\bar{u} P_{\alpha} d)(\bar{\ell} P_R N)\,, \\
    \mathcal{O}^V_{\alpha R} &= (\bar{u} \gamma^{\mu} P_{\alpha} d)(\bar{\ell} \gamma_{\mu} P_R N)\, \\
    \mathcal{O}^T_{\alpha R} &= (\bar{u} \sigma^{\mu\nu} P_{\alpha} d)(\bar{\ell} \sigma_{\mu\nu} P_R N)\,,
\end{align}
where $u$ and $d$ represent the relevant up type and down type quark flavors (e.g., $b$ and $c$ for $B$ meson decays), $\ell$ denotes the charged lepton, and $P_{\alpha} = (1 \pm \gamma_5)/2$ are the chirality projection operators. These effective operators arise naturally from the SMNEFT framework. The most relevant dimension-six SMNEFT operators that contribute to RHN production are:
\begin{align}
    \mathcal{O}_{LNuQ} &= (\bar{L}_p N_r)(\bar{u}_s Q_t^j)\,, \\
    \mathcal{O}_{Nedu} &= (\bar{N}_p \gamma^{\mu} e_r)(\bar{d}_s \gamma_{\mu} u_t)\,, \\
    \mathcal{O}^{(3)}_{LNQd} &= (\bar{L}_p^j \sigma^{\mu\nu} N_r)\epsilon_{jk}(\bar{Q}_s^k \sigma_{\mu\nu} d_t)\,,
\end{align}
where $L = (\nu_L, \ell_L)^T$ and $Q = (u_L, d_L)^T$ are the left-handed lepton and quark $SU(2)_L$ doublets, respectively, while $e$, $u$, $d$, and $N$ are $SU(2)_L$ singlets. The flavor indices $p, r, s, t$ and $SU(2)_L$ indices $j, k$ will be suppressed for clarity, and $\epsilon_{jk}$ is the antisymmetric tensor with $\epsilon_{12} = 1$. In the following, we examine the specific decay channels and their associated branching ratios, which depend sensitively on the SMNEFT operator structure. The correspondence between the SMNEFT operators and those defined in Eq.~(\ref{eq:lagrangian_smneft}) is as follows: 
\begin{align}
    \mathcal{O}_{LNuQ} &\leftrightarrow \mathcal{O}^S_{LR}\,, \\
    \mathcal{O}_{Nedu} &\leftrightarrow  \mathcal{O}^V_{RR}\,, \\
    \mathcal{O}^{3}_{LNQd} &\leftrightarrow  \mathcal{O}^T_{RR}\,. 
\end{align}

\subsubsection{Two-body decays}

The production of $\bar{N}$ in the SMNEFT framework occurs  through leptonic and semileptonic meson decays. The purely leptonic production channel is the two-body decay process $M^{\pm} \to \ell^{\pm} \; (N/\bar{N})$, where $M$ represents a charged meson (such as $B^{\pm}$, $D^{\pm}$, $K^{\pm}$ and $\pi^\pm$) and $\ell^{\pm}$ denotes a charged lepton ($e$ in our analysis). The decay rate for a scalar interaction is
\begin{equation}\label{eq:2bd_scalar}
	\Gamma_M (C^S_{RR, LR}) = \vert C^S_{RR, LR} \vert^2 \frac{f_M^2 G_F^2 \sqrt{\lambda \left( m_M^2, m_N^2, m_{\ell}^2\right)}}{8 \pi} \frac{m_M\left(m_M^2 - m_{\ell}^2 - m_N^2\right)}{(m_q + m_{q'})^2}\,.
\end{equation}
The decay rate for a vector interaction is
\begin{equation}\label{eq:2bd_vector}
	\Gamma_M (C^V_{RR}) = \vert C^V_{RR} \vert^2 \frac{f_M^2 G_F^2 \sqrt{\lambda \left( m_M^2, m_N^2, m_{\ell}^2\right)}}{8 \pi} \frac{m_M^2 \left(m_{\ell}^2 + m_N^2 \right) - \left(m_{\ell}^2 - m_N^2\right)^2}{m_M^3}\,.
\end{equation}
Note that a tensor interaction does not contribute to two-body decays. 
The key components of these expressions are
\begin{itemize}
    \item The 3-body kinematic function, $\lambda(a,b,c) = a^2 + b^2 + c^2 - 2ab -2ac -2bc$.

    \item The meson decay constant $f_M$. The relevant decay constants are provided in Table \ref{tab:meson_decay_constants}.

    \item The constituent quark masses of the meson, $m_q, m_{q'}$.

    \item $m_M$ is the mass of the parent meson, $m_N$ is the mass of RHN and, $m_{\ell}$ is the mass of the lepton.
\end{itemize}

\begin{figure}[t]
    \centering
    \includegraphics[scale=0.40]{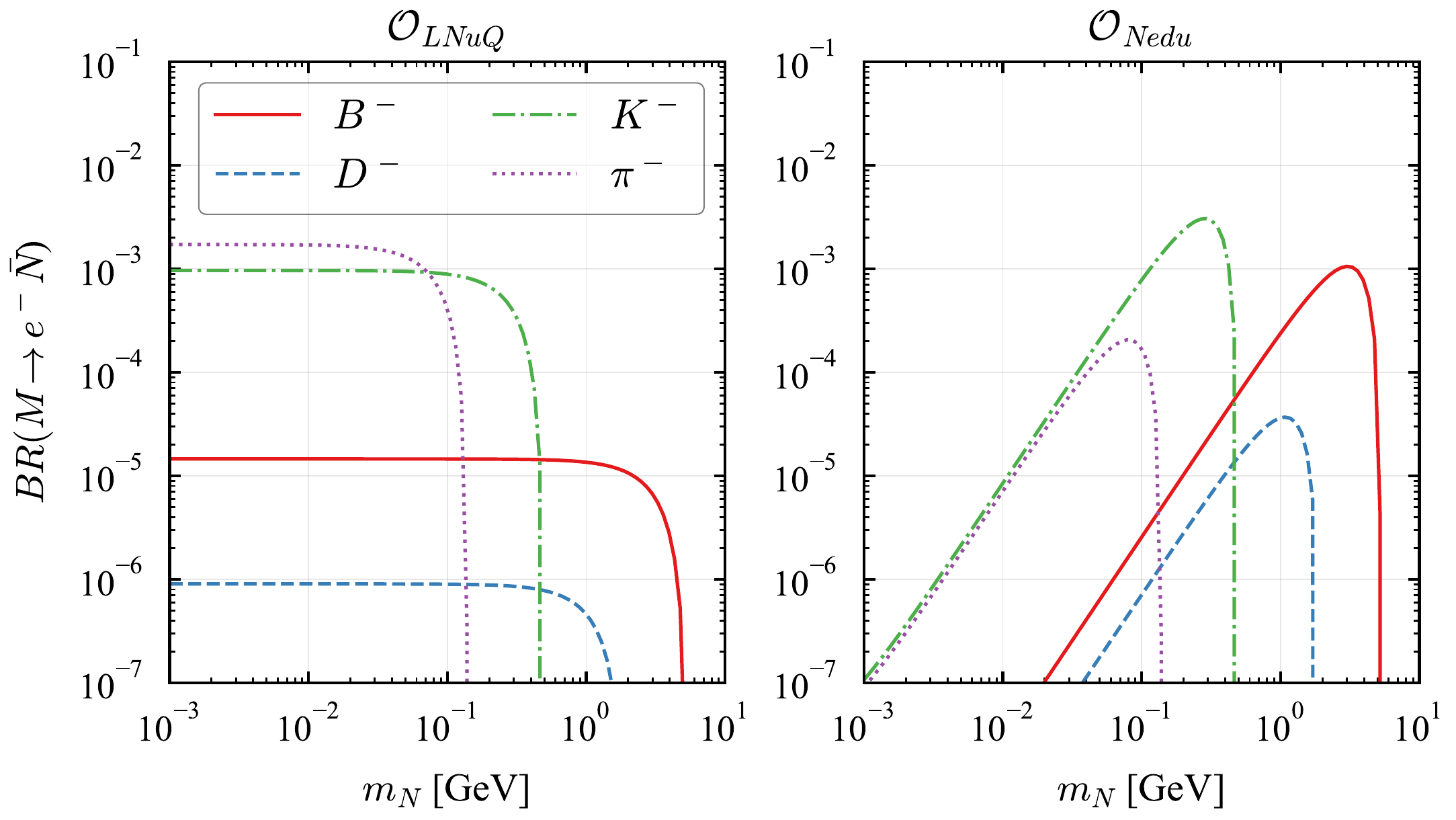}
    \caption{Branching ratios for two-body decays of $B$, $D$, $K$ and $\pi$ mesons to $\bar{N}$ as a function of the RHN mass for different SMNEFT operators (with Wilson coefficients set to $C=10^{-2}$). For Dirac neutrinos, the plot shows only the $M^{-} \to \ell^{-} \bar{N}$ channel; in the Majorana case, the total branching ratio is twice this value.}
    \label{fig:2bd_production_branching_ratios}
\end{figure}

Figure~\ref{fig:2bd_production_branching_ratios} shows the branching ratios for two-body decays of $B$, $D$, $K$, and $\pi$ mesons mediated by the operators $\mathcal{O}_{LNuQ}$ and $\mathcal{O}_{Nedu}$ over the relevant RHN mass range. For the $\mathcal{O}_{LNuQ}$ operator, the branching ratio tends to saturate at lower $m_N$ values, resulting in a relatively higher yield. In contrast, the branching ratio associated with $\mathcal{O}_{Nedu}$ decreases sharply for smaller $m_N$  smaller, leading to a reduced production rate. For $\mathcal{O}_{Nedu}$, the two-body decay branching ratios are significant only if the RHN mass is close to that of the parent meson. 

\begin{table}[t]
    \centering
    \begin{tabular}{|c|c|c|c|}
        \hline
        \textbf{Meson} & \textbf{$f_M$} ({GeV}) \\
        \hline
        $B$ & $ 0.19 $ \\
        $D$ & $ 0.212 $ \\
        $K$ & $ 0.13 $ \\
        $\pi$ & $ 0.155 $ \\
        \hline
    \end{tabular}
    \caption{Relevant meson decay constants for 
    two body meson decay~\cite{ParticleDataGroup:2020ssz}.}
    \label{tab:meson_decay_constants}
\end{table}

\subsubsection{Three-body decays}

Semi-leptonic three-body decays of neutral mesons provide complementary channels for RHN production and are particularly important for heavier mesons where phase space considerations favor multi-body decays. The primary 
three-body decay channel is $(M/\bar{M}^0) \to P^{\mp} \ell^{\pm} (N/\bar{N})$, where $P$ denotes the daughter meson. This decay process is  mediated by effective operators in the SMNEFT framework, with the decay kinematics determined by the masses of the involved particles and the specific operator structure. The differential decay width for this process is given by \cite{Datta:2022czw}
\begin{align}\label{eq:3bd_bd}
	\frac{d^2 \Gamma}{dq^2 d\cos\theta_{\ell}} = \frac{\sqrt{\lambda\left(m_M^2, m_P^2, q^2\right) \; \lambda\left(q^2, m_{\ell}^2, m_N^2\right)}}{512 \pi^2 m_M^3 q^2} \sum_{\lambda_{\ell},\lambda_{\bar{N}}} \vert \bar{\mathcal{M}} \vert^2\,,
\end{align}
where $q^2 = (p_{\ell} + p_{N})^2$ is the invariant mass squared of the lepton-RHN system, $\theta_{\ell}$ is the angle between the charged lepton momentum in the $\ell N$ rest frame and the direction of the daughter meson in the parent meson rest frame, and $\lambda_{\ell}, \lambda_{\bar{N}}$ denote the lepton and RHN helicity respectively. For detailed definitions of these kinematic variables, we refer to Ref.~\cite{Datta:2022czw}. In addition to the primary decay channel, there is another three-body decay channel $\overline{M}^0 \to P^{(*)\mp} \ell^{\pm} \bar{N}$, where $P^*$ denotes a vector meson (e.g., $D^*$), the differential decay width must account for the subsequent decay $P^* \to P \pi$. The decay width for this process is given by
\begin{align}\label{eq:3bd_bdstar}
	\frac{d^2 \Gamma}{dq^2 d\cos\theta_{\ell}} = 3\frac{\sqrt{\lambda\left(m_M^2, m_{P^*}^2, q^2\right) \; \lambda\left(q^2, m_{\ell}^2, m_N^2\right)}}{2048 \pi^4 m_M^3 q^2} \mathcal{B}\left(P^* \to P \pi\right) \sum_{\lambda_{\ell},\lambda_{\bar{N}}} \sum_{\lambda_{P^*}} \overline{\vert\mathcal{M}} \vert^2\,.
\end{align}
Here, $\mathcal{B}\left(P^* \to P \pi\right)$ represents the branching fraction of the vector meson decay, $\lambda_{\ell}, \lambda_{\bar{N}}, \lambda_{P^*}$ denote the lepton, RHN and vector meson helicity, respectively. The matrix element now includes a sum over the vector meson polarizations. The differential distributions in $q^2$ and $\cos\theta_{\ell}$ offer valuable kinematic signatures for distinguishing SMNEFT contributions from background processes and for probing the structure of the underlying effective operators~\cite{Datta:2022czw}, although we do not consider them in this analysis. For $B$ meson decays, we use the form factors used in Ref.~\cite{Datta:2022czw}. For $D$, $K$ and $\pi$ meson decays, we use the form factors of Refs.~\cite{Lubicz:2017syv, Koponen:2012di, Aoki:2017spo}.

\begin{figure}[t]
    \centering
    \includegraphics[scale=0.34]{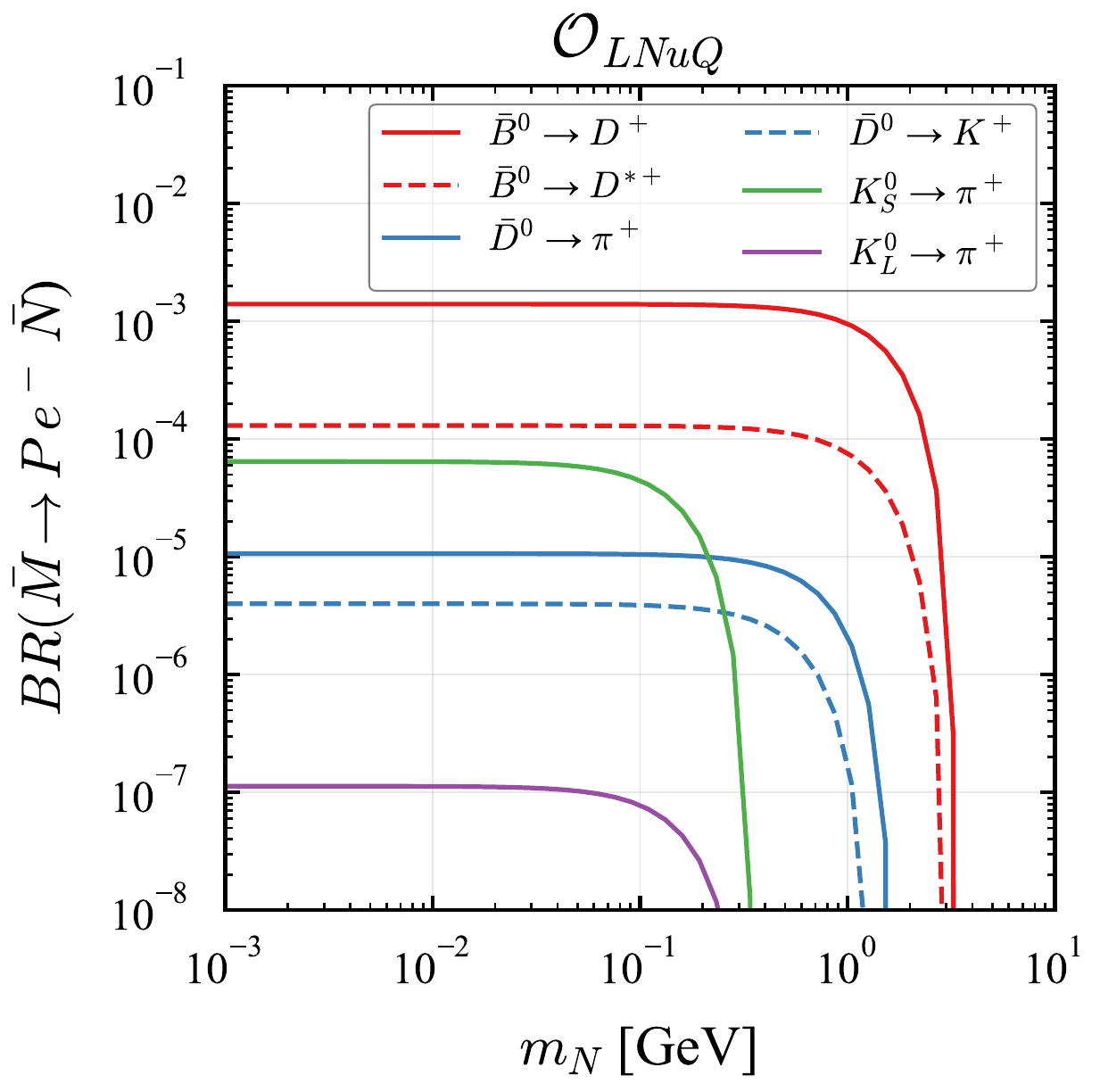}
    \includegraphics[scale=0.34]{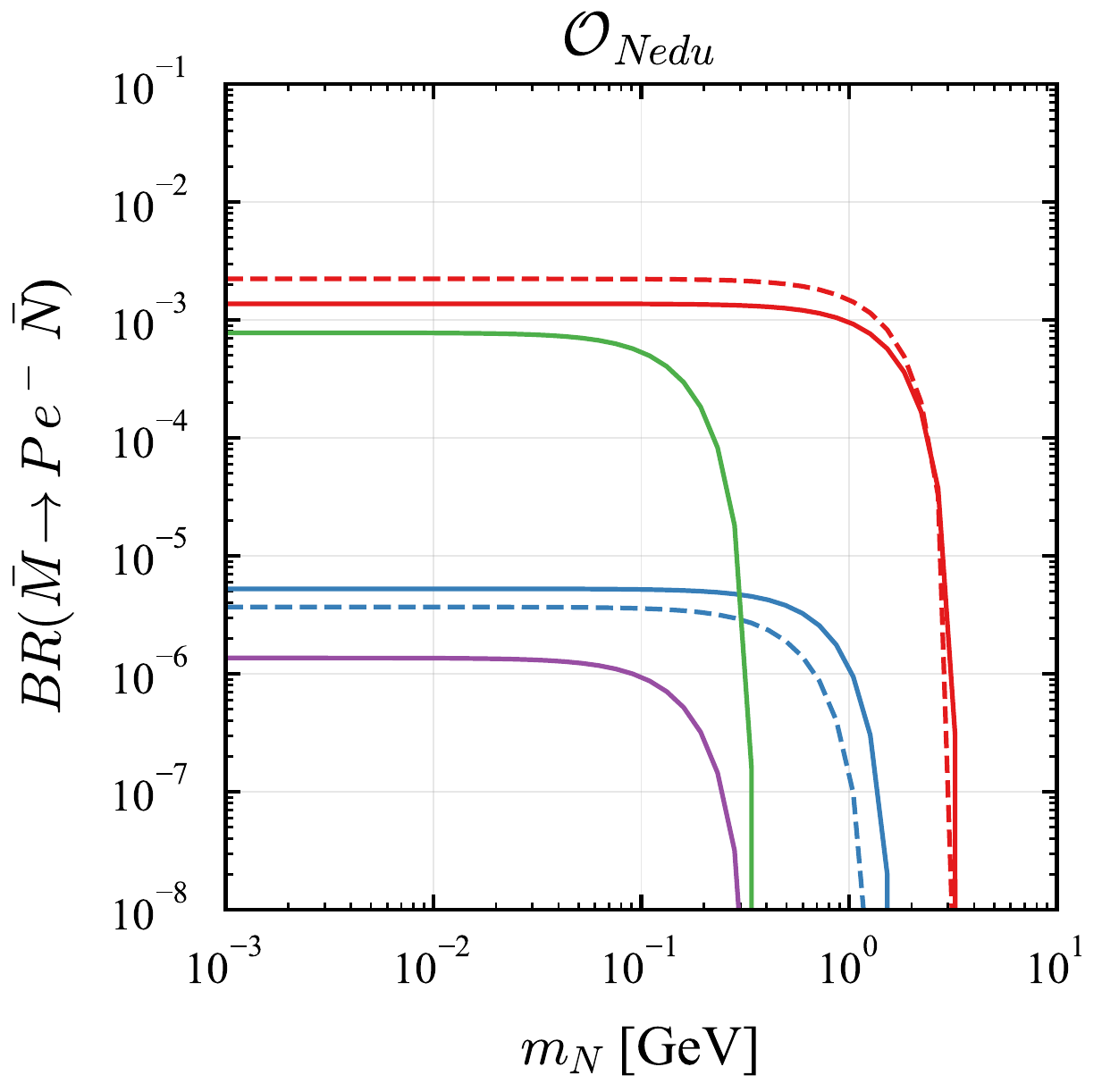}
    \includegraphics[scale=0.34]{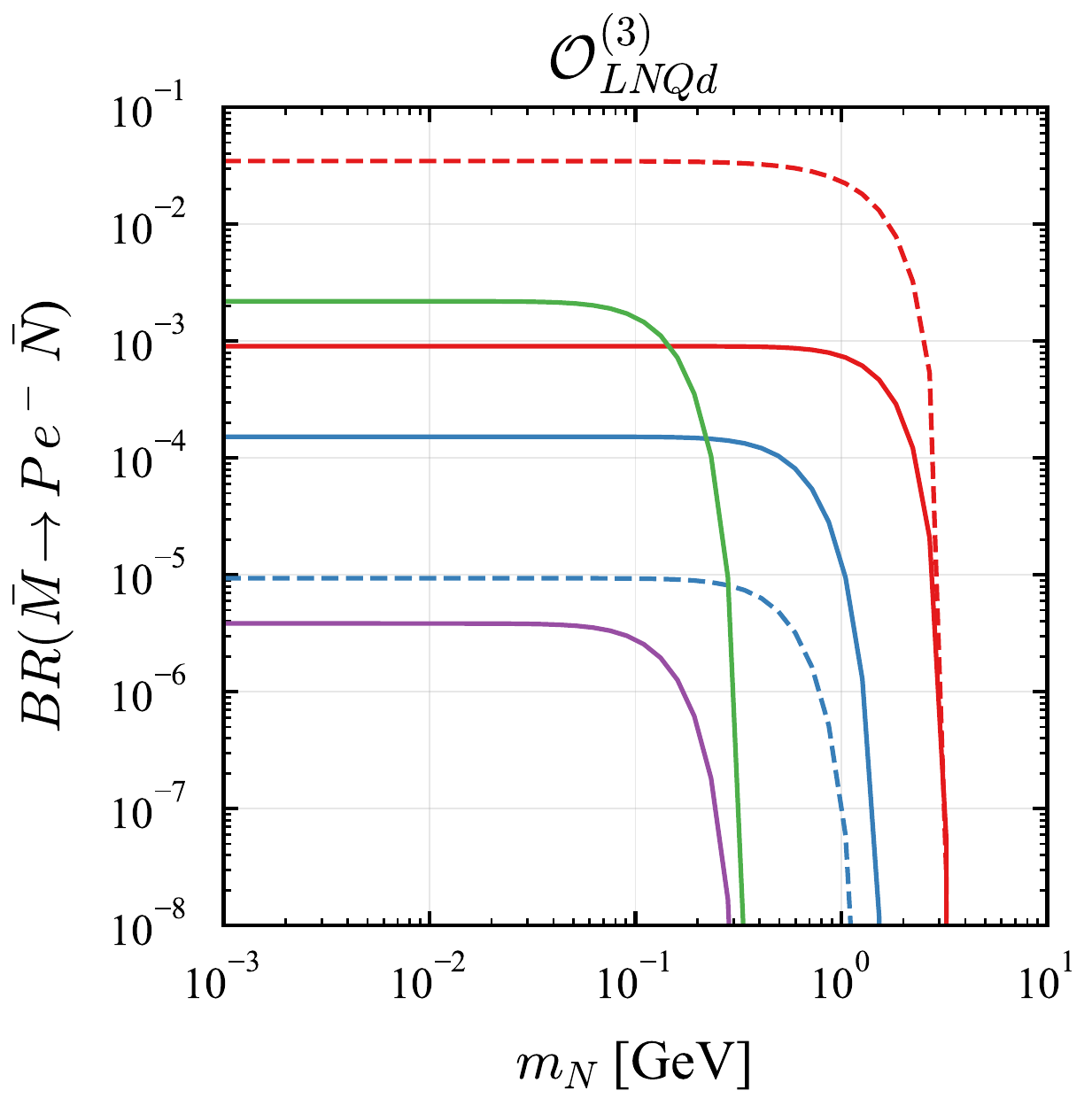}
    \caption{Branching ratios for three-body decays of mesons as a function of the RHN mass with $C = 10^{-2}$ for $\mathcal{O}_{LNuQ}$ (top left), $\mathcal{O}_{Nedu}$ (top right), and $\mathcal{O}^3_{LNQd}$ (bottom).}
    \label{fig:3bd_production_branching_ratios}
\end{figure}

To illustrate the impact of SMNEFT operators on RHN production, in Fig.~\ref{fig:3bd_production_branching_ratios} we show the branching ratios for three-body decays of $B, D, K$ mesons as a function of $m_N$ for the three production operators. These plots highlight how the production rates depend on both the operator structure and the RHN mass, and indicate which operators dominate in different mass regions. Several key features are to be noted from these plots. First, for all operators, the branching ratios approach constant values for light RHNs, and the $K^0_{L}$ channel has the smallest branching ratio. Second, the tensor operator $\mathcal{O}^{(3)}_{LNQd}$ contributes significantly to the $\bar{B}^0 \to D^{* +} e^- \bar{N}$ channel, with branching ratios approximately one order of magnitude larger than those for  the other channels and operators. This enhancement increases event statistics at FASER, making this channel particularly promising for experimental searches. 

\subsubsection{RHN polarization}

An important quantity for distinguishing Dirac from Majorana RHNs is the polarization of RHNs produced in meson decays. We define the polarization as
\begin{equation} \label{eq:rhn_polarization_production}
    P(N/\bar{N}) = \frac{\Gamma_{\pm 1/2} - \Gamma_{\mp 1/2}}{\Gamma_{\mp 1/2} + \Gamma_{\pm 1/2}}\,.
\end{equation}
where $\Gamma_{-1/2}$ and $\Gamma_{1/2}$ are the production rates for RHNs with helicities $-1/2$ and $1/2$, respectively. This definition ensures that $P = 1$ corresponds to a purely right-handed helicity for both $N$ ($1/2$) and $\bar{N}$ ($-1/2$) production. The polarization of $N$ and $\bar{N}$ in the Dirac case are identical for the definition in Eq.~(\ref{eq:rhn_polarization_production}). 

\begin{figure}[t]
    \centering
    \includegraphics[scale=0.35]{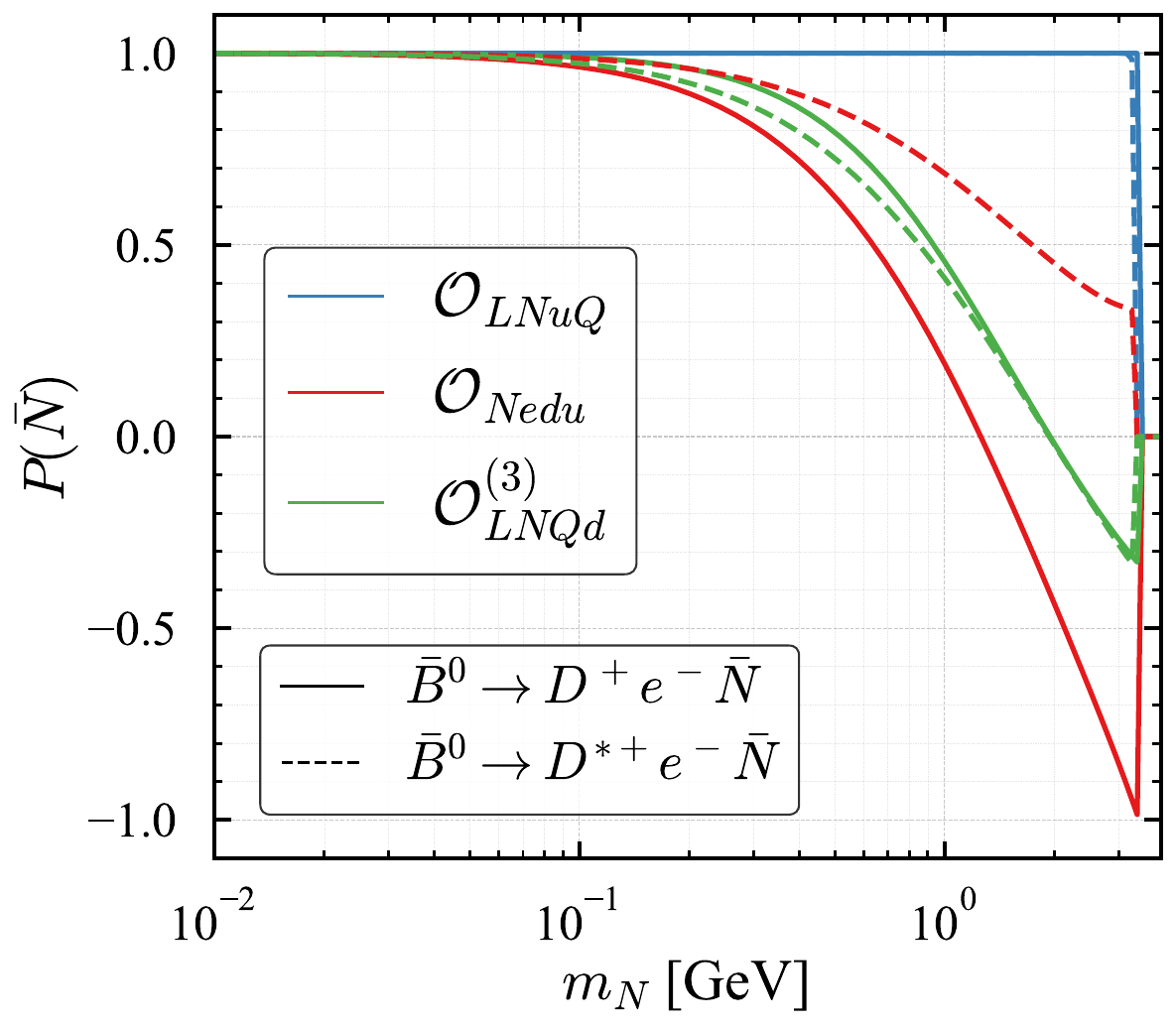}
    \includegraphics[scale=0.35]{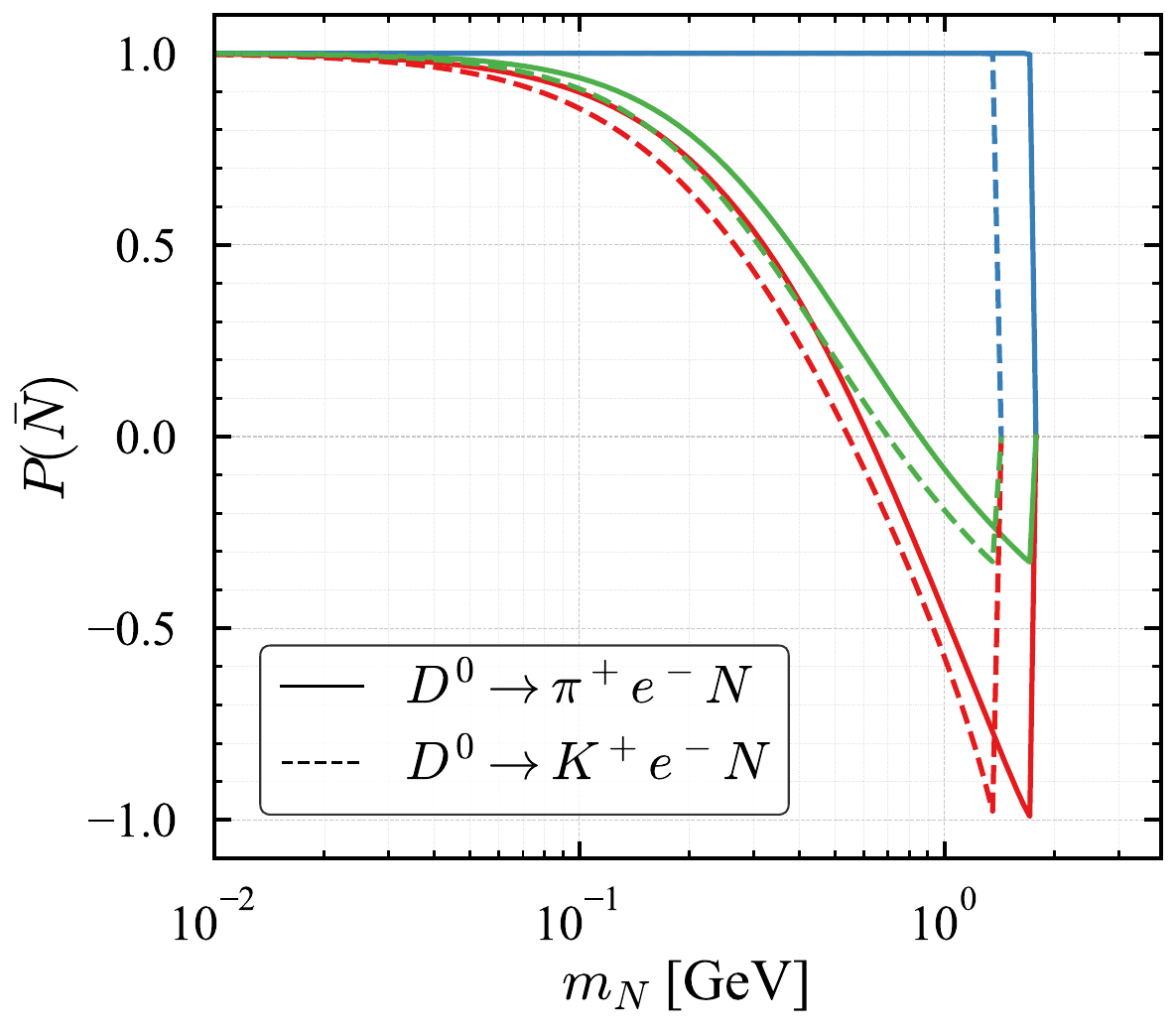}
    \includegraphics[scale=0.35 ]{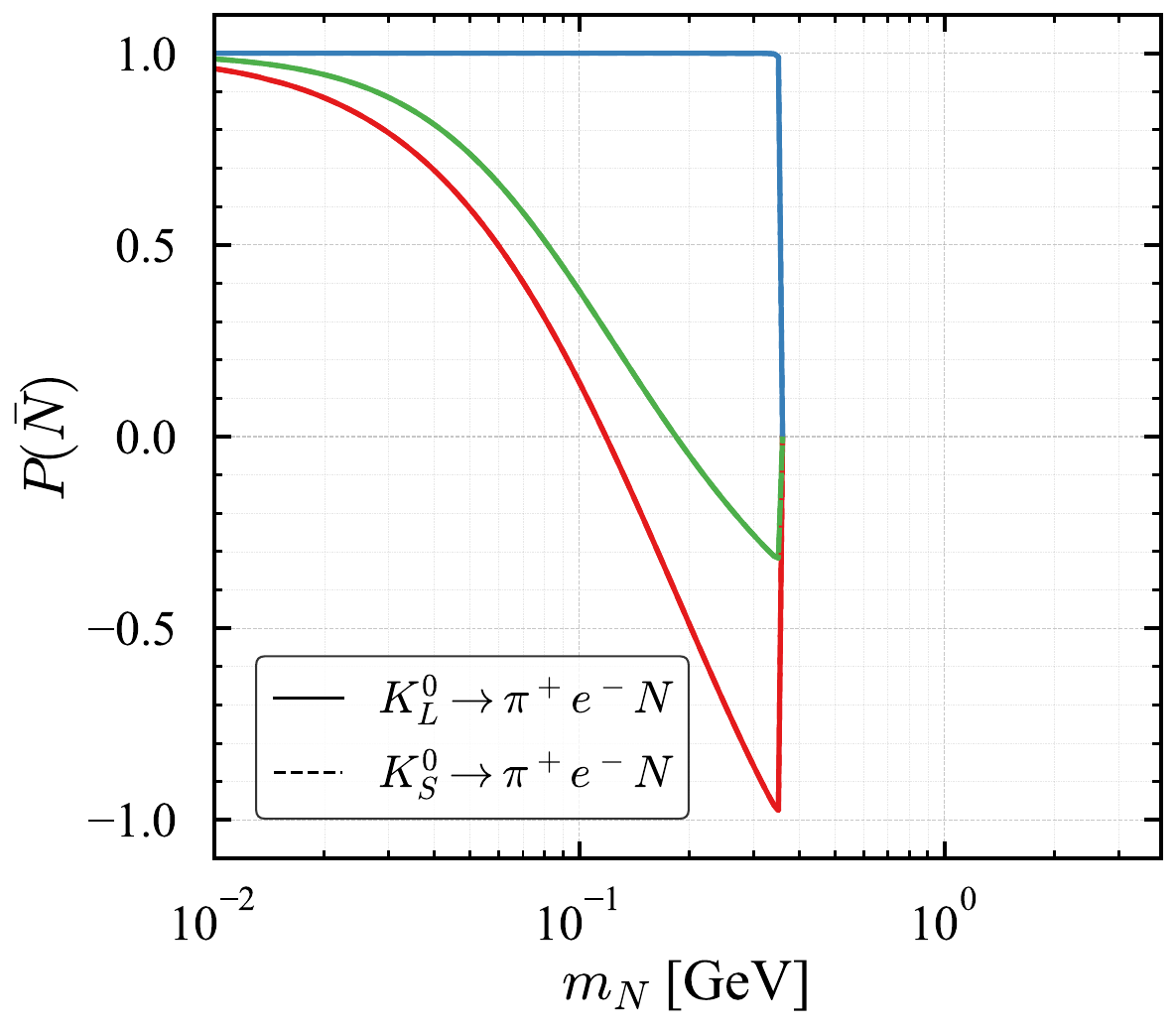}
    \caption{Polarization of $\bar{N}$ produced through three-body meson decays for $B$ (top left), $D$ (top right), and $K$ (bottom) mesons. Results are shown for $\mathcal{O}_{LNuQ}$ (blue), $\mathcal{O}_{Nedu}$ (orange), and $\mathcal{O}^{(3)}_{LNQd}$ (green). For all operators, the polarization approaches unity as $m_N$ decreases, which enhances differences in the angular distributions for Dirac and Majorana RHNs.}
    \label{fig:pol_bmeson}
\end{figure}

For two-body meson decays, across all operators considered in our analysis, the RHN polarization consistently approaches unity, so we do not display these results separately. Figure~\ref{fig:pol_bmeson} shows the polarization $P(\bar{N})$ as a function of $m_{N}$ for three-body decays of $B$, $D$, $K$ mesons for the different SMNEFT operators. The polarization depends strongly on both the RHN mass and the specific production channel. Note that for all operators the polarization approaches unity as the RHN mass decreases, since in this limit the RHN becomes effectively massless compared to the parent meson, and approaches a pure helicity eigenstate. This behavior significantly enhances the observable differences in the angular distributions of decay products, making the Dirac/Majorana discrimination easier in the low-mass regime. Conversely, as the RHN mass approaches the parent meson mass, the polarization deviates substantially from unity and approaches zero in the kinematic limit. This reduction in polarization diminishes the ability to distinguish between Dirac and Majorana RHNs, as we demonstrate below. 

\subsubsection{Constraints on Wilson coefficients}

In the absence of direct experimental constraints on long-lived RHNs in the mass range considered, we derive indirect limits on the SMNEFT Wilson coefficients by leveraging existing measurements of meson branching ratios. Our methodology requires that RHN production channels do not contribute to the total decay width of parent mesons in excess of current experimental uncertainties. We impose the condition that the branching ratio for any RHN production channel does not exceed the experimental uncertainty on the measured branching ratio of the corresponding meson. This conservative approach ensures that our predictions remain consistent with existing measurements, while allowing for potentially observable RHN signals. Table~\ref{tab:meson_branching_ratios} lists the relevant mesons, the decay channels, and their experimental branching ratios used to constrain the Wilson coefficients.

\begin{table}[t]
    \centering
    \begin{tabular}{|c|c|c|}
        \hline
        & \textbf{Decay channel} & \textbf{Branching ratio} \\
        \hline \hline

        \multirow{1}{*}{$B$ meson}
            & $B^{\pm} \to \ell^+ \nu_{\ell} X$ & $0.109 \pm 0.0028$ \\
        \hline

        \multirow{3}{*}{$D$ meson}
            & $D^{\pm} \to e^{\pm} \nu_e$& $< 8.8 \times 10^{-6}$ \\
            & $\overline{D}^0 \to K^- e^+ \nu$& $0.035 \pm 0.00017$ \\
            & $\overline{D}^0 \to \pi^- e^+ \nu$ & $(2.91 \pm 0.003) \times 10^{-3}$ \\
        \hline

        $K$ meson & $K^{\pm} \to e^{\pm} \nu_e$ & $(1.582 \pm 0.007) \times 10^{-5}$ \\
        \hline

        $\pi$ meson & $\pi^{\pm} \to e^{\pm} \nu_e$ & $(1.23 \pm 0.004) \times 10^{-4}$ \\
        \hline
    \end{tabular}
    \caption{Meson decay channels and their experimental branching ratios~\cite{ParticleDataGroup:2024cfk} used to constrain the SMNEFT Wilson coefficients. 
    }
    \label{tab:meson_branching_ratios}
\end{table}

\begin{figure}[t]
    \centering
    \includegraphics[scale=0.37]{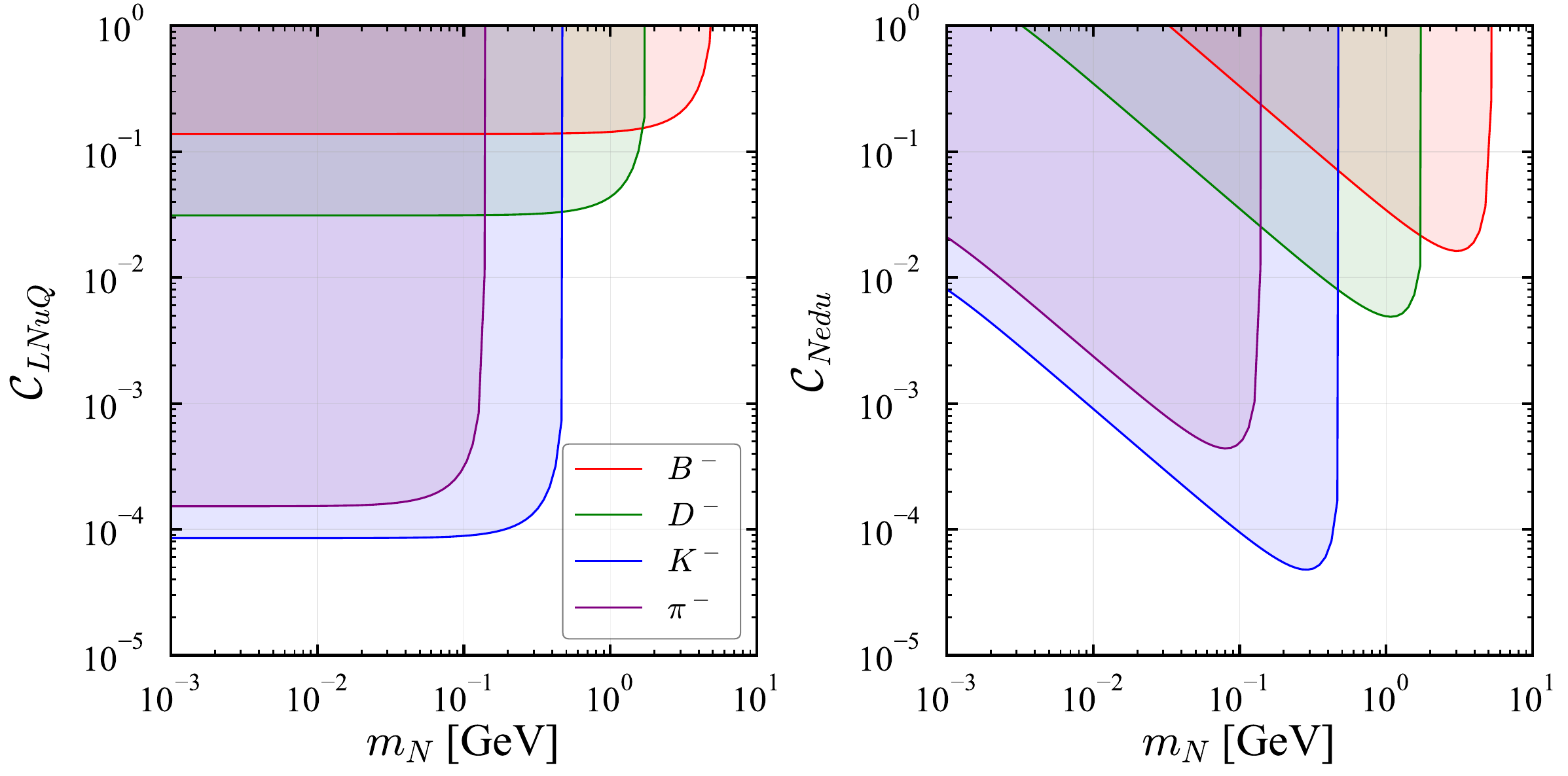}
    \caption{Constraints on the Wilson coefficients from two-body leptonic decays of $B$, $D$, $K$ and $\pi$ mesons. The shaded regions are ruled out.}
    \label{fig:wilson_coefficient_limits_2bd}
\end{figure}

\begin{figure}[t]
    \centering
    \includegraphics[scale=0.32]{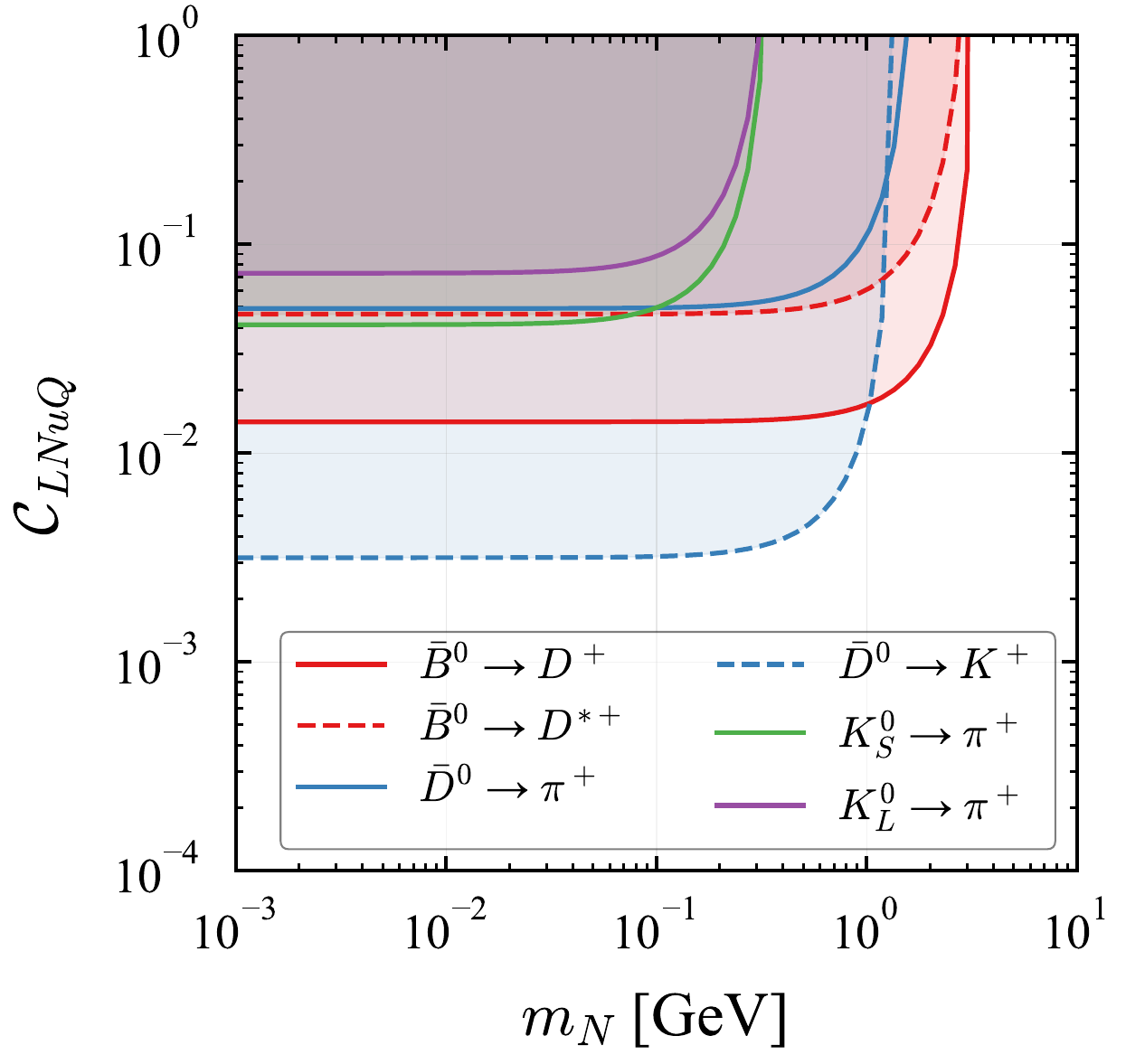}
    \includegraphics[scale=0.32]{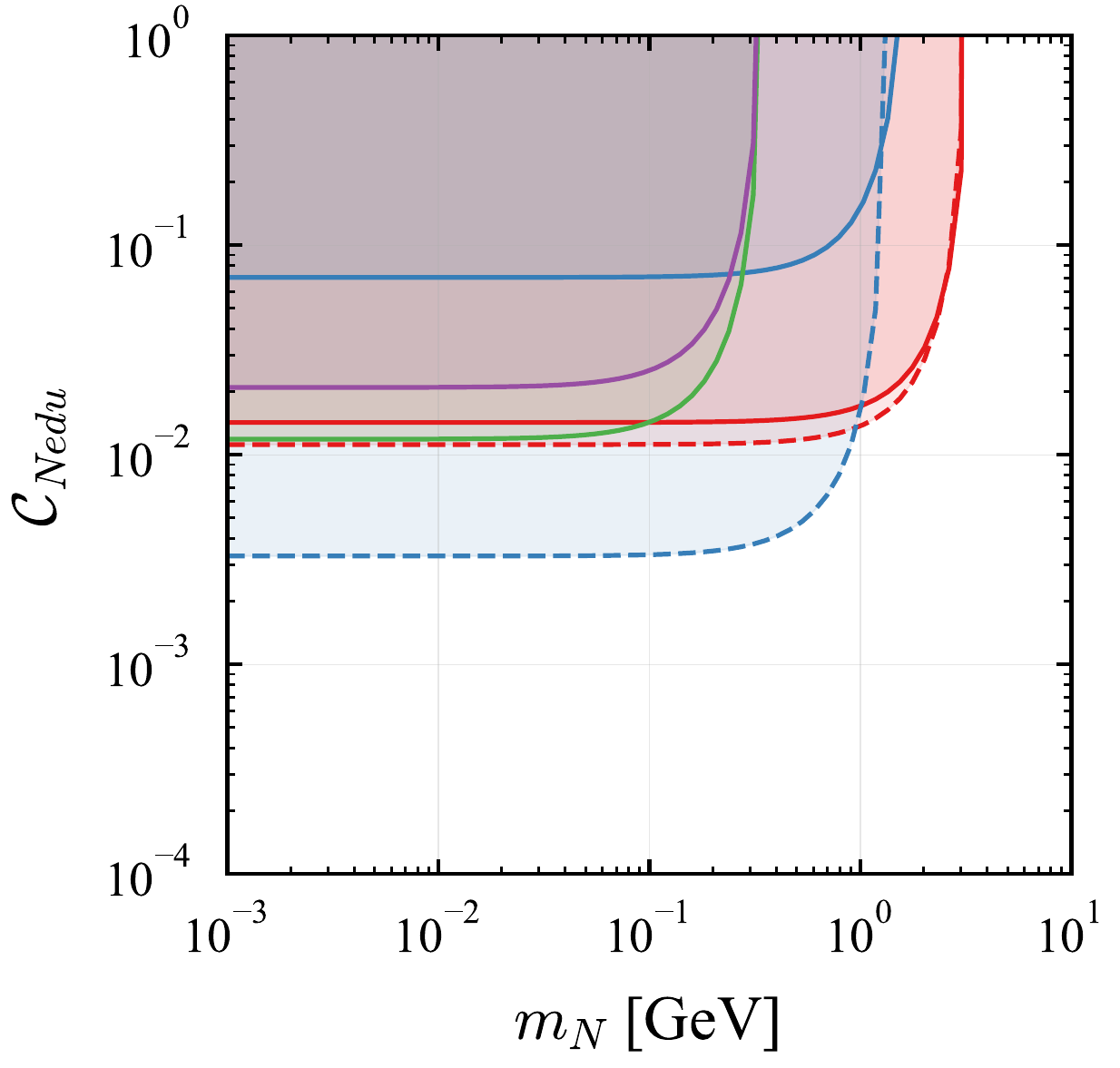}
    \includegraphics[scale=0.32]{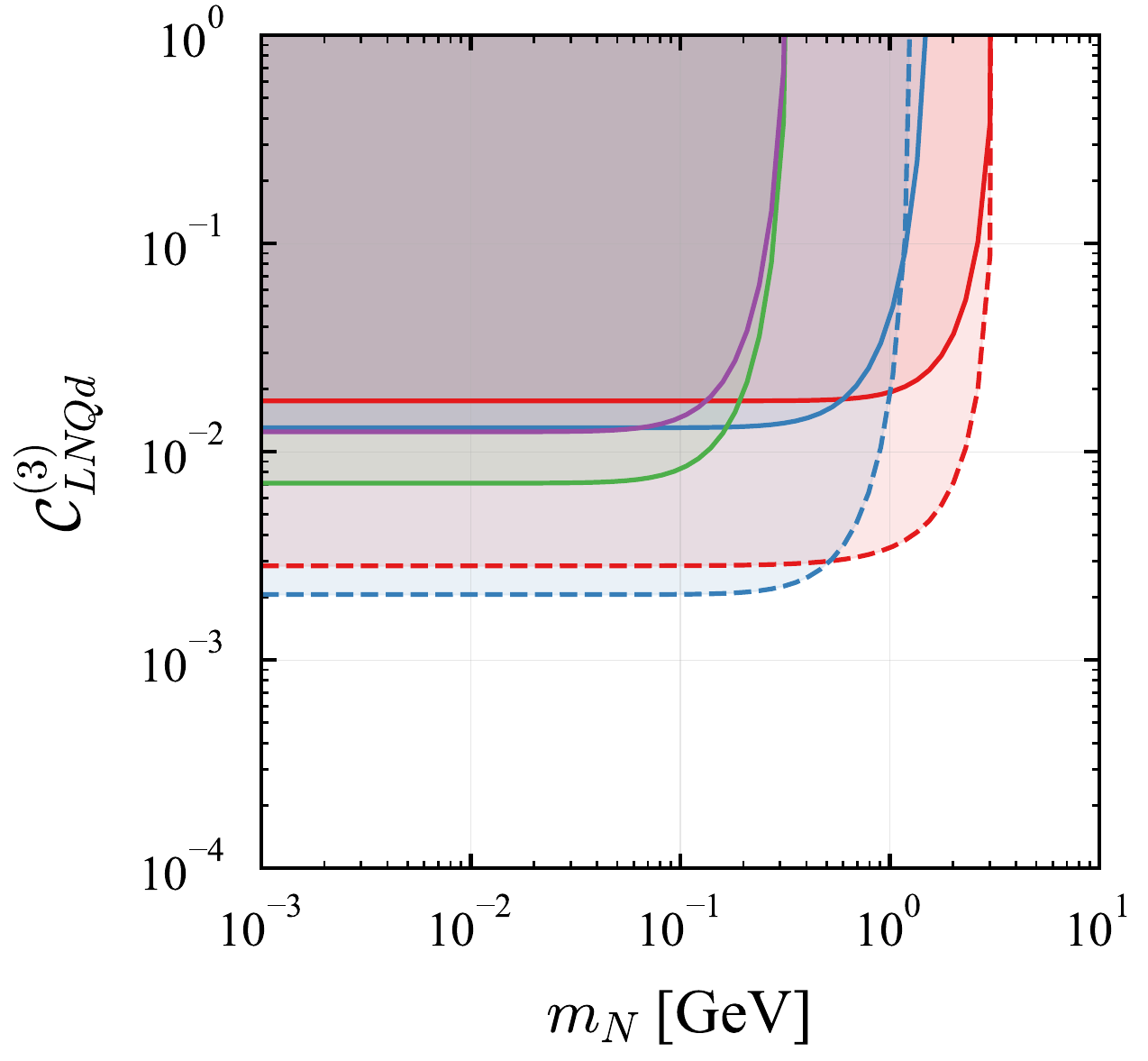}
    \caption{Constraints on the Wilson coefficients from  three-body semileptonic meson decays. The shaded regions are ruled out. }
    \label{fig:wilson_coefficient_limits_3bd}
\end{figure}

Figures~\ref{fig:wilson_coefficient_limits_2bd} and \ref{fig:wilson_coefficient_limits_3bd} illustrate the resulting constraints on the WCs from two-body and three-body meson decays respectively, where the shaded regions are ruled out. These constraints establish the parameter space in which our analysis is both theoretically motivated and experimentally accessible. 

\subsection{RHN decay}
\subsubsection{Effective operators for RHN decay}

The decays, $N \to \nu \ell^+ \ell^-$ and $N \to N^\prime \ell^+ \ell^-$, where $\nu$ is a SM neutrino and $N^\prime$ is a RHN state lighter than $N$, is governed by the effective Lagrangian,
\begin{equation}
    -\mathcal{L}_{\rm {decay}} \supset 2\sqrt{2}G_F \left( \mathcal{C}_{LNLe} \mathcal{O}_{LNLe} + \mathcal{C}_{Ne} \mathcal{O}_{Ne} + \mathcal{C}_{LN} \mathcal{O}_{LN}\right)\,.
\end{equation}
Here $\mathcal{C}$ are the Wilson coefficients that parameterize the strength of the decay operators,
\begin{align}
	\label{eq:op_lnle}
    \mathcal{O}_{LNLe} &= \left(\bar{L}_jN\right)\epsilon_{jk}\left(\bar{L}_k \ell_R\right),
    \\
	\label{eq:op_ne}
	   \mathcal{O}_{Ne} &= \left(\bar{N'}\gamma_{\mu}N\right) \left(\bar{\ell}_R\gamma^{\mu}\ell_R\right),
	   \\
	\label{eq:op_ln}
	   \mathcal{O}_{LN} &= \left(\bar{N'}\gamma_{\mu}N\right) \left(\bar{L}\gamma^{\mu}L\right)\,,
\end{align}
where $L = (\nu_L, \ell_L)^T$ is the $SU(2)_L$ lepton doublet, $\ell_R$ is the right-handed charged lepton singlet, and $N,N'$ denote the RHNs. The flavor indices are implicit in these expressions. To connect the SMNEFT operators to the low-energy regime, we employ the General Neutrino Interaction (GNI) Lagrangian. This framework encapsulates all possible Lorentz structures to describe the decay of RHNs, 
\begin{equation}
    - \mathcal{L}_{\rm{GNI}} = \sum_{i,j} \left[ G_{ij} \left(\chi \gamma_i N\right)\left(\bar{\ell}\gamma_{j}\ell\right) + \overline{G}_{ij} \left(\bar{N}\gamma_i \chi \right)\left(\bar{\ell}\gamma_{j}\ell\right)   \right] +  \rm{h.c.}\,.
\end{equation}
Here, the indices $i,j$ span the complete set of Lorentz structures: scalar (S), pseudoscalar (P), vector (V), axial vector (A), and tensor (T). The $\chi$ represents either a SM neutrino $\nu$ or a light RHN $N^\prime$ depending on the operator. The coefficients $G_{ij}$ and $\overline{G}_{ij}$ are the interaction strengths, and are generally independent of each other~\cite{deGouvea:2021ual}. We assume $G_{ij} = 0$
when matching the SMNEFT operators. The connection between the GNI framework and the SMNEFT description is established through tree-level matching at the electroweak scale, yielding~\cite{Han:2022uho}
\begin{align}
    \overline{G}_{SS} &= -\overline{G}_{SP} = -\overline{G}_{PS} = \overline{G}_{PP} = \frac{3}{8} C_{LNLe}\,,
    \\
    \overline{G}_{VV} &= \overline{G}_{AV} = \frac{1}{4} \left( C_{LN} + C_{Ne} \right)\,,
    \\
    \overline{G}_{VA} &= \overline{G}_{AA} = \frac{1}{4} \left( C_{LN} - C_{Ne} \right)\,,
    \\
    \overline{G}_{TT} &= \frac{1}{16} C_{LNLe}\,.
\end{align}
With these matching relations the decay width of RHNs can be expressed in terms of either SMNEFT Wilson coefficients or GNI coefficients.

\subsubsection{Kinematics of RHN decay}

\begin{figure}
    \centering
    \includegraphics[scale=0.33] {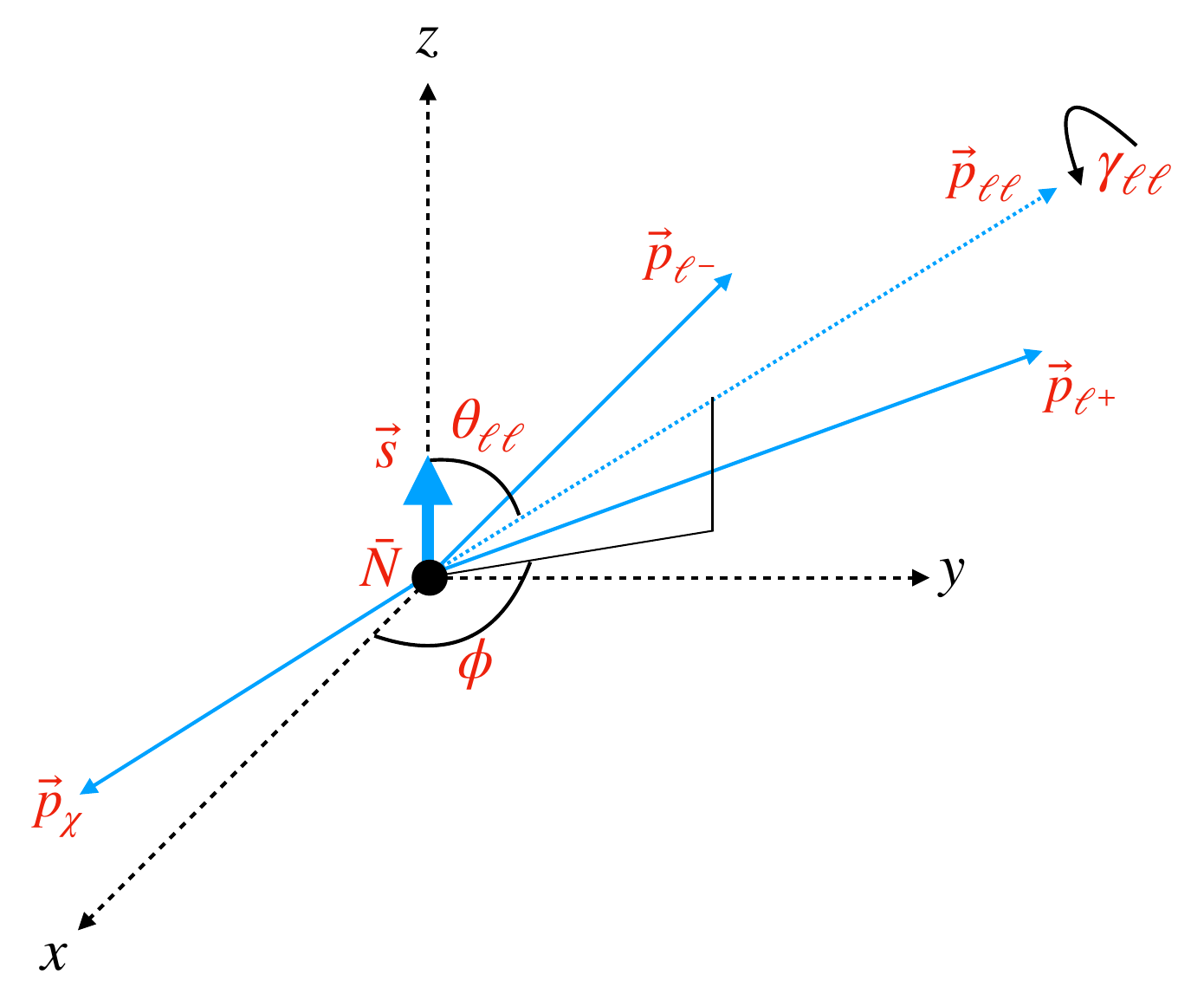}
    \caption{Schematic of the spin vector, momentum directions, and angular definitions relevant to the decay $\bar{N} \to \chi\ell^+ \ell^-$.}
    \label{fig:rhn_coordinate}
\end{figure}

We concentrate on the primary decay channel in the $m_N$ range under consideration: $\bar{N} \to \chi \ell^- \ell^+$ (where $\chi$ can be either $\nu$ or $N^\prime$). The fully differential decay width can be written in terms of invariant mass and angular variables as
\begin{equation}\label{eq:rhn_decay}
    \frac{d\Gamma \left(\bar{N} \to \chi \ell^- \ell^+ \right)}{dz_{\ell\ell} dz_{\chi \ell^-} d\cos\theta_{\ell\ell} d\gamma_{\ell\ell} d\phi} = \frac{1}{(2\pi)^5} \frac{1}{64 m_N^3} \sum_{j=1}^{13} C_j K_j\,,
\end{equation}
where $z_{\ell\ell} \equiv m_{\ell\ell}^2/m_N^2$ and $z_{\chi \ell^-} \equiv m_{\chi \ell^-}^2/m_N^2$ are the normalized invariant masses of the dilepton and neutrino-lepton systems, respectively, with e.g., the dilepton invariant mass defined as $m_{\ell\ell}^2 \equiv (p_{\ell^+} + p_{\ell^-})^2$. The coupling coefficients  $C_j$ are functions of the GNI coefficients, and  $K_j$ are Lorentz-invariant combinations of momenta and spin. Detailed expressions of $C_j$ and $K_j$ can be found in Ref.~\cite{deGouvea:2021ual}. The angular variables $\theta_{\ell\ell}$ and $\gamma_{\ell\ell}$ characterize the orientation of the dilepton system relative to the RHN spin, as illustrated in Fig.~\ref{fig:rhn_coordinate}. Specifically, 
$\theta_{\ell\ell}$ is the angle between the dilepton momentum vector $p_{\ell\ell} \equiv \vec{p}_{\ell^+}+\vec{p}_{\ell^-}$ and the z-axis, and $\gamma_{\ell\ell}$ is the angle of rotation of the lepton momenta $\vec{p}_{\ell^-}$ and $\vec{p}_{\ell^+}$ around the $\vec{p}_{\ell\ell}$ axis. Since the decay width is independent of the azimuthal angle $\phi$, we fix $\phi = \pi/2$ for simplicity in this section. The momenta of the decay products in the RHN rest frame are given by
\begin{align}
    p_{\chi} &= \left[E_{\chi} , 0, -E_{\chi}\sin\theta_{\ell\ell}, -E_{\chi}\cos\theta_{\ell\ell}\right]\,,
    \label{eq:neutrino_momentum}
    \\
    p_{\ell^-} &= \left[E_{\ell^-}, \vert \vec{p}_{\ell^-} \vert \sin\theta_{\ell\ell} \sin\gamma_{\ell\ell}, -\vert \vec{p}_{\ell^-} \vert\left(\cos\theta_{\chi \ell^-} \sin\theta_{\ell\ell} + \sin\theta_{\chi \ell^-} \cos\gamma_{\ell\ell} \cos\theta_{\ell\ell}\right), \right. \nonumber
    \\
    & \left. \vert \vec{p}_{\ell^-} \vert \left( - \cos\theta_{\chi \ell^-} \cos\theta_{\ell\ell} + \sin\theta_{\chi \ell^-} \cos\gamma_{\ell\ell} \sin\theta_{\ell\ell}\right)
    \right]\,, \label{eq:electron_momentum}
    \\
    p_{\ell^+} & = \left[m_N - E_{\chi} - E_{\ell^-} , -p_{\chi}^x - p_{\ell^-}^x, -p_{\chi}^y - p_{\ell^-}^y, -p_{\chi}^z - p_{\ell^-}^z \right]\,,\label{eq:positron_momentum}
\end{align}
where 
\begin{align}
    E_{\chi} &= \frac{m_N^2 - m_{\ell\ell}^2}{2 m_N}\,,
    \label{eq:neutrino_energy}
    \\
    E_{\ell^-} &= \frac{m_{\ell\ell}^2 +m_{\chi \ell^-}^2 + m_{\ell^+}^2}{2 m_N}\,,
    \label{eq:electron_energy}
    \\
    E_{\ell^+} &= \frac{m_{N}^2 - m_{\chi \ell^-}^2 - m_{\ell^+}^2}{2 m_N}\,,
    \label{eq:positron_energy}
    \\
    \cos \theta_{\chi \ell^-} &= \frac{E_\chi (m_N^2 - m_{\ell\ell}^2) - m_N(m_{\chi \ell^-}^2 - m_{\ell^-}^2)}{(m_N^2 - m_{\ell\ell}^2)\sqrt{E_{\ell^-}^2 - m_{\ell^-}^2}}\,.
    \label{eq:neutrino_electron_angle}
\end{align}
Two main aspects of the decay kinematics are particularly relevant:
\begin{itemize}
    \item \textbf{Angular distributions:} The angular correlations in $\cos\theta_{\ell\ell}$ and $\gamma_{\ell\ell}$ differ significantly between Dirac and Majorana RHNs. These differences arise from the operator structure of the decay and the nature of the RHN, making angular distributions a direct probe of the Dirac/Majorana character.
    \item \textbf{Energy and momentum distributions:} The available phase space and the resulting energy spectra of the decay products depend strongly on the RHN mass. This mass dependence affects both the shape and the range of kinematic variables and, as we will see, plays a crucial role in the sensitivity to Dirac/Majorana discrimination. Differences in the energy spectra between the Dirac and Majorana RHNs also contribute to their experimental distinguishability.
\end{itemize}

\begin{figure}[t]
    \centering
    \includegraphics[scale=0.43]{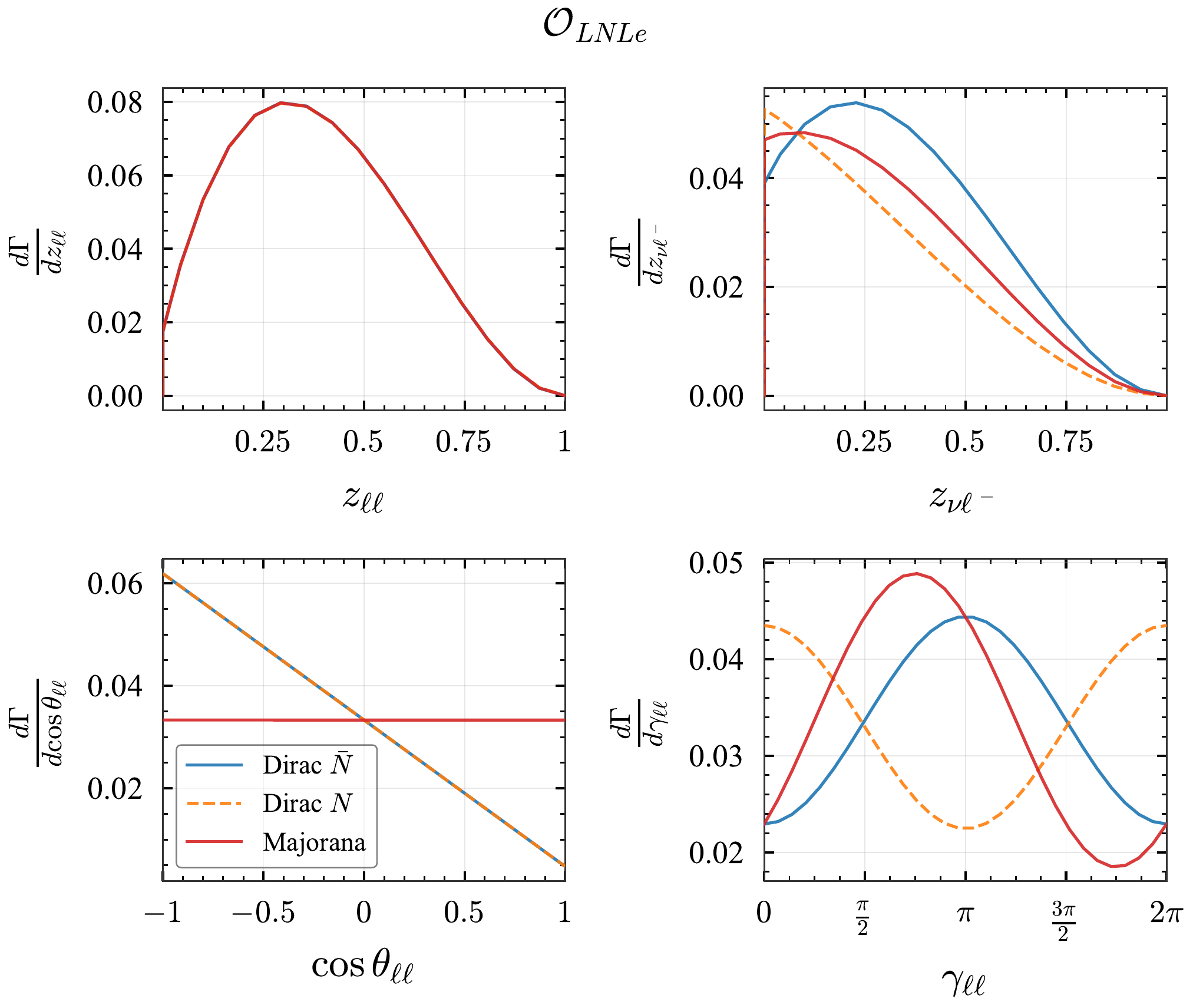}
    \caption{Differential decay distributions for $\bar{N} \to \chi e^+ e^-$ for $\mathcal{O}_{LNLe}$ and $m_N = 1$~GeV, for Dirac and Majorana RHNs with unit polarization. The panels display normalized one-dimensional distributions (d$\Gamma$/dX) for $z_{\ell\ell}$, $z_{\nu\ell^-}$, $\cos\theta_{\ell\ell}$ and 
    $\gamma_{\ell\ell}$. Dirac $N$ distributions are shown in dashed orange, Dirac $\bar{N}$ distributions are shown in solid blue, and Majorana neutrino distributions are shown in solid red.}
    \label{fig:rhn_decay_projections_lnle}
\end{figure}

\begin{figure}[t]
    \centering
    \includegraphics[scale=0.43]{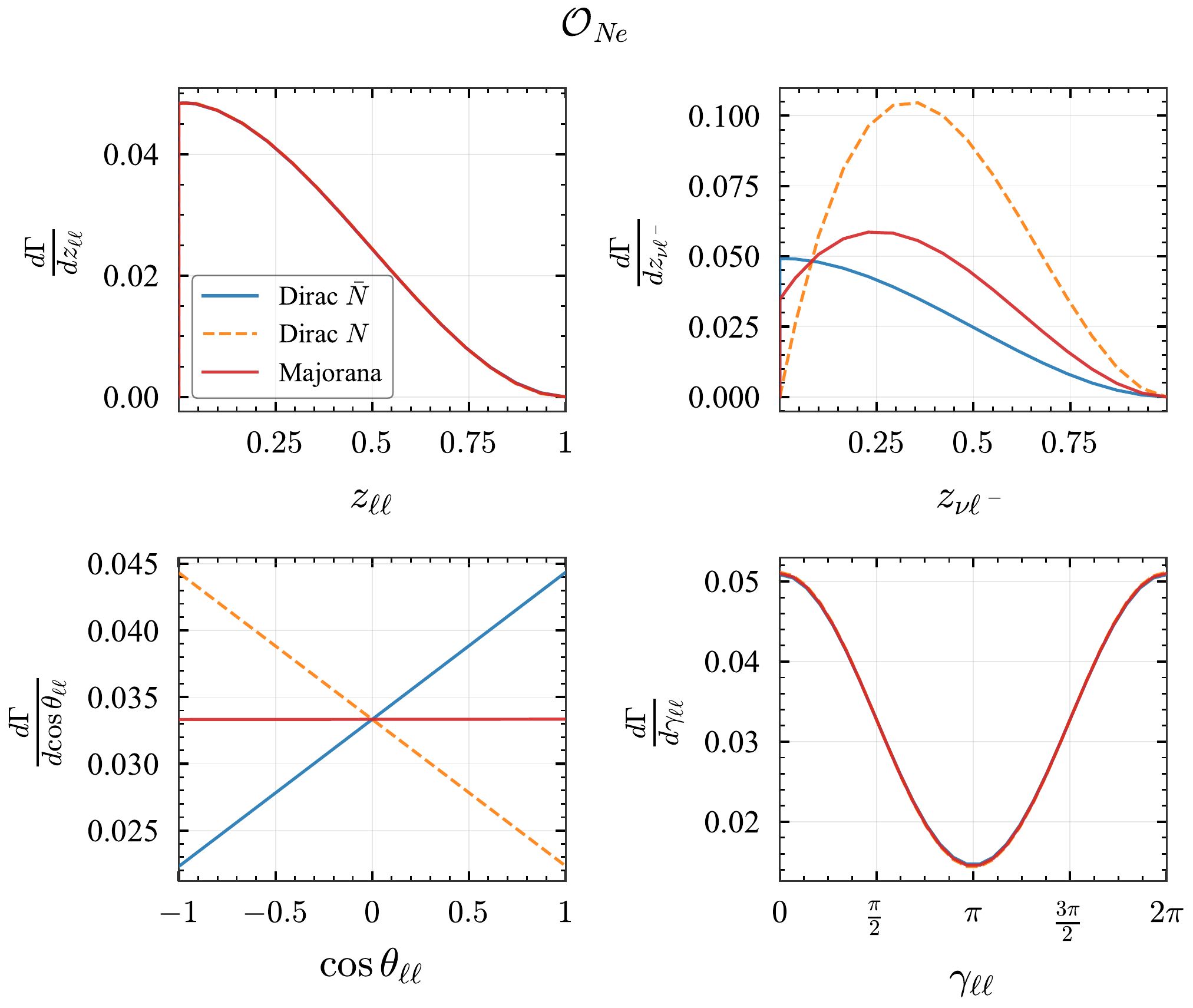}
    \caption{Similar to Fig.~\ref{fig:rhn_decay_projections_lnle}, for $\mathcal{O}_{Ne}$.}
    \label{fig:rhn_decay_projections_ne}
\end{figure}

\begin{figure}[t]
    \centering
    \includegraphics[scale=0.43]{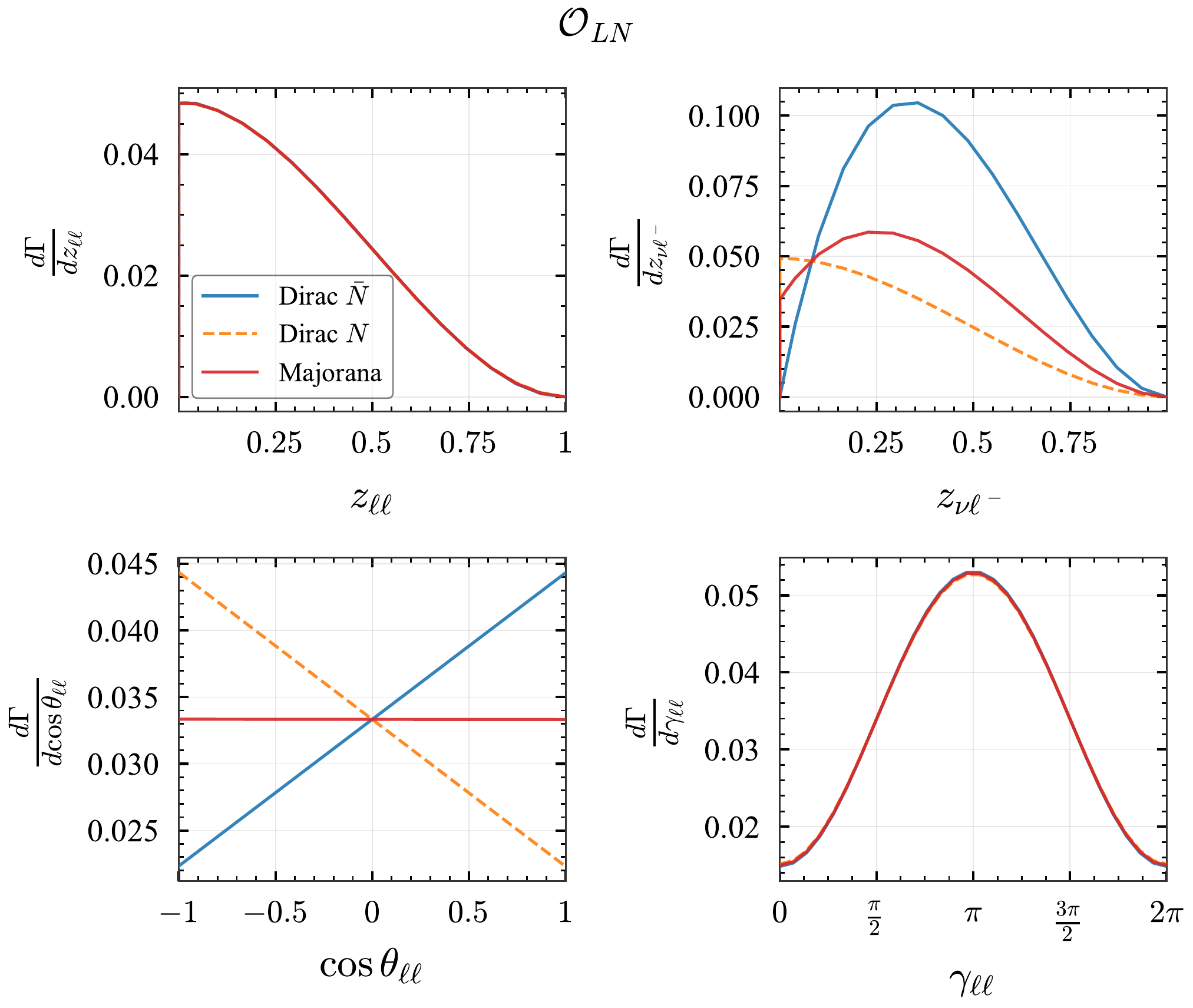}
    \caption{Similar to Fig.~\ref{fig:rhn_decay_projections_lnle}, for $\mathcal{O}_{LN}$.}
    \label{fig:rhn_decay_projections_ln}
\end{figure}

To illustrate the physical implications of these kinematic variables, in Figs.~\ref{fig:rhn_decay_projections_lnle}, \ref{fig:rhn_decay_projections_ne} and \ref{fig:rhn_decay_projections_ln} we show the relevant kinematic variable distributions for $\bar{N} \to \chi e^- e^+$ for 
$m_N = 1$~GeV, for Dirac $N$, $\bar{N}$ and Majorana $N$ for $\mathcal{O}_{LNLe}$, $\mathcal{O}_{LN}$, $\mathcal{O}_{Ne}$, respectively. We focus on four variables: $z_{\ell\ell}$, $z_{\nu \ell^-}$, $\cos\theta_{\ell\ell}$ and $\gamma_{\ell\ell}$. The panels show the one-dimensional projections (i.e., $d\Gamma/dX$ for each variable). In Eq.~(\ref{eq:rhn_decay}), the kinematic factors $K_7$ to $K_{13}$ depend on the spin and thus contribute to the asymmetry in the angular distributions. Since the polarization of the RHN affects these distributions, we incorporate the polarization calculated using Eq.~(\ref{eq:rhn_polarization_production}) by rescaling the kinematic factors $K_7$ to $K_{13}$ by $P \in [-1, 1]$. In 
Figs. \ref{fig:rhn_decay_projections_lnle}, \ref{fig:rhn_decay_projections_ne} and \ref{fig:rhn_decay_projections_ln} we assume $P = 1$ for simplicity, but we account for the polarization dependence in the final simulations. Several important features emerge from these figures:

\begin{itemize}
    \item \textbf{$z_{\ell\ell}$ distribution:} The invariant mass of the dilepton system, $z_{\ell\ell}$, peaks strongly at low values for Dirac and Majorana RHNs. This is a generic feature of three-body decays, reflecting the available phase space. The neutrino energy is proportional to $1 - z_{\ell\ell}$, so the neutrino tends to be energetic, while the $e^-$ energy is proportional to $z_{\ell\ell}$ and thus tends to be softer. The $e^+$, by energy conservation, typically carries the remaining energy, resulting in a harder spectrum similar to that of the neutrino.
    \item \textbf{$z_{\nu\ell^-}$ distribution:} Similarly, the invariant mass of the $\nu e^-$ system, $z_{\nu \ell^-}$, also peaks at low values, reinforcing the observation that the $e^-$ is generally soft in these decays.
    \item \textbf{Angular distributions ($\cos\theta_{\ell\ell}$ and $\gamma_{\ell\ell}$):} The angular variables are where the Dirac/Majorana distinction is most pronounced. For Majorana RHNs, the $\cos\theta_{\ell\ell}$ distribution is always flat, reflecting the symmetry of the decay (zero forward-backward asymmetry). For Dirac $\bar{N}$, the distribution is operator-dependent: for $\mathcal{O}_{LNLe}$, decays are relatively enhanced near $\cos\theta_{\ell\ell} = -1$, while for $\mathcal{O}_{Ne}$ and $\mathcal{O}_{LN}$, the relative enhancement is near $\cos\theta_{\ell\ell} = 1$. The 
    $\gamma_{\ell\ell}$ distribution is nearly identical for Dirac and Majorana RHNs for most operators, except for a phase shift in $\mathcal{O}_{LNLe}$, which does not significantly affect experimental sensitivity.
\end{itemize}


In summary, the kinematic structure of $\bar{N} \to \chi \ell^- \ell^+$ decay is rich and highly sensitive to both the RHN mass and the Dirac/Majorana nature of the RHN. The angular distributions, in particular, provide a robust way to distinguish between the two scenarios, and these differences are ultimately reflected in the spatial distributions of the decay products at the detector. In the following sections, we will show how these kinematic features translate into observable signatures at FASER and how they can be exploited to experimentally probe the nature of RHNs.

\section{Simulations}
\label{Section3}

We describe the FASER experiment, the simulation framework, and the methodology used to model RHN production, propagation, and decay at the LHC. We also outline the analysis strategy for distinguishing Dirac and Majorana RHNs at FASER. 

\subsection{FASER}

The Forward Search Experiment (FASER) is a dedicated detector at the LHC designed to search for long-lived particles (LLPs) produced far-forward region of the proton-proton collision region~\cite{FASER:2022hcn}. Located 480 meters downstream from the ATLAS interaction point, FASER is strategically positioned to exploit the high flux of light, weakly-coupled particles produced in the forward direction due to the large cross sections of meson production. Its compact geometry and precise instrumentation enable efficient detection of LLP decays while maintaining an exceptionally low background environment.

The FASER detector consists of a decay volume where LLPs can decay, high-resolution tracking stations for precise trajectory reconstruction, a magnetic field to separate charged tracks, and an electromagnetic calorimeter for energy measurements. This setup provides excellent spatial and timing resolution, which is crucial for identifying displaced vertices from LLP decays.

FASER is particularly well suited for studying low-mass RHNs, as the forward region at the LHC is rich in $B$, $D$, $K$, and $\pi$ mesons, which can decay to RHNs. The large flux of highly-boosted mesons in the forward direction enhances the RHN yield within the FASER detector. The low-background setting, combined with FASER's sensitivity to displaced vertices, provides a unique opportunity to study RHN decays and distinguish between Dirac and Majorana RHNs by analyzing the kinematics of their decay products. 

Several FASER and FASER2 detector configurations have been proposed. Table~\ref{tab:detector_configurations} summarizes the main parameters of the FASER detector and its upgrades (FASER2) at the Forward Physics Facility. The ``Distance" refers to the separation between the decay volume and the ATLAS interaction point, measured along the beamline. The ``Decay volume length" is the length of the region within the detector where RHNs or other LLPs can decay. The ``Detector cross section" indicates the transverse cross sectional shape and size of the detector: both FASER and the cavern-based FASER2 have circular cross sections with radii of 10~cm and 1~m, respectively, while the latest FASER2 (2025) design adopts a rectangular cross section measuring 3~m in width and 1~m in height. "Magnetic field" specifies the field strength (in Tesla) and the corresponding region along the beam axis ($z$-axis) where it is applied. In the cavern-based FASER2 setup, the magnetic field spans a 20~m region ($z=0$~m to 20~m), fully covering the 10~m decay volume and extending into downstream detector components, to the final tracking station and the end calorimeter located at $z = 20.5$~m. In contrast, the FASER2 (2025) configuration features a localized 4~m magnetic region ($z=11.5$~m to 15.5~m), positioned after the decay volume but before the final tracking and calorimetry stations. Finally, the "Integrated luminosity" is the total expected proton-proton luminosity to be recorded at each detector location over the course of LHC operation. We focus on the most recent ``FASER2 (2025)" configuration~\cite{Salin:2927003}, but we also comment on how different detector designs could affect the ability to discriminate between Dirac and Majorana RHNs. 

\begin{table}[t]
    \small
    \centering
    \begin{tabular}{|l|c|c|c|c|c|}
        \hline
        \textbf{} & \textbf{Distance} & \textbf{Decay} & \textbf{Detector} & \textbf{Magnetic} & \textbf{Integrated} \\
        \textbf{} & \textbf{(m)} & \textbf{ volume} & \textbf{cross} & \textbf{field (T)} & \textbf{luminosity} \\
        \textbf{} & \textbf{} & \textbf{length (m)} & \textbf{section} & \textbf{} & \textbf{} \\
        \hline
        FASER & 480 & 1.5 & Circular, & 0.6 & 150 fb$^{-1}$ \\
         & & & 10~cm radius & & \\
        \hline
        FASER2 & 620 & 10 & Circular, & 1.0& 3 ab$^{-1}$ \\
        (Cavern) & & & 1~m radius&  ($z$ = 0 $\to$ 20~m) & \\
        \hline
        FASER2 & 620 & 10 & Rectangular, & 1.0& 3 ab$^{-1}$ \\
        (2025)& & & 3 m $\times$ 1 m & ($z$ = 11.5 $\to$ 15.5~m) & \\
        \hline
    \end{tabular}
    \caption{Comparison of the main detector parameters for FASER and its planned upgrade FASER2 at the Forward Physics Facility.  }
    \label{tab:detector_configurations}
\end{table}

\subsection{RHN production and propagation}

We simulate the complete chain of RHN phenomenology at the LHC, from meson production in the forward region, through RHN generation and propagation, to decay within the FASER detector and signal reconstruction. The approach uses the FORESEE framework~\cite{Kling:2021fwx}, with additional optimization using NuMojo~\cite{numojo} and HepJo~\cite{hepjo} tailored for this analysis.

To accurately model RHN production, we use hadron production spectra from FORESEE~\cite{Kling:2021fwx}, which include detailed meson kinematics and cross section data in the forward region. We then simulate the decay of these mesons to RHNs, incorporating the relevant branching ratios, form factors, and operator-dependent kinematics. This step accounts for theoretical uncertainties from hadronic physics and modelling of the forward region.

\begin{figure}[t]
    \centering
    \includegraphics[scale=0.35]{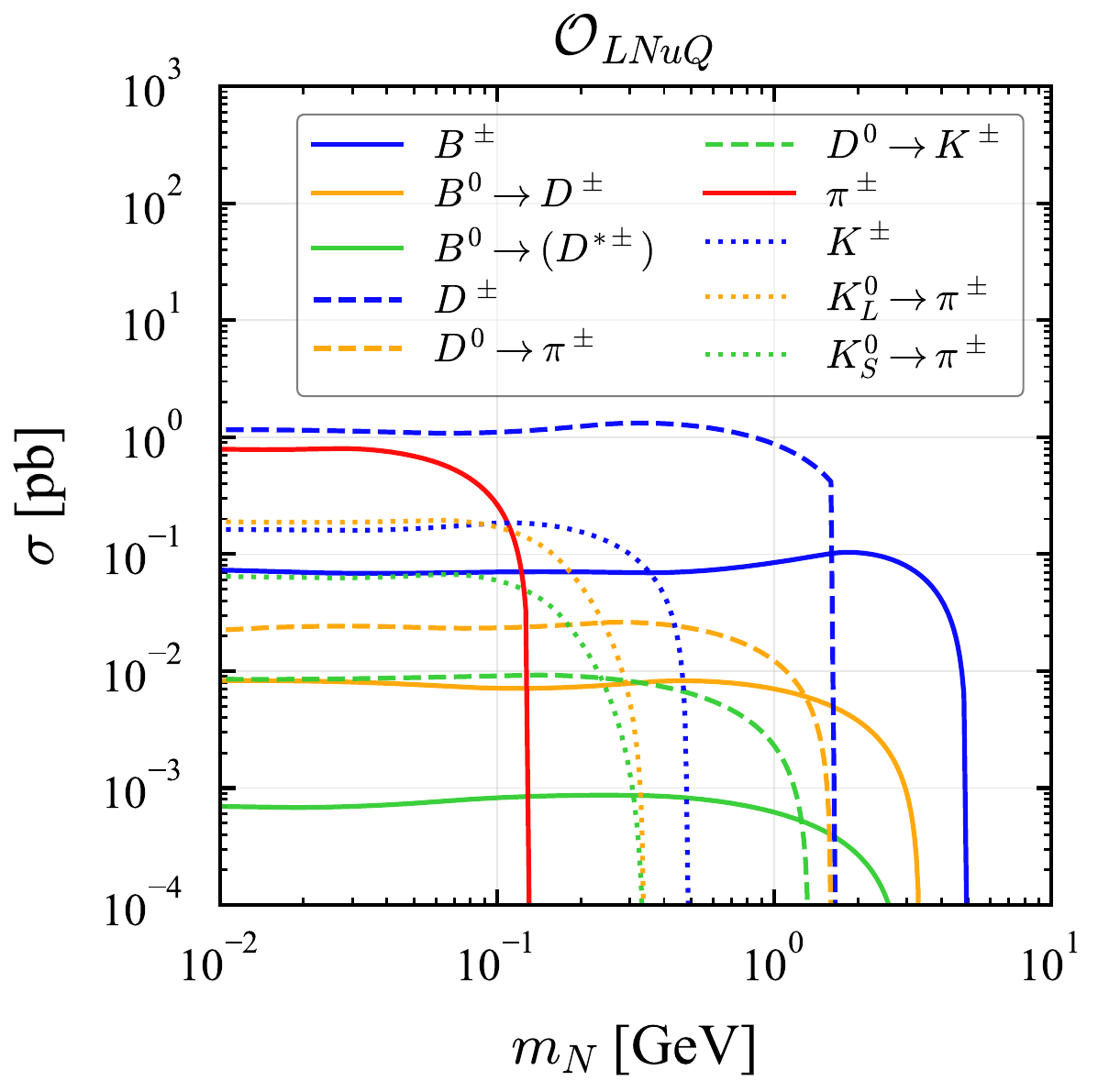}
    \includegraphics[scale=0.35]{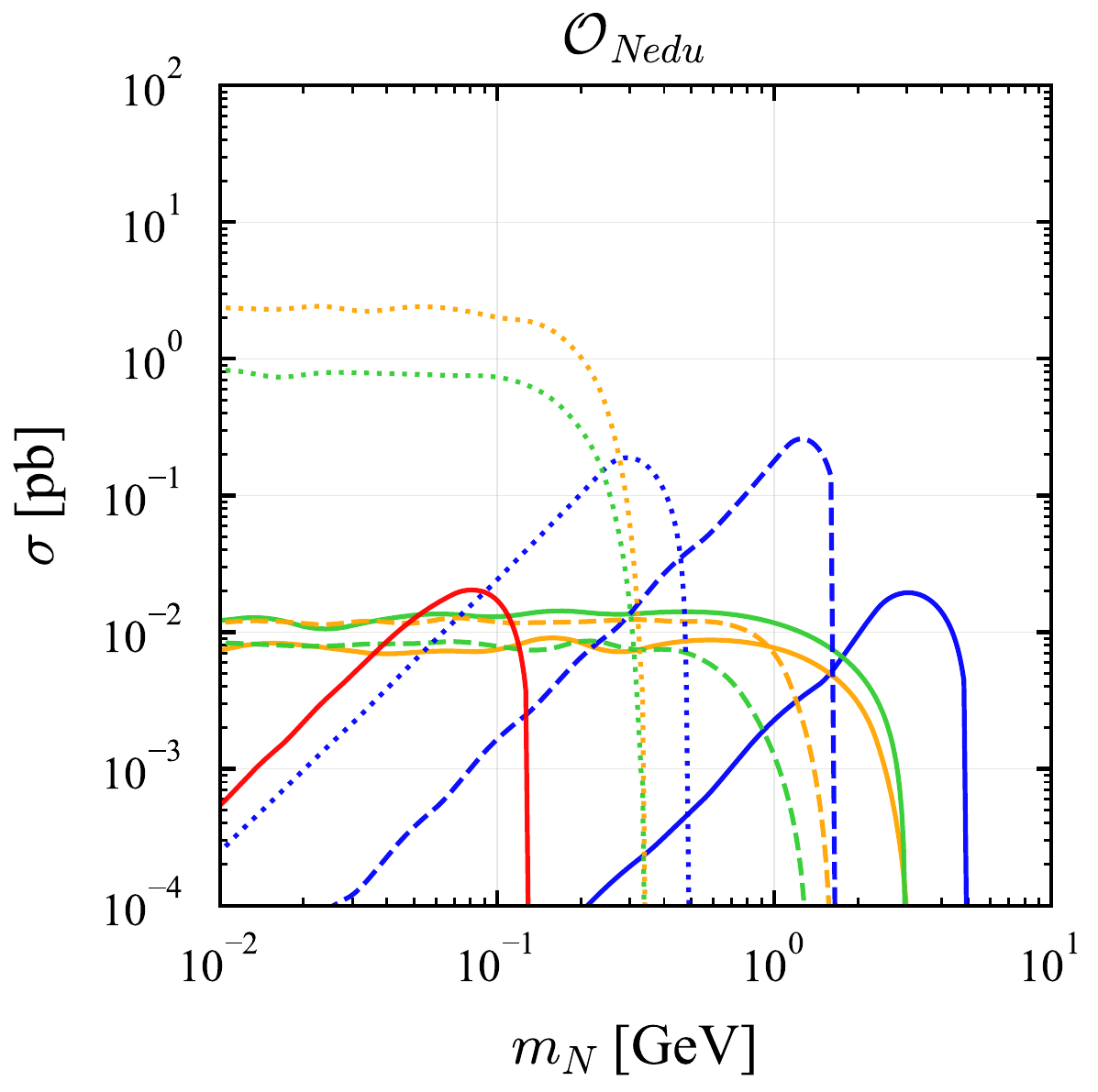}
    \includegraphics[scale=0.4]{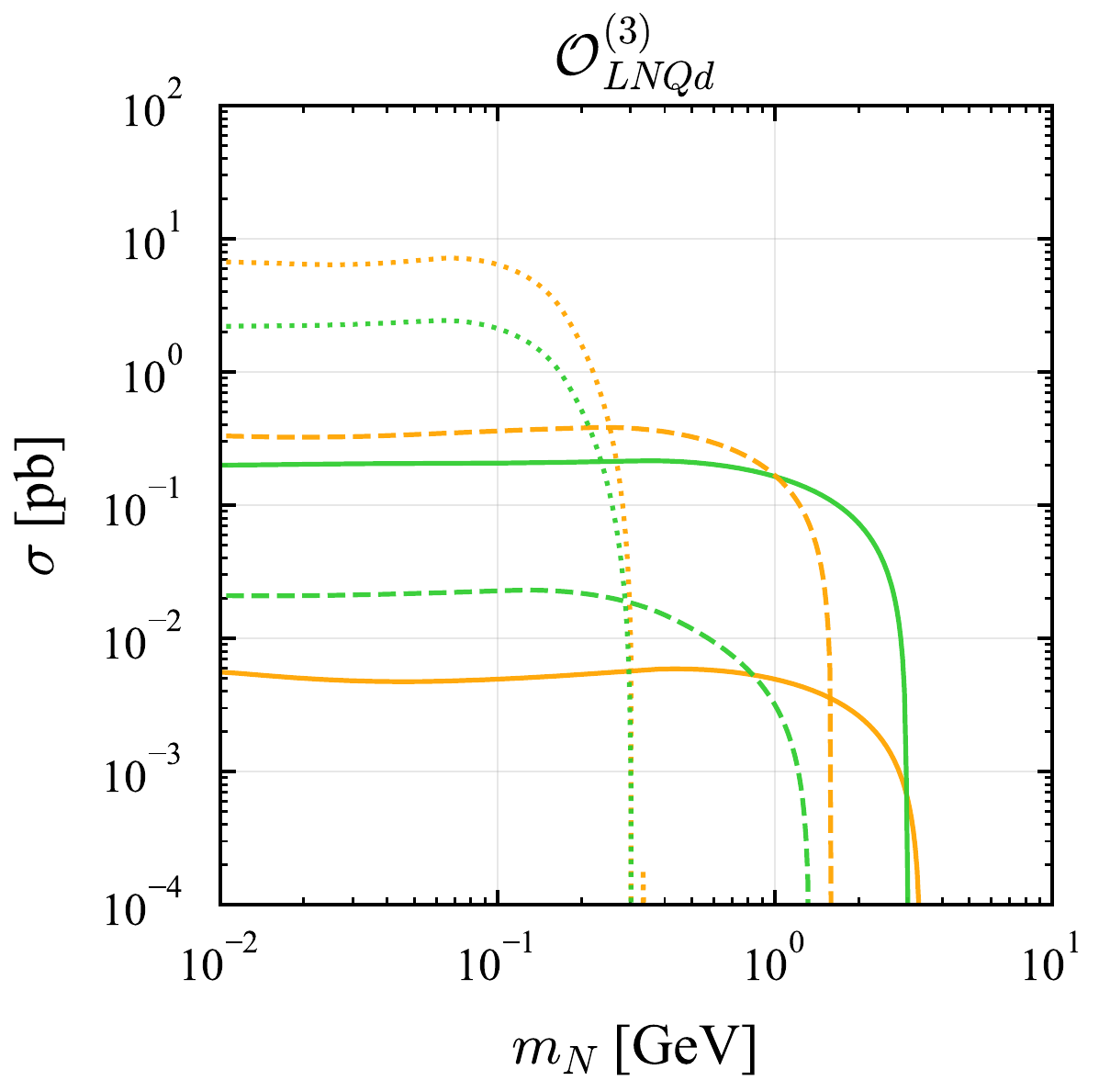}
    \caption{Predicted production cross sections for Dirac 
    $\bar{N}$ via meson decays at the LHC as a function of RHN mass for $\mathcal{O}_{LNuQ}$ (top left), $\mathcal{O}_{Nedu}$ (top right), and $\mathcal{O}^{(3)}_{LNQd}$ (bottom). The Wilson coefficients are set to their maximum allowed experimental values.}
    \label{fig:production_cross_sections}
\end{figure}

Figure~\ref{fig:production_cross_sections} shows the predicted production cross sections for Dirac $\bar{N}$ at the LHC as a function of the RHN mass. For Majorana RHNs, the production cross section is exactly twice that of the Dirac $\bar{N}$. These results incorporate the constraints on Wilson coefficients discussed earlier. As expected, the cross sections closely follow the behavior of the branching ratios, with saturated production rates for lighter RHNs and operator-specific variations. 

It is important to note that there are substantial theoretical uncertainties in the absolute production rates, arising from the choice of hadronic event generator (e.g., Pythia8~\cite{Sjostrand:2014zea}, EPOSLHC~\cite{Pierog:2013ria}), as the forward region is not well understood. While these uncertainties are not shown in Fig.~\ref{fig:production_cross_sections} for clarity, they are fully accounted for in our statistical analysis.

After production, RHNs are propagated toward the FASER detector. Only RHNs that are sufficiently long-lived and have the appropriate kinematics to reach the detector volume are considered in the subsequent analysis. The propagation accounts for the RHN's lifetime, boost factor, and angular distribution.

\subsection{RHN decay and detector simulation}

Once produced and propagated toward the FASER detector, RHNs may decay within the detector volume via the process $\bar{N} \to \chi e^+ e^-$. To accurately model the experimental signatures, we simulate these decays using the full three-body kinematics, incorporating the effects of the relevant SMNEFT operators on both angular and energy distributions. The simulation also accounts for the polarization of the RHN, which plays a crucial role in shaping the angular distributions of the decay products.

The FASER2 detector is modeled as a 10-meter-long decay volume, equipped with high-resolution tracking stations and a magnetic field (1~T in the latest plans). We fix the $z$-axis to be along the beam axis of the FASER detector. For each RHN decay occurring within the detector, we simulate the trajectories of the $e^+$ and $e^-$, including the effects of magnetic bending, and record their intersection points with the final tracking station located along 
the $z$-axis at $z = 20.5$~m. This enables the reconstruction of the vertical separation between the electron and positron tracks and their displacements from the beam axis, which are crucial elements of our analysis.

We employ a fixed binning scheme for all distributions, with bin widths chosen to be larger than the spatial resolution of the FASER tracking system, approximately $\mathcal{O}(100)\,\mu$m. We divide the range $x = 10^{-3}$~m to $x = 3$~m into 30 logarithmically uniform bins. While Gaussian smearing could be applied to account for finite resolution in the bins with a small width, we find its impact to be negligible since event rates in these regions are extremely low. 


\section{Results}
\label{Section4}

\subsection{Distinguishing Dirac and Majorana RHNs}

We present the main results of our analysis, focusing on how spatial observables in the FASER detector can be used to distinguish between Dirac and Majorana RHNs. Our approach centers on three spatial observables (defined along the $x$-axis), each measured at the final tracking station located at $z = 20.5$~m from the front of the FASER2 detector.
\begin{itemize}
    \item {The total horizontal separation (along the $x$-axis) between the electron and positron tracks $X(e^+ - e^-) $, which reflects the combined displacement of the two leptons in the $x$-direction}.
    \item The horizontal displacement of the electron track from the beam axis $X(e^-)$.
    \item The horizontal displacement of the positron track from the beam axis $X(e^+)$.
\end{itemize}
All three distances are defined to be positive.
These observables are particularly sensitive to the underlying angular distributions of the decay products, which differ for Dirac and Majorana RHNs. The magnetic field $B_y$ of the FASER detector separates the $e^+$ and $e^-$ tracks along the $x$-axis, making this direction optimal for resolving the spatial differences between the two cases. These laboratory-frame observables are directly measurable at FASER unlike analyses in the RHN rest frame, such as those performed in the context of Belle~II~\cite{Han:2022uho}.  

\begin{figure}[t]
    \centering
    \includegraphics[scale=0.3]{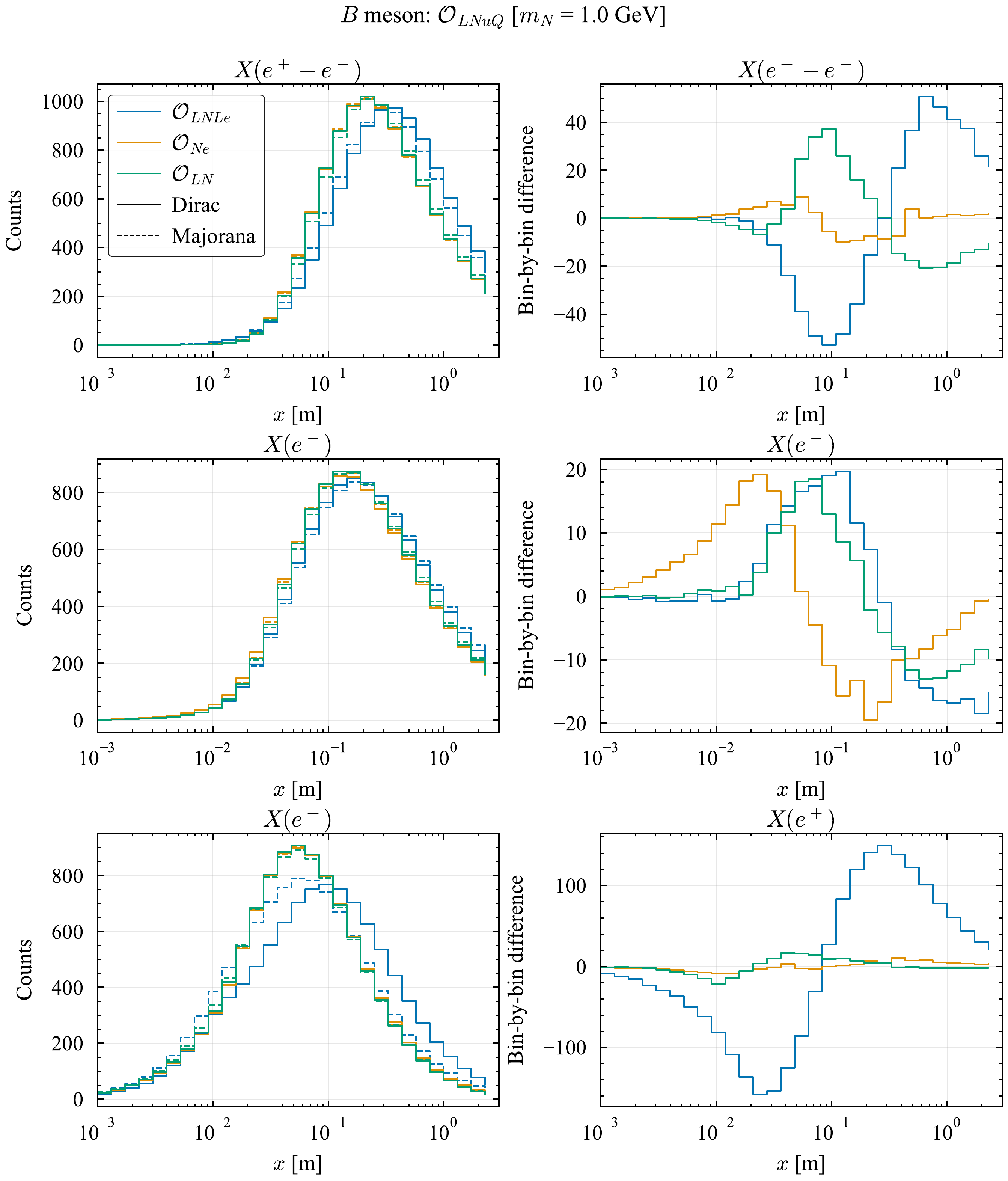}
    \caption{Comparison of spatial distributions for three observables: the total electron-positron separation $X(e^+ - e^-)$ (top), electron displacement $X(e^-)$ (middle), and positron displacement $X(e^+)$ (bottom) for
    $B$ mesons produced via $\mathcal{O}_{LNuQ}$ and $m_N = 1.0$~GeV. $x$ is the horizontal displacement from the beam axis. The left panels show event counts for Dirac (solid lines) and Majorana (dashed lines) RHNs, while the right panels display the bin-by-bin difference between the Dirac and Majorana event distributions.  The results are based on $10^4$ simulated events. }
    \label{fig:combined_vertical_distance}
\end{figure}

Figure~\ref{fig:combined_vertical_distance} shows a representative example of these spatial distributions for $m_N = 1$~GeV for $B$ mesons produced via $\mathcal{O}_{LNuQ}$. The left panels show the event counts for the Dirac (solid lines) and Majorana (dashed lines) RHNs, while the right panels show their corresponding bin-by-bin differences computed as Dirac minus Majorana. Given that the combined vertical distribution $X(e^+ - e^-)$ and the electron distribution $X(e^-)$ often exhibit less sensitivity to distinguish Dirac and Majorana RHNs compared to the positron distribution $X(e^+)$, we focus primarily on the positron distributions to streamline the discussion. However, for completeness, we present the $\chi^2$ results for all three $X$ distributions to enable a comprehensive comparison. 

From Fig.~\ref{fig:combined_vertical_distance}, we observe that the amplitude of the difference between Dirac and Majorana distributions is largest for $ X(e^+)$ compared to $X(e^-)$ and
$X(e^+ - e^-)$. 
The reduced sensitivity of electron distributions arises from the complex dependence of the electron momentum on multiple angular variables, as evident from Eq.~(\ref{eq:electron_momentum}). The electron momentum depends on various angular variables and the interference between these contributions effectively dilutes any information distinguishing Dirac from Majorana RHNs that originates from differences in the $\cos\theta_{\ell\ell}$ distributions. In contrast, the positron distribution exhibits enhanced sensitivity due to its more direct correlation with the angular distribution of the decay products. This correlation becomes particularly pronounced for $m_N = 1$ GeV, where the neutrino typically carries significantly more energy than the electron. To understand this behavior, we examine the momentum equations in the RHN rest frame. For $m_N = 1$~GeV, the electron is typically soft ($E_{e^-} \ll E_\nu$) so that the positron momentum simplifies to
\begin{align}
    p_{e^+} &\approx \left[m_N - E_{\nu}, 0, E_{\nu} \sin\theta_{\ell\ell},  E_{\nu} \cos\theta_{\ell\ell} \right]\,.
\end{align}
Since the positron's longitudinal momentum scales as $p_{e^+}^z \propto \cos\theta_{\ell\ell}$, it directly probes the angular asymmetry in RHN decay. When the boosted positrons propagate through the FASER detector's magnetic field, this angular information is preserved and manifests as spatial separation in the $x$-direction at the tracking stations. 

Thus, the momentum distribution of positrons ($p_{e^+}$) along the beam axis provides a direct window into the $\cos\theta_{\ell\ell}$ distribution, revealing differences between Dirac and Majorana RHNs. In the Dirac case, $\mathcal{O}_{LNLe}$ produces a strongly asymmetric $\cos\theta_{\ell\ell}$ distribution, while the Majorana case always yields a flat distribution as shown in the Fig.~\ref{fig:rhn_decay_projections_lnle}. To understand the shape of the positron distributions for $\mathcal{O}_{LNLe}$, consider the following: Because the dileptons produced from RHN decay are boosted along the beam axis, there are more events at $\cos\theta_{\ell\ell} = -1$ (corresponding to larger values of $x$ at the tracker) compared to $\cos\theta_{\ell\ell} = 1$ (closer to the beam axis and hence smaller values of $x$ at the tracker) for Dirac RHNs, while the Majorana case maintains a flat distribution across all $\cos\theta_{\ell\ell}$ values (and hence across the $x$-axis at the tracker). 
Consequently, when we examine the difference between Dirac and Majorana distributions in the bottom right panel of Fig.~\ref{fig:combined_vertical_distance}, we observe that at smaller values of $x$ (closer to the beam axis), the Dirac distribution has fewer events than the Majorana distribution, resulting in a dip in the difference. Conversely, at larger values of $x$ (away from the beam axis), the Dirac distribution has more events than the Majorana distribution, resulting in a peak in the difference. This explains the substantial modulation in the spatial distribution of positrons. 

For $\mathcal{O}_{Ne}$ and $\mathcal{O}_{LN}$, the Dirac and Majorana distributions are more similar because the corresponding $\cos\theta_{\ell\ell}$ distributions are similar. An important consideration is that in the Dirac case, $N$ and $\bar{N}$ are produced with $\cos\theta_{\ell\ell}$ distributions with slopes of opposite signs.  When these contributions are combined, they partially cancel, resulting in a distribution that more closely resembles the flat Majorana distribution. The cancelation in the spatial distribution is not complete because as shown in Fig.~\ref{fig:rhn_decay_projections_lnle}, the 
\(z_{\nu\ell^-}\) distributions differ between Dirac \(N\) and \(\bar{N}\), resulting in distinct momentum spectra of daughter particles from their decays. 
As these particles propagate through the magnetic field and reach the tracking region, the  differences in momentum lead to corresponding variations in the spatial distributions. As a result, the combined spatial distributions of Dirac \(N\) and \(\bar{N}\) appear nearly flat in Fig.~\ref{fig:combined_vertical_distance}, but retain small deviations due to these residual effects.

\begin{figure}[t]
    \centering
    \includegraphics[scale=0.245]{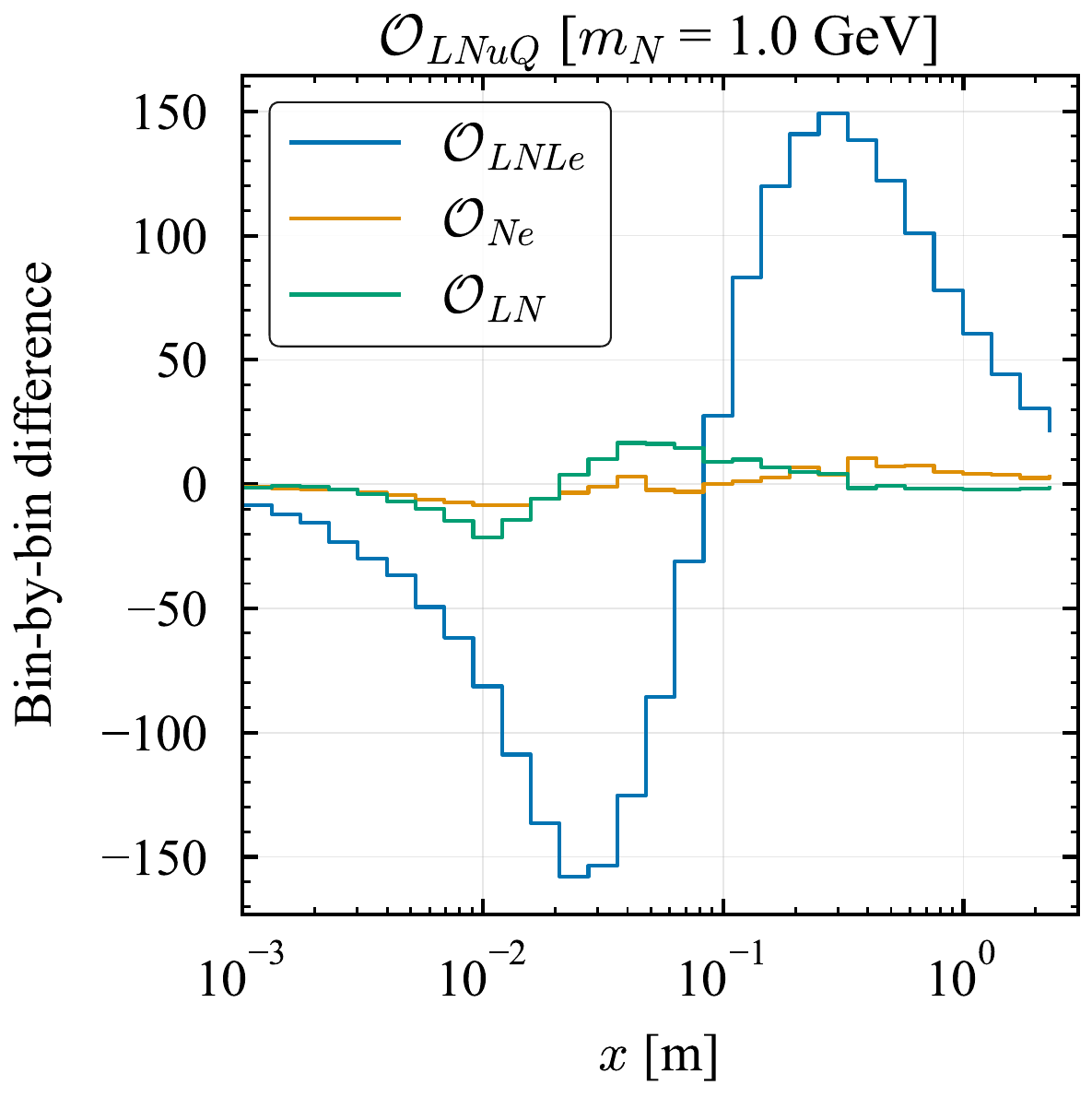}
    \includegraphics[scale=0.245]{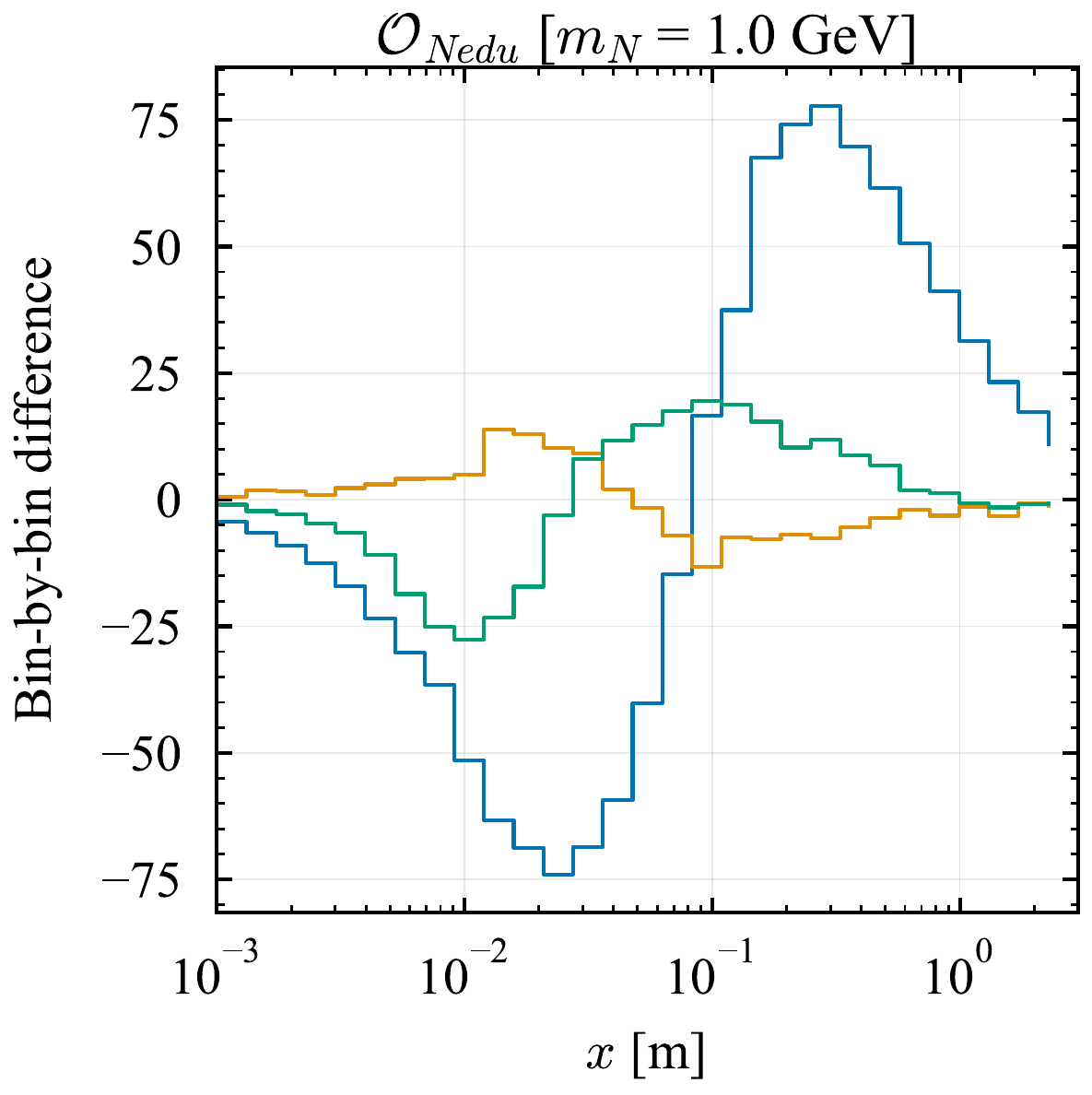}
    \includegraphics[scale=0.245]{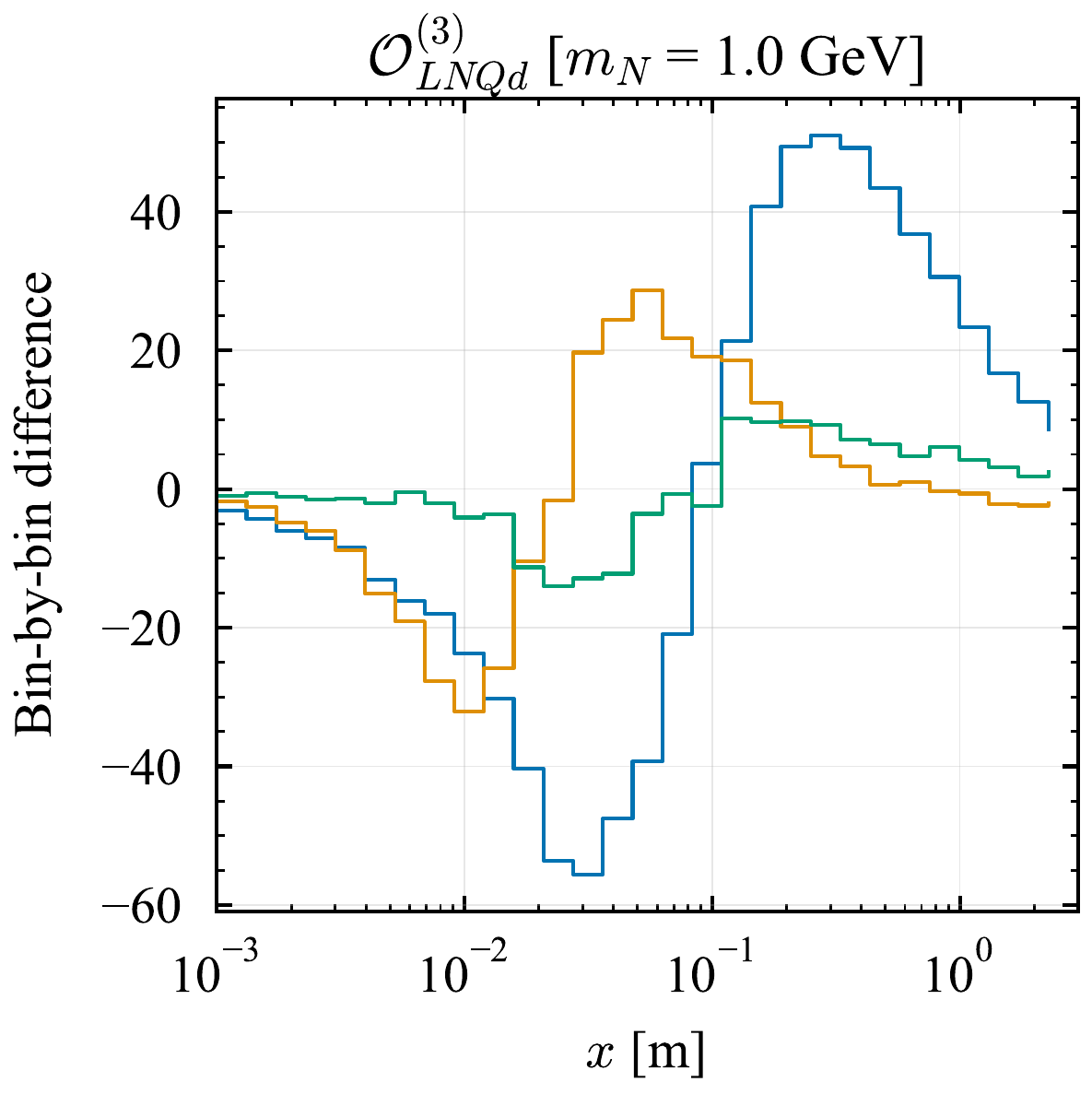}

    \includegraphics[scale=0.245]{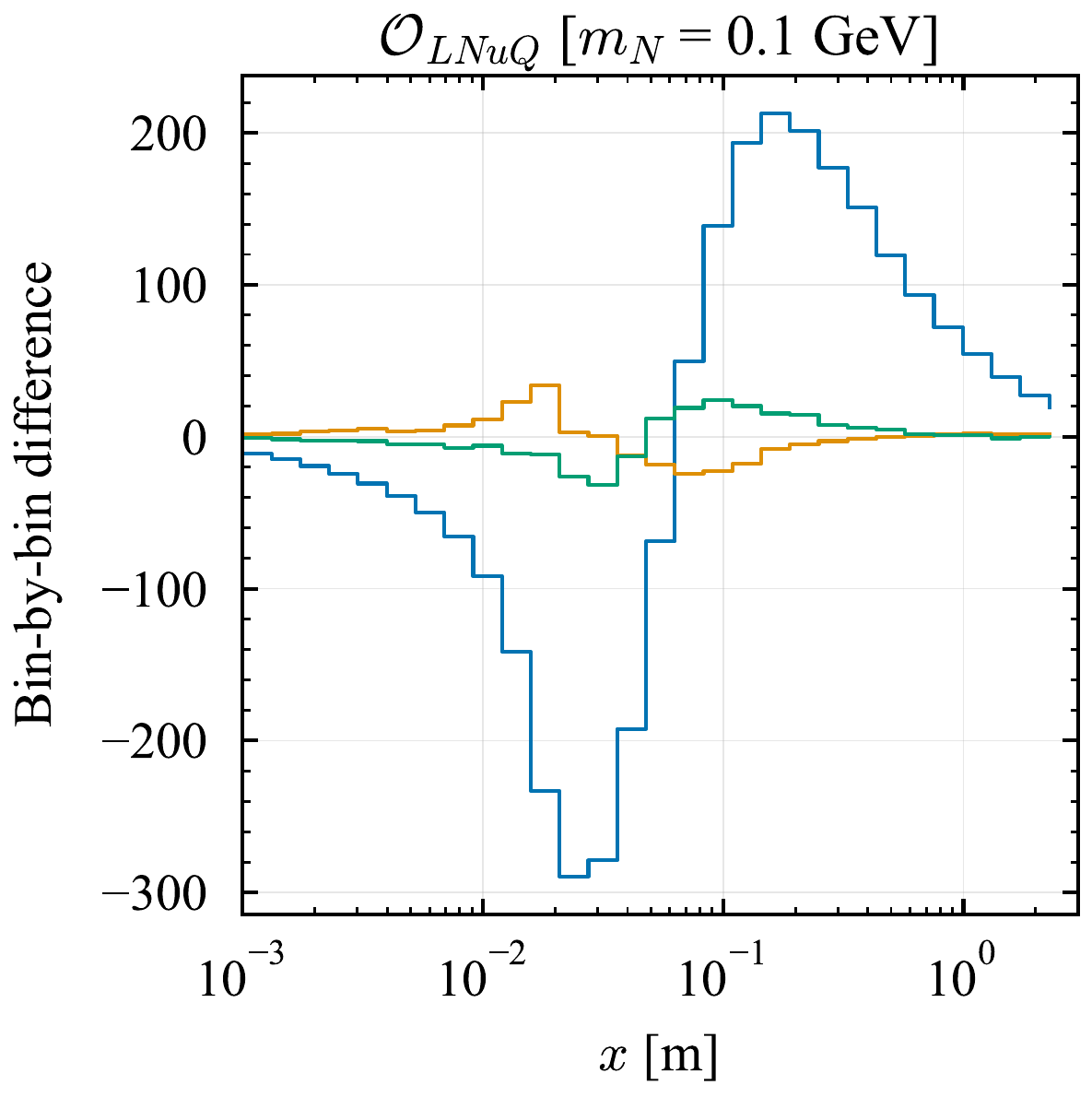}
    \includegraphics[scale=0.245]{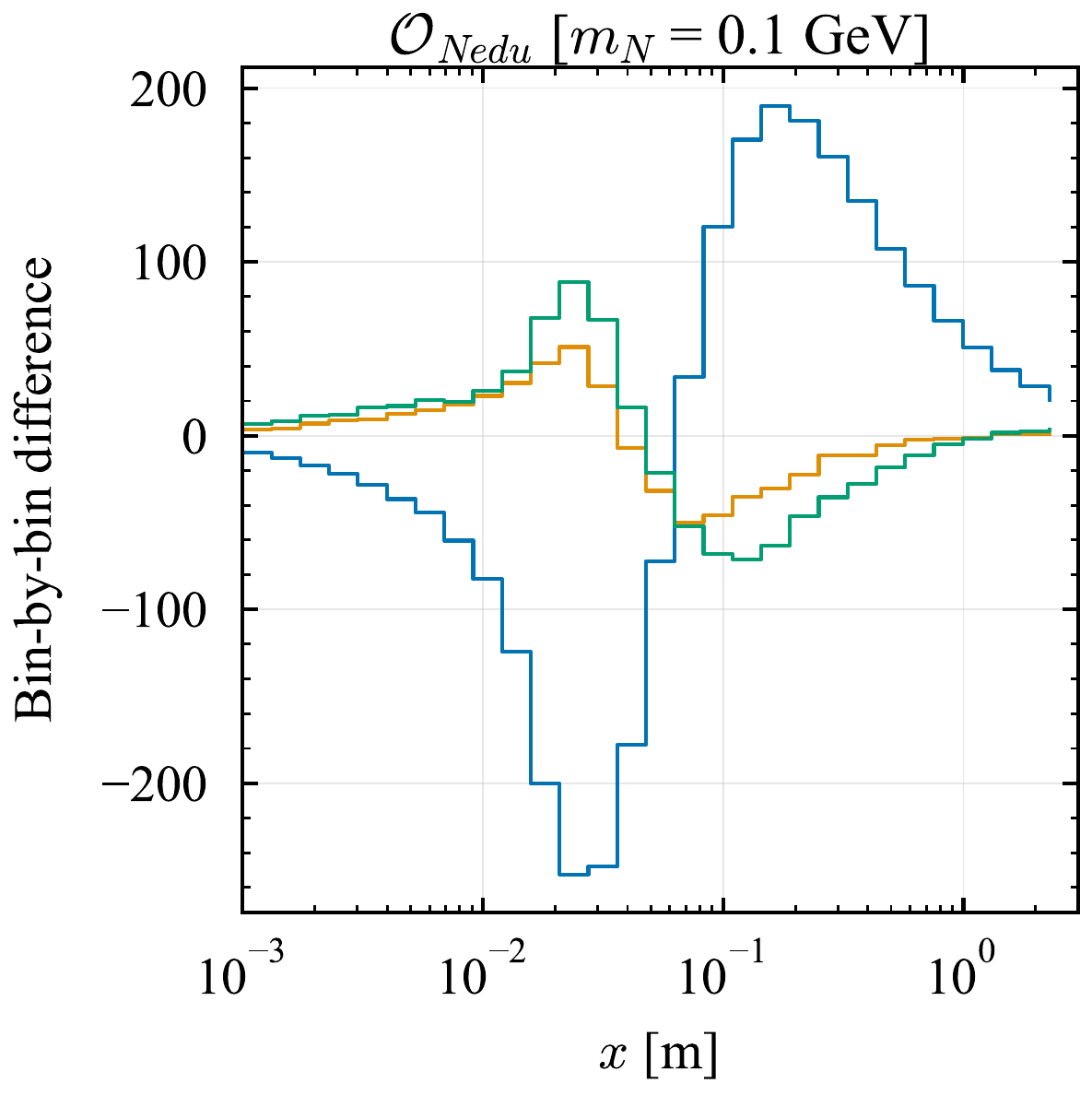}
    \includegraphics[scale=0.245]{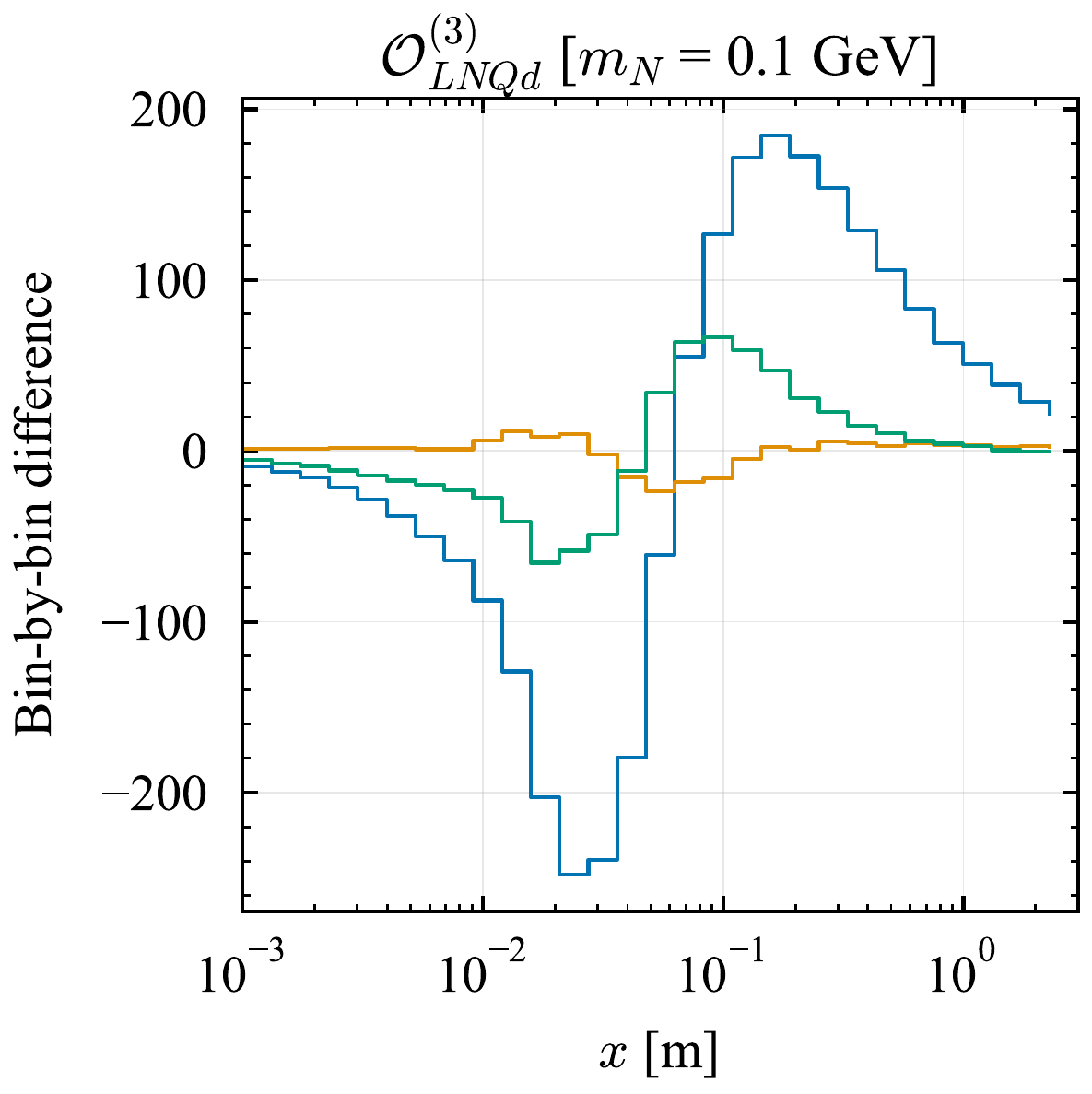}

    \includegraphics[scale=0.245]{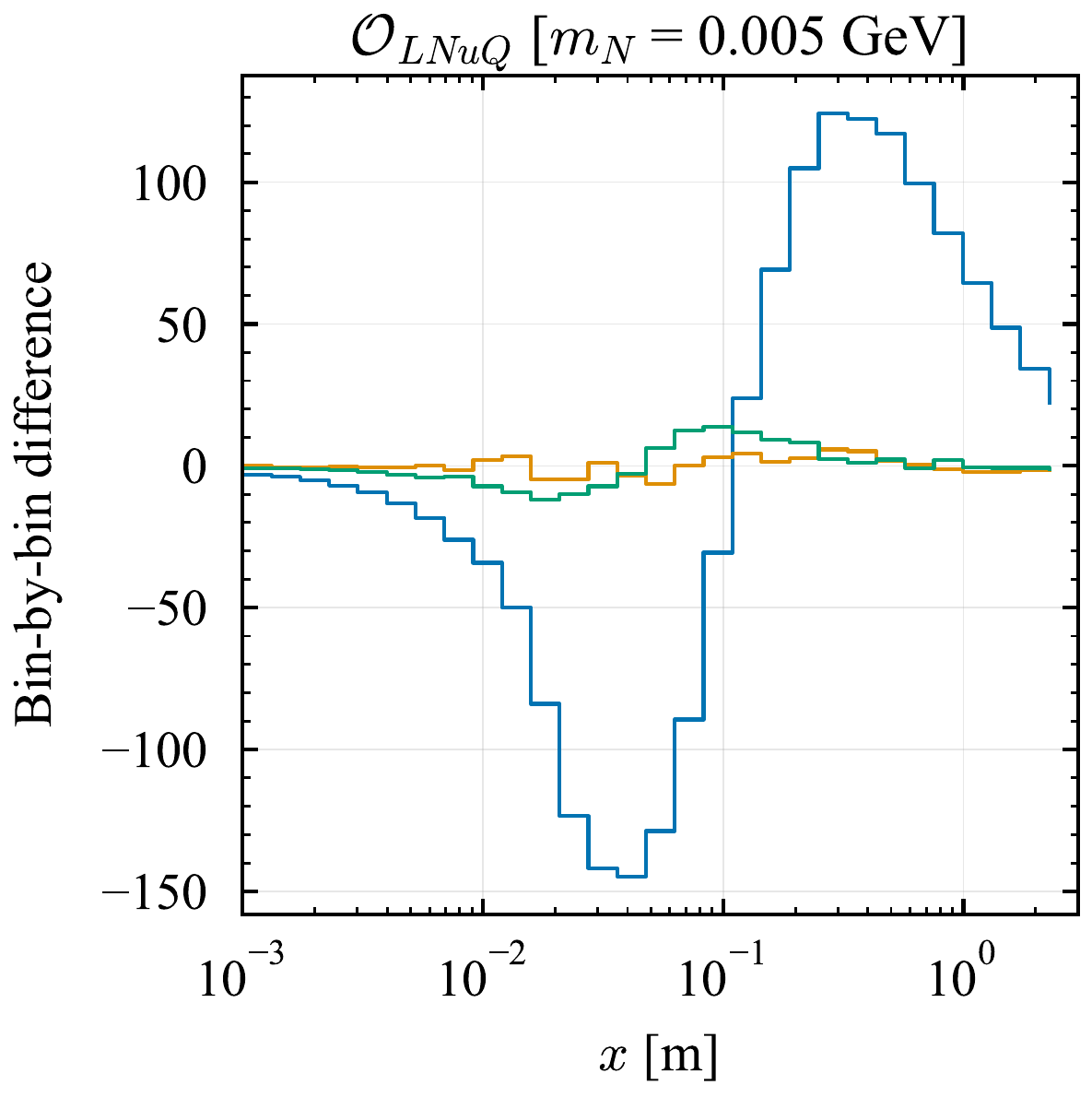}
    \includegraphics[scale=0.245]{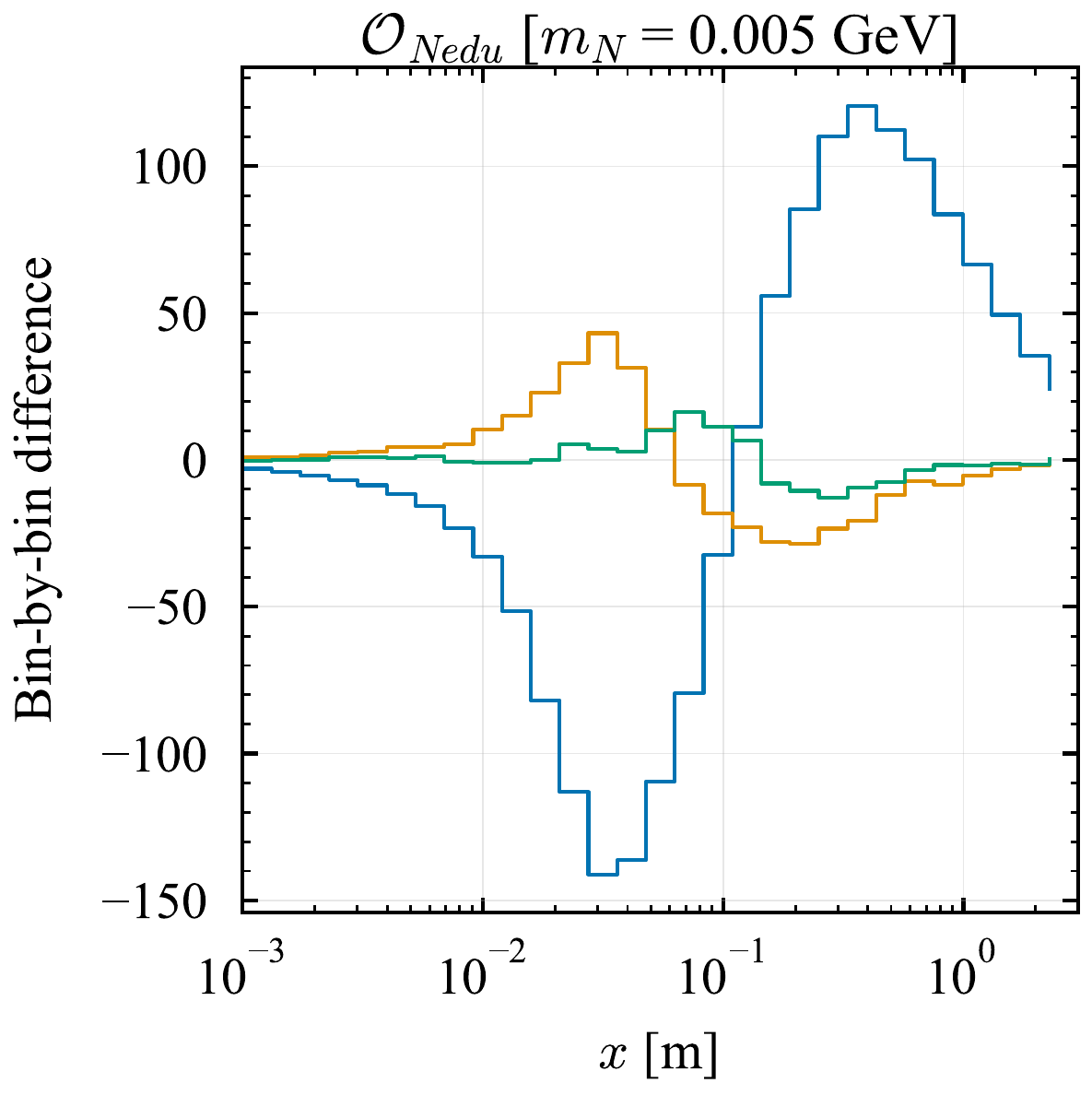}
    \includegraphics[scale=0.245]{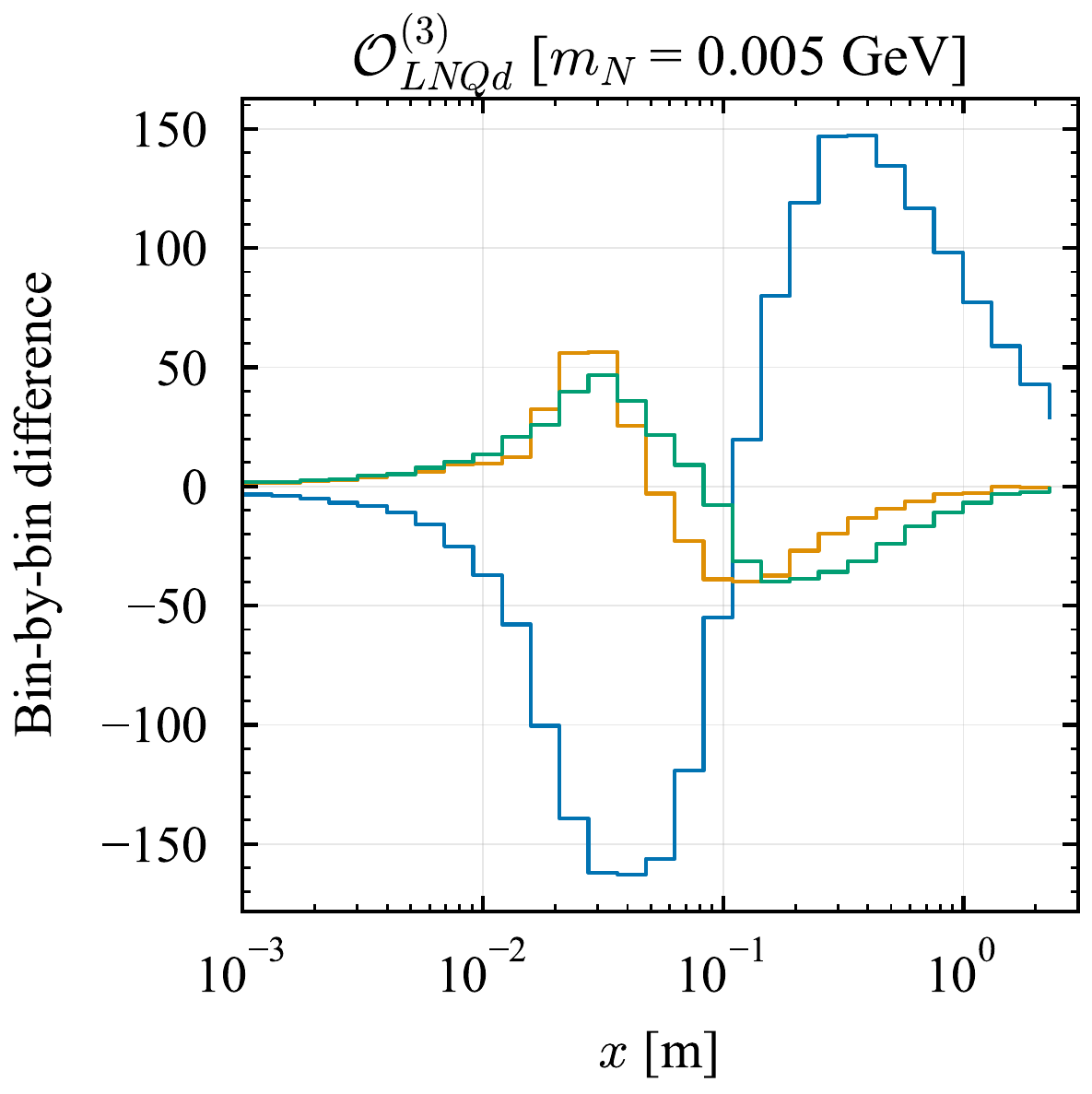}
    \caption{Difference in the spatial distributions of positrons (Dirac minus Majorana) for RHNs produced in $B$ meson decay as a function of horizontal displacement from the beam axis. The panels are arranged by RHN mass (rows: 1~GeV, 100~MeV, 5~MeV) and SMNEFT operator (columns: $\mathcal{O}_{LNuQ}$, $\mathcal{O}_{Nedu}$, $\mathcal{O}^{(3)}_{LNQd}$).}
    \label{fig:vertical_distance_positron_B}
\end{figure}

\begin{figure}[!htbp]
    \centering
    \includegraphics[scale=0.24]{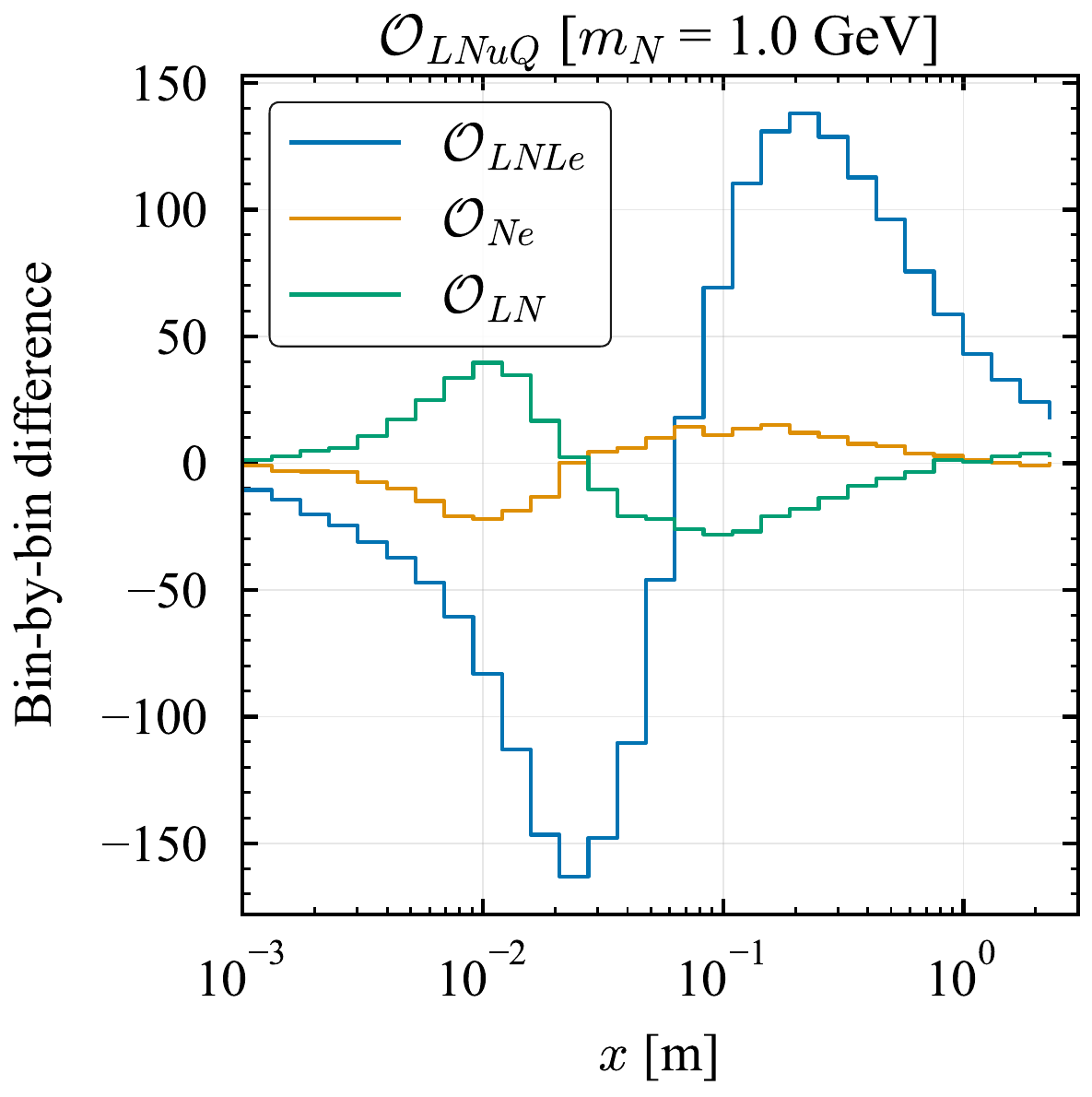}
    \includegraphics[scale=0.24]{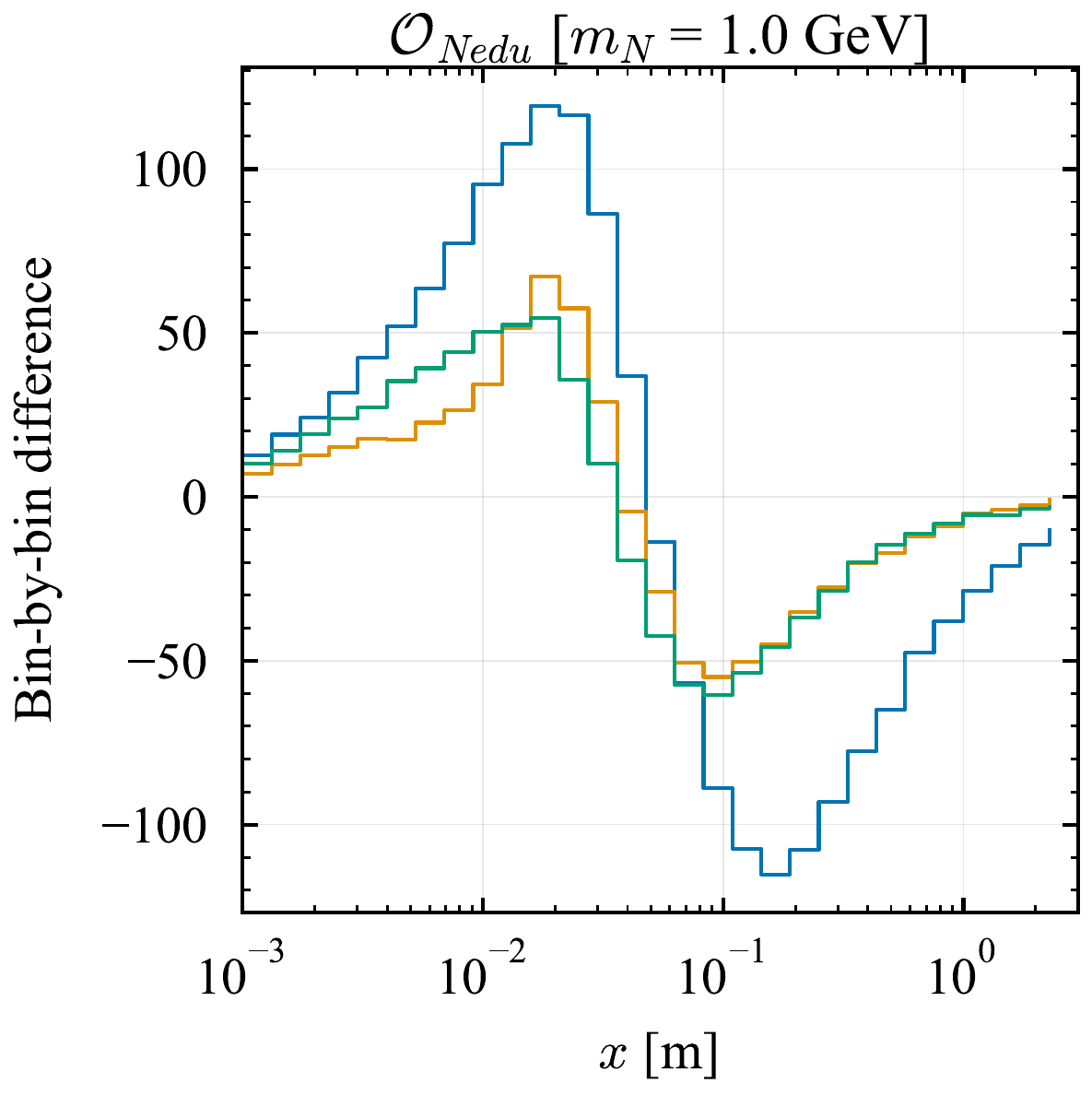}
    \includegraphics[scale=0.24]{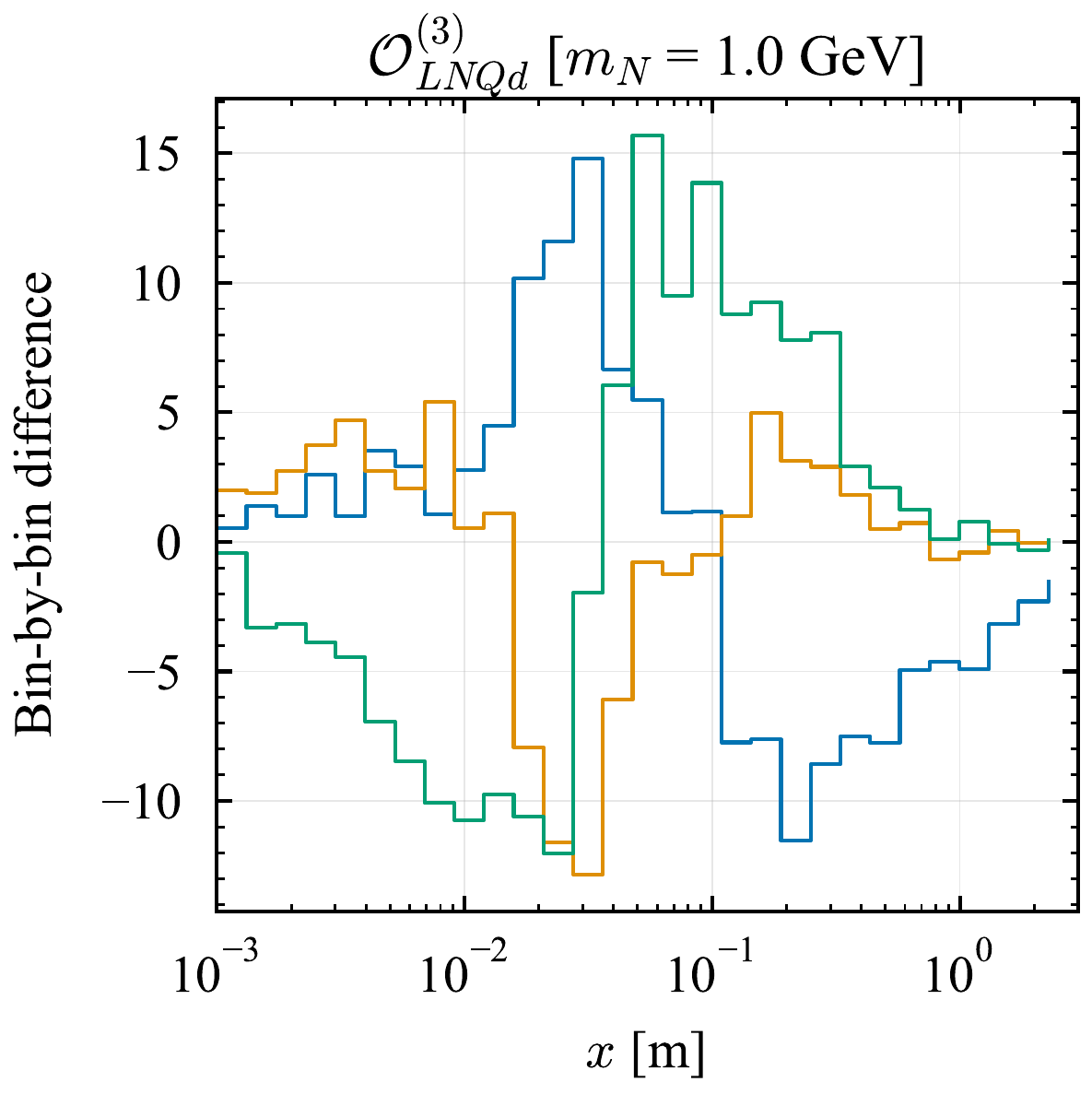}

    \includegraphics[scale=0.24]{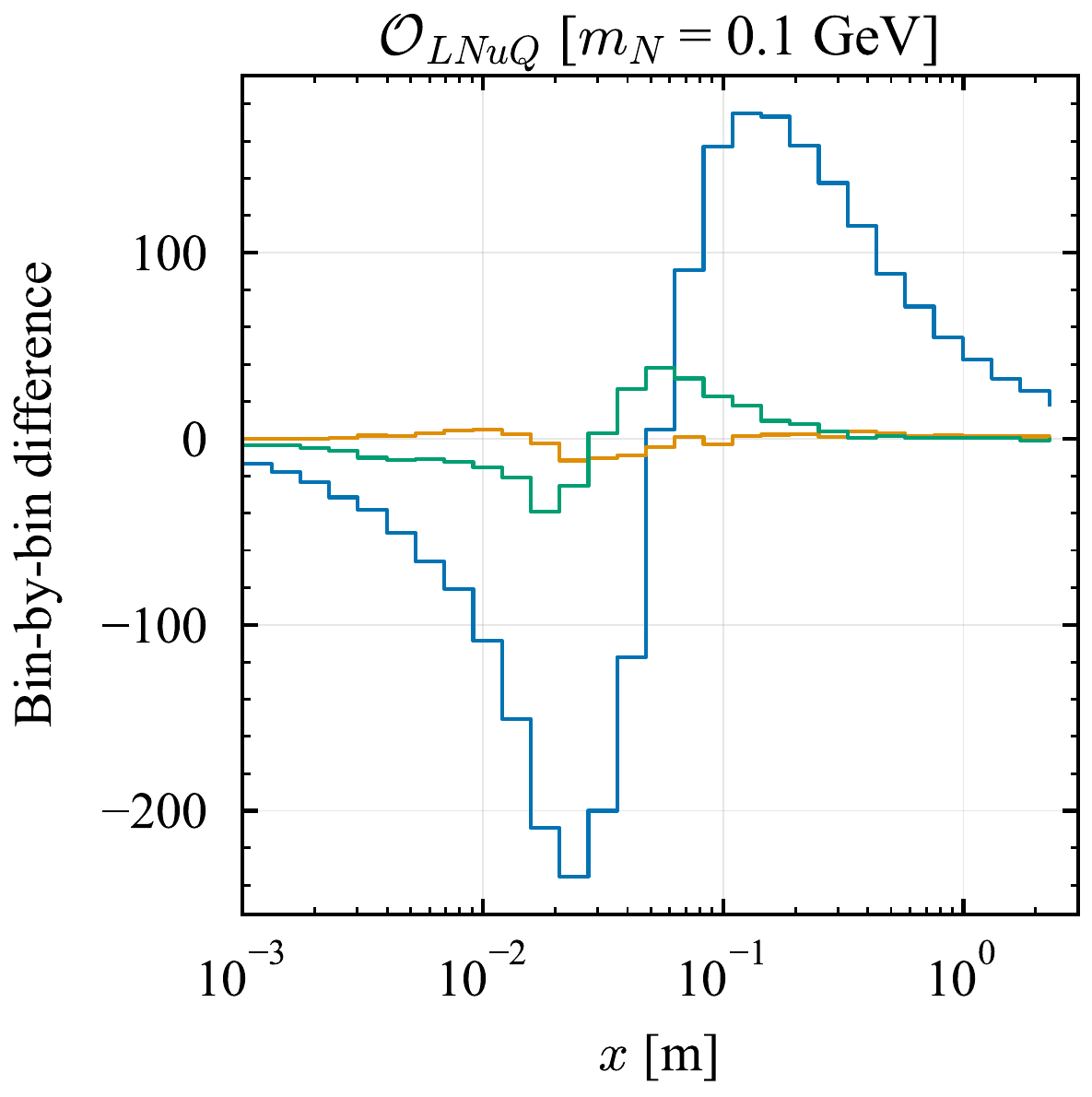}
    \includegraphics[scale=0.24]{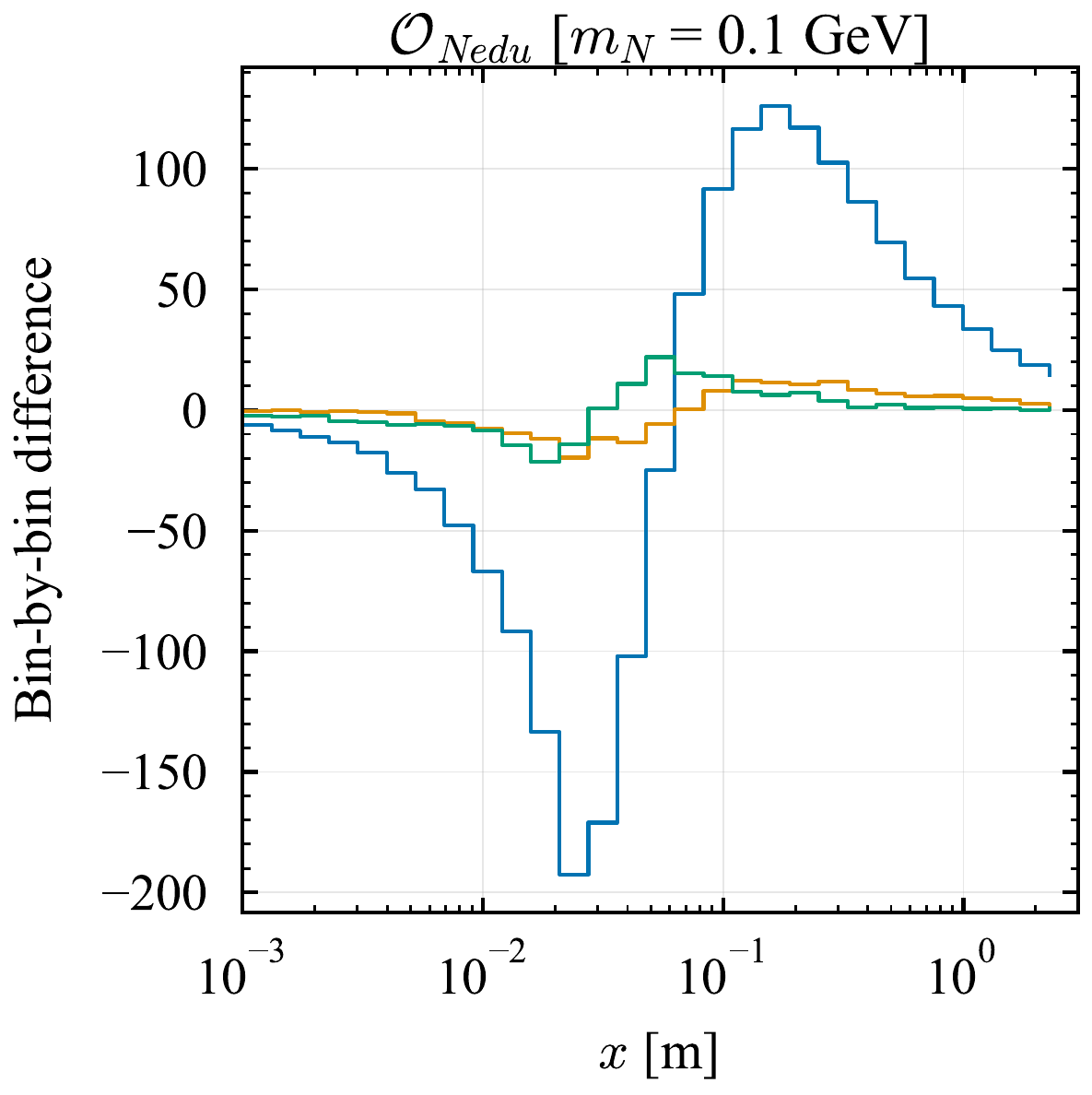}
    \includegraphics[scale=0.24]{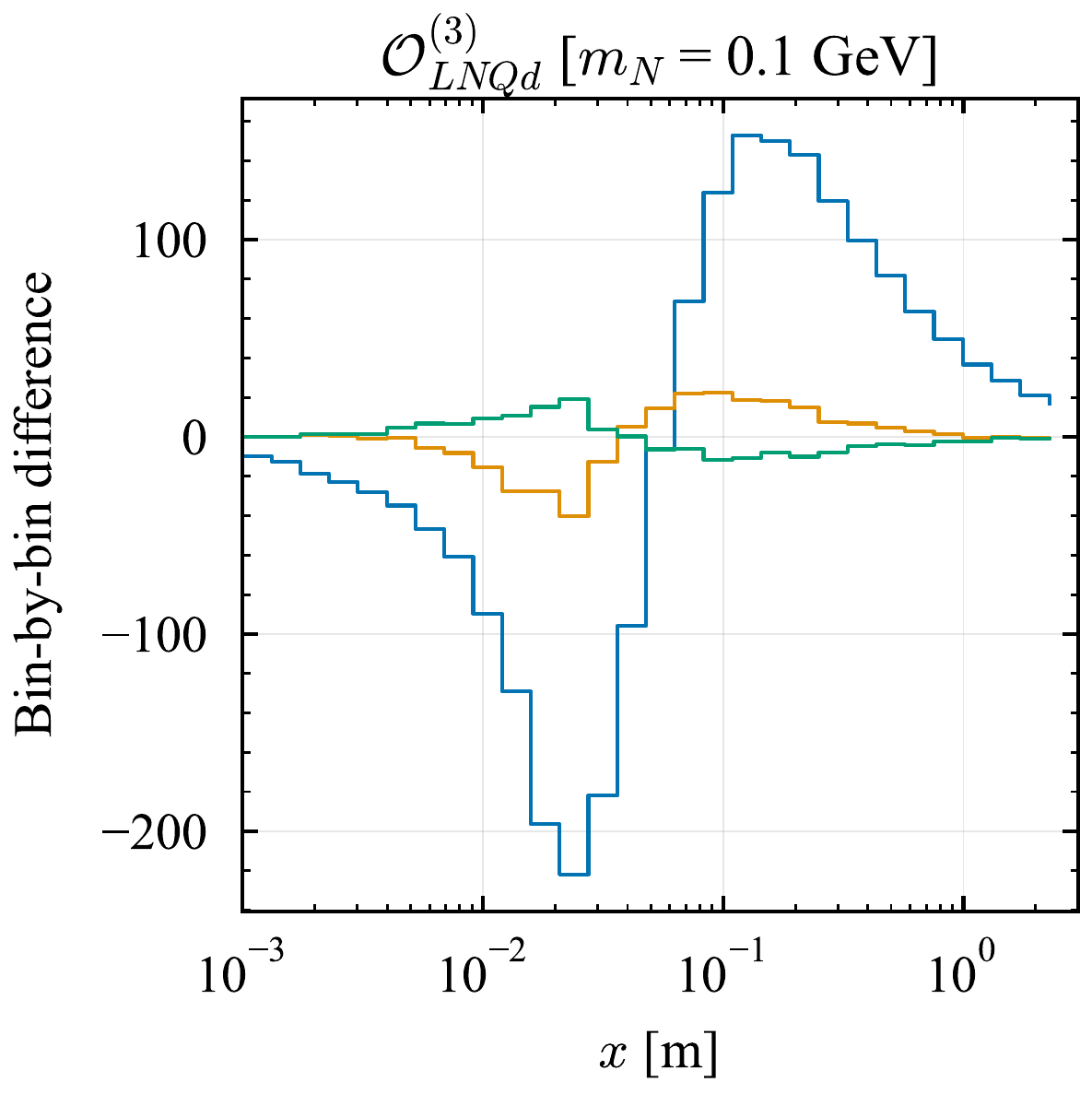}

    \includegraphics[scale=0.24]{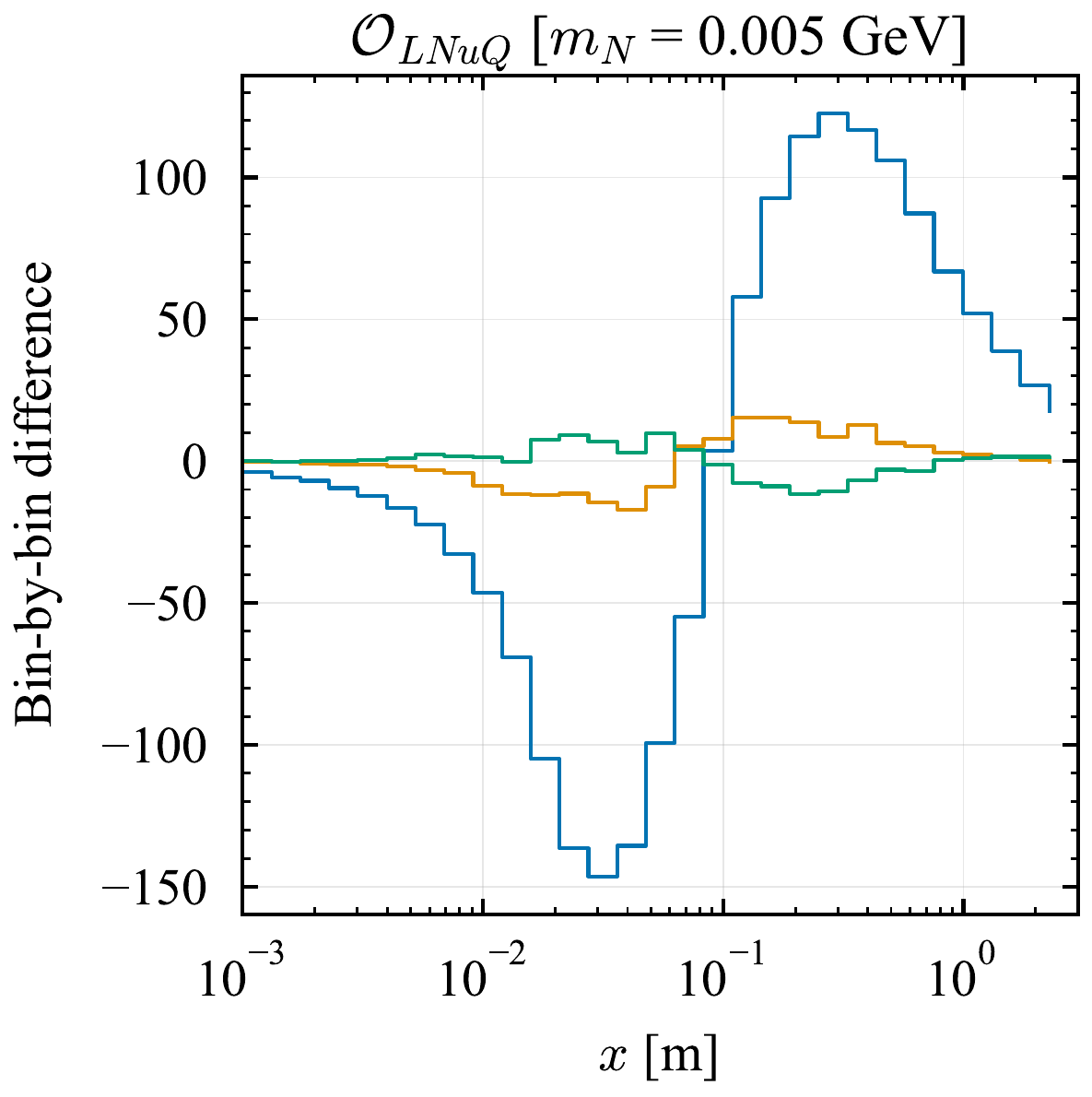}
    \includegraphics[scale=0.24]{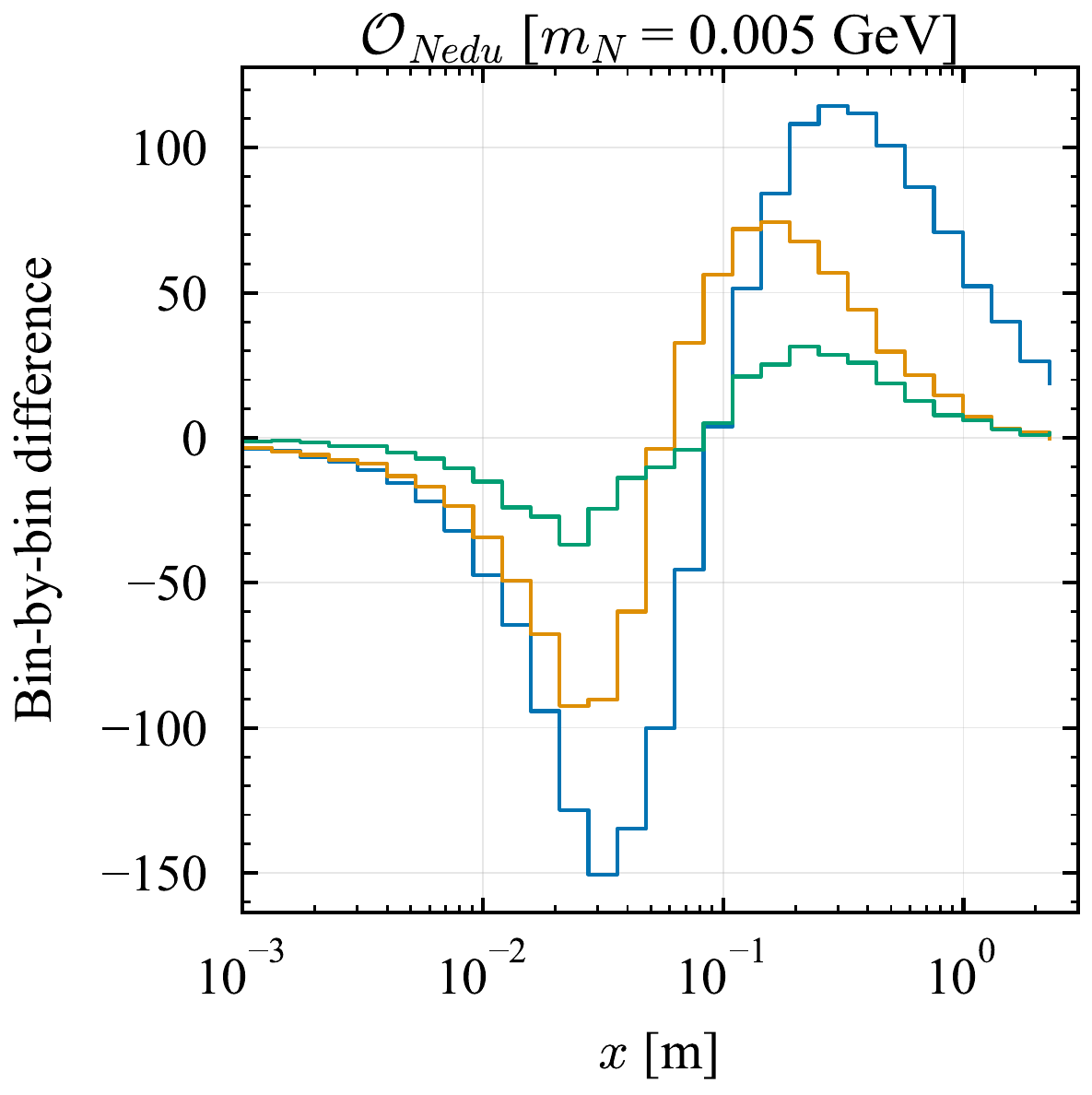}
    \includegraphics[scale=0.24]{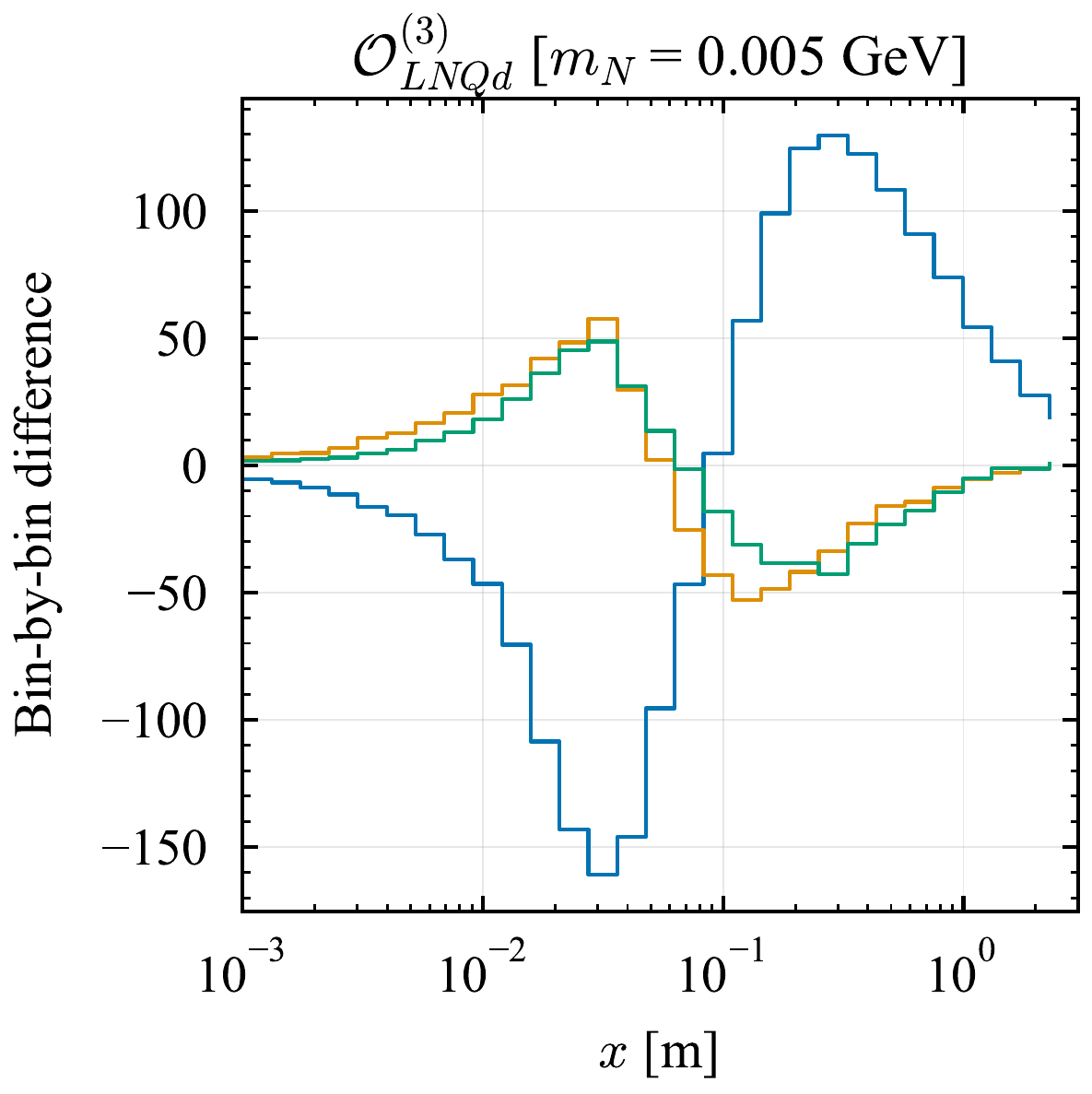}
    \caption{Similar to Fig.~\ref{fig:vertical_distance_positron_B} for RHNs produced in $D$ meson decay.}
    \label{fig:vertical_distance_positron_D}
\end{figure}

\begin{figure}[!htbp]
    \centering
    \includegraphics[scale=0.24]{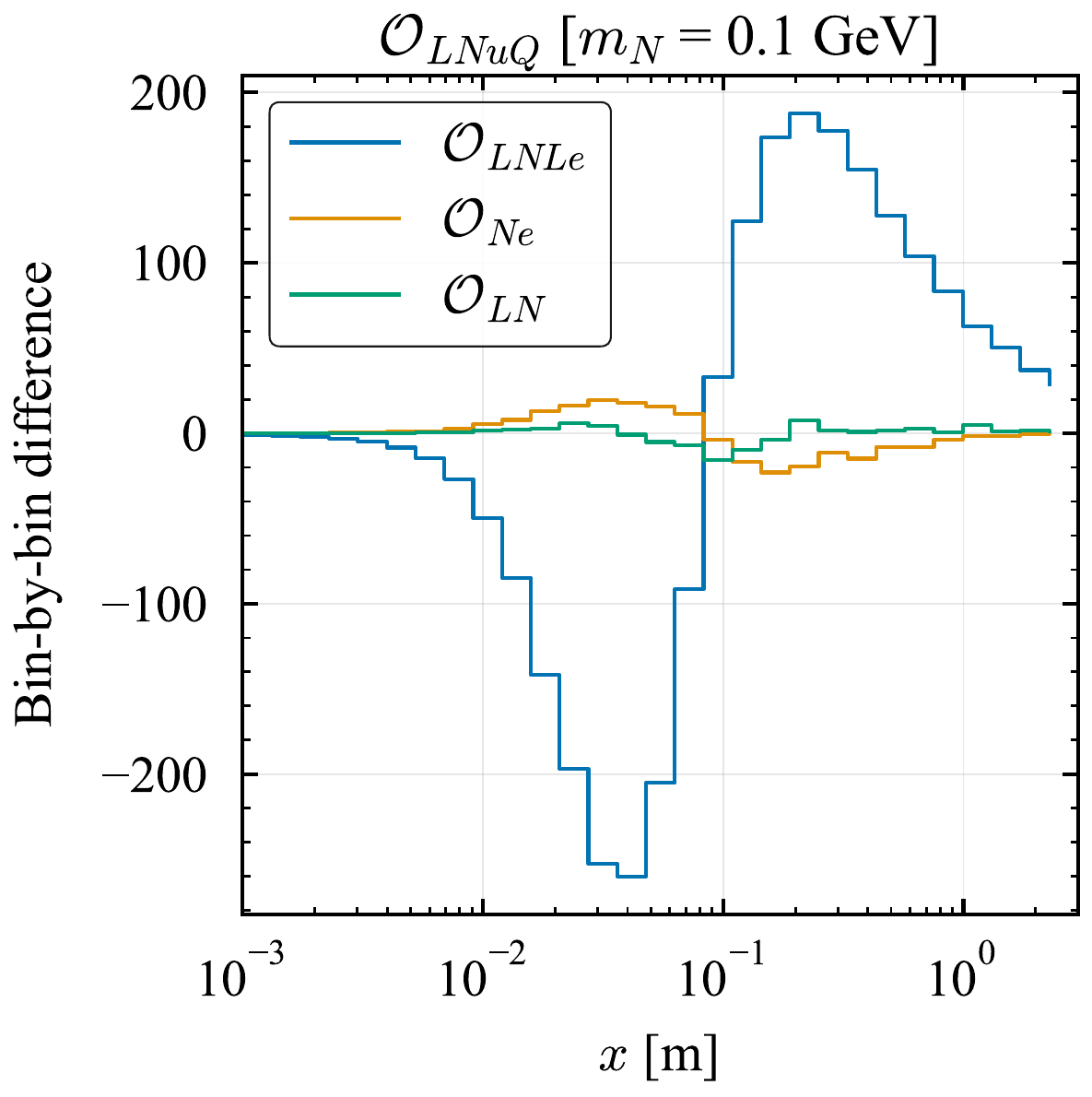}
    \includegraphics[scale=0.24]{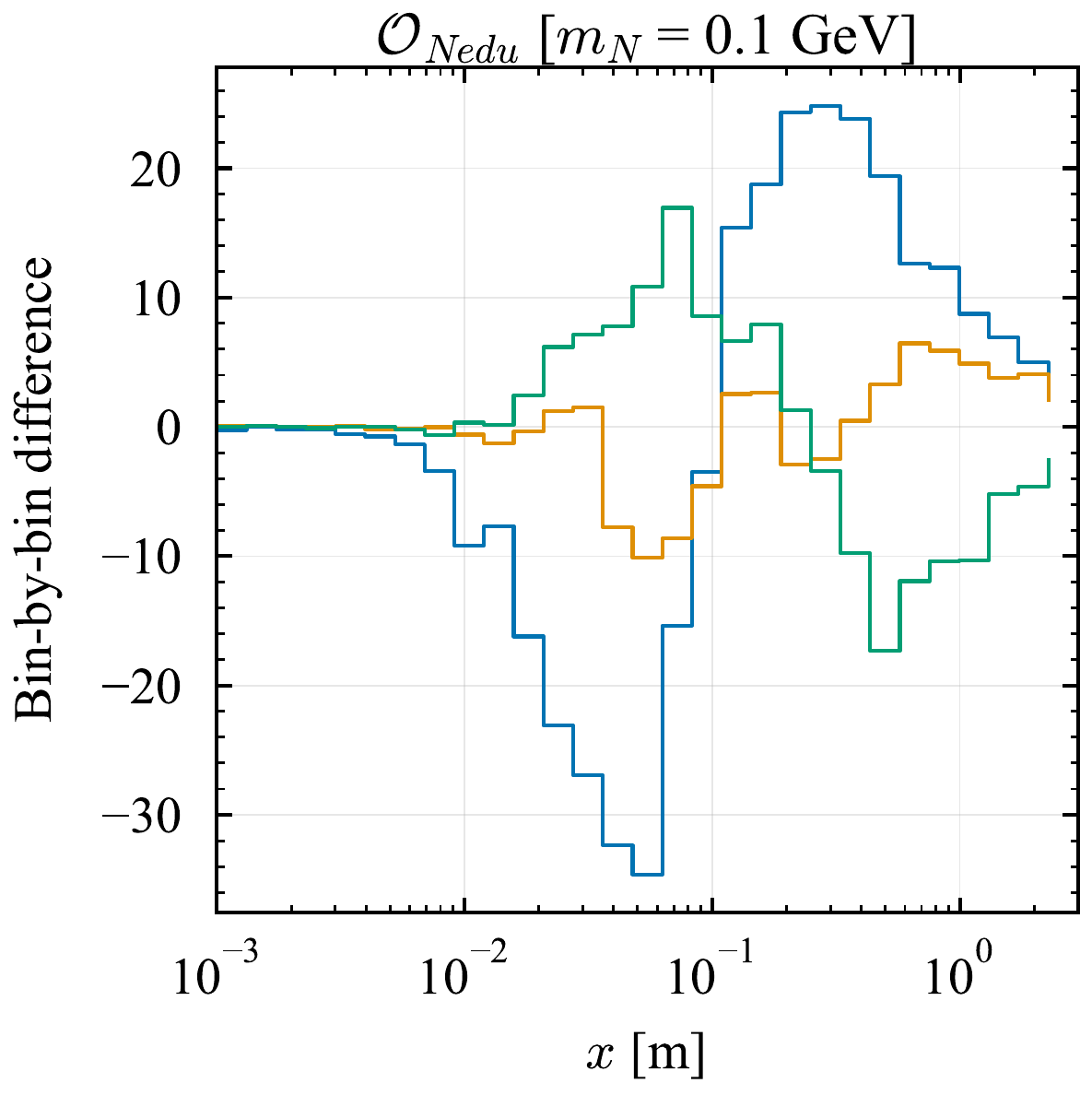}
    \includegraphics[scale=0.24]{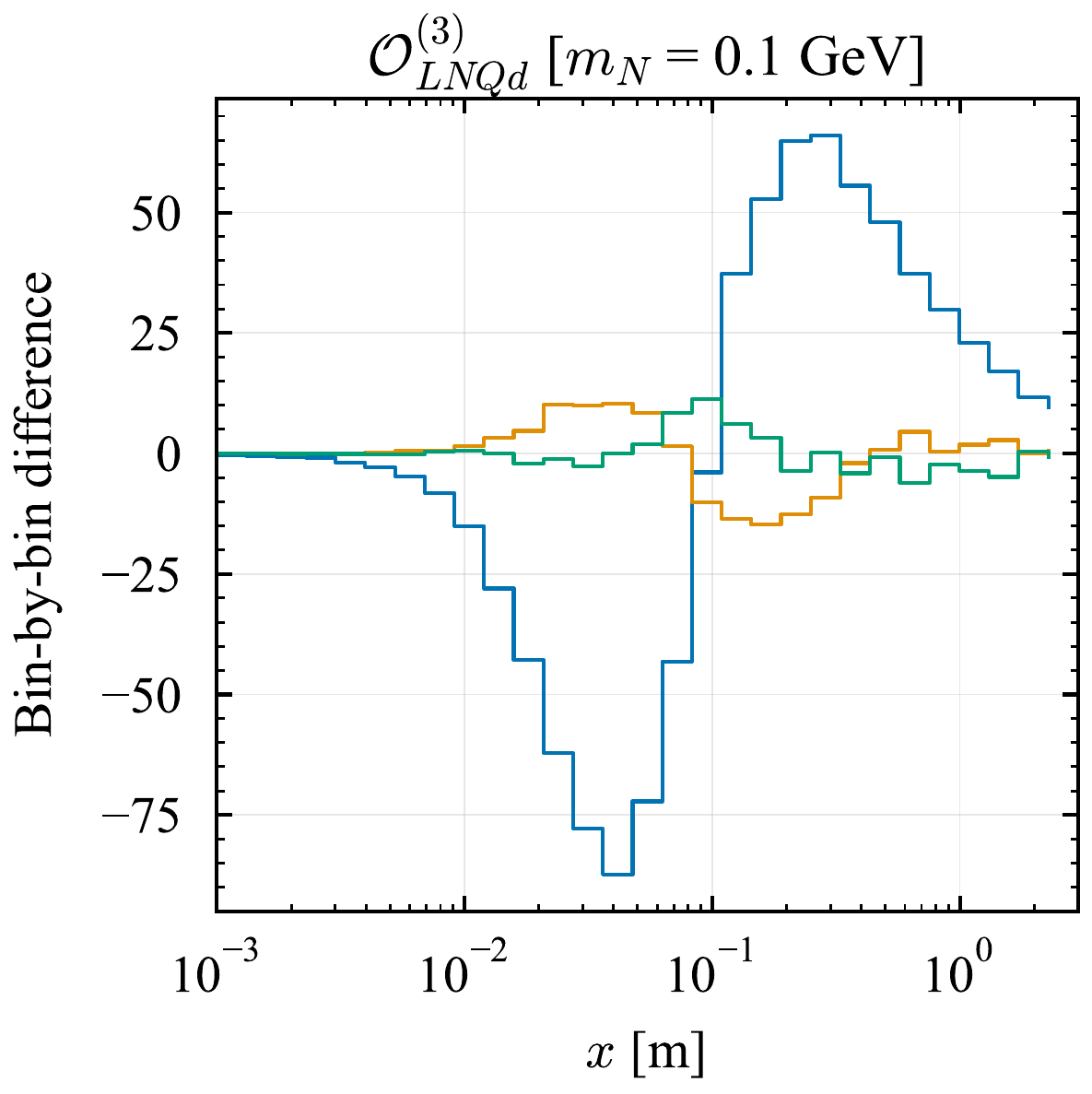}

    \includegraphics[scale=0.24]{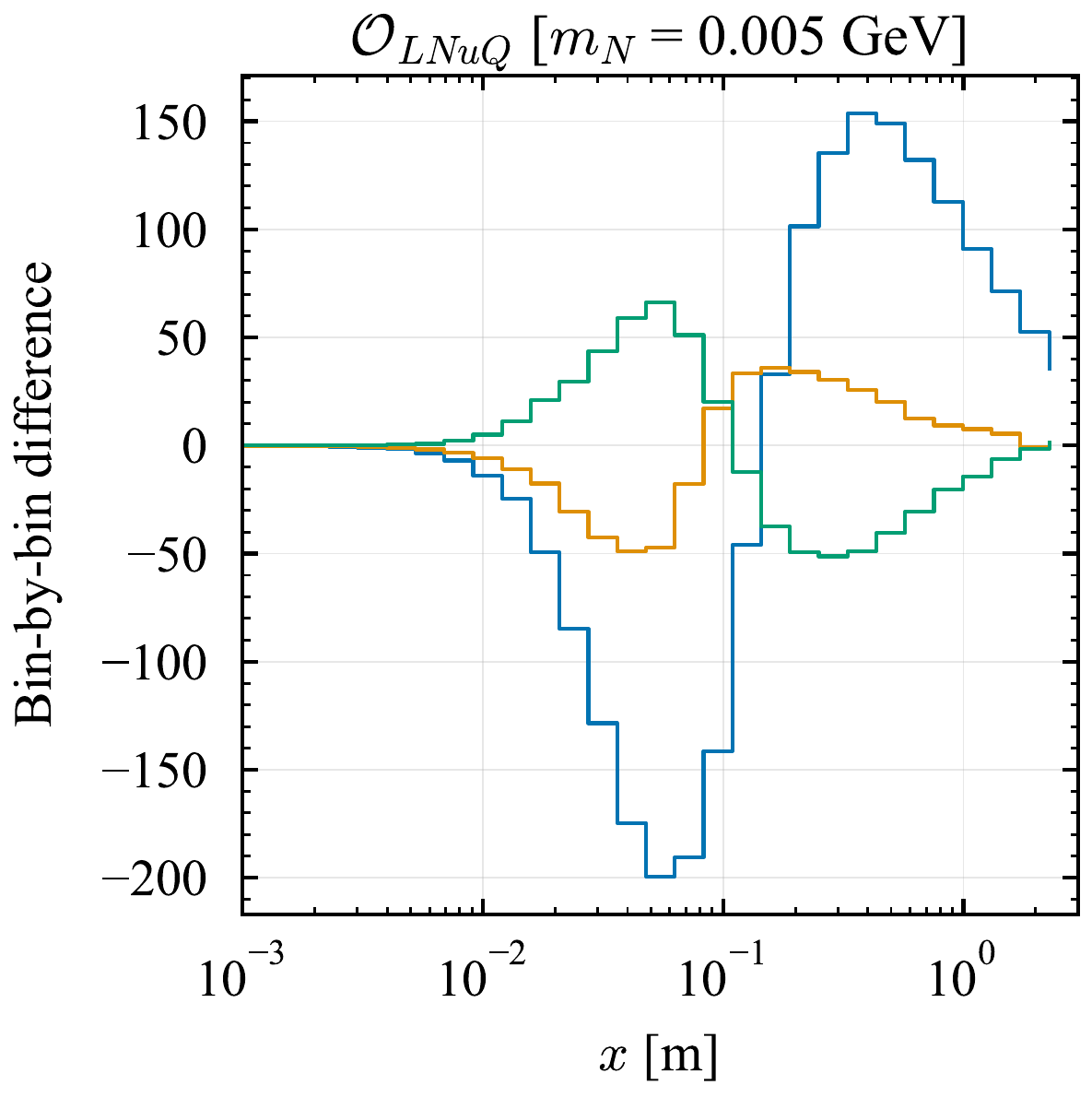}
    \includegraphics[scale=0.24]{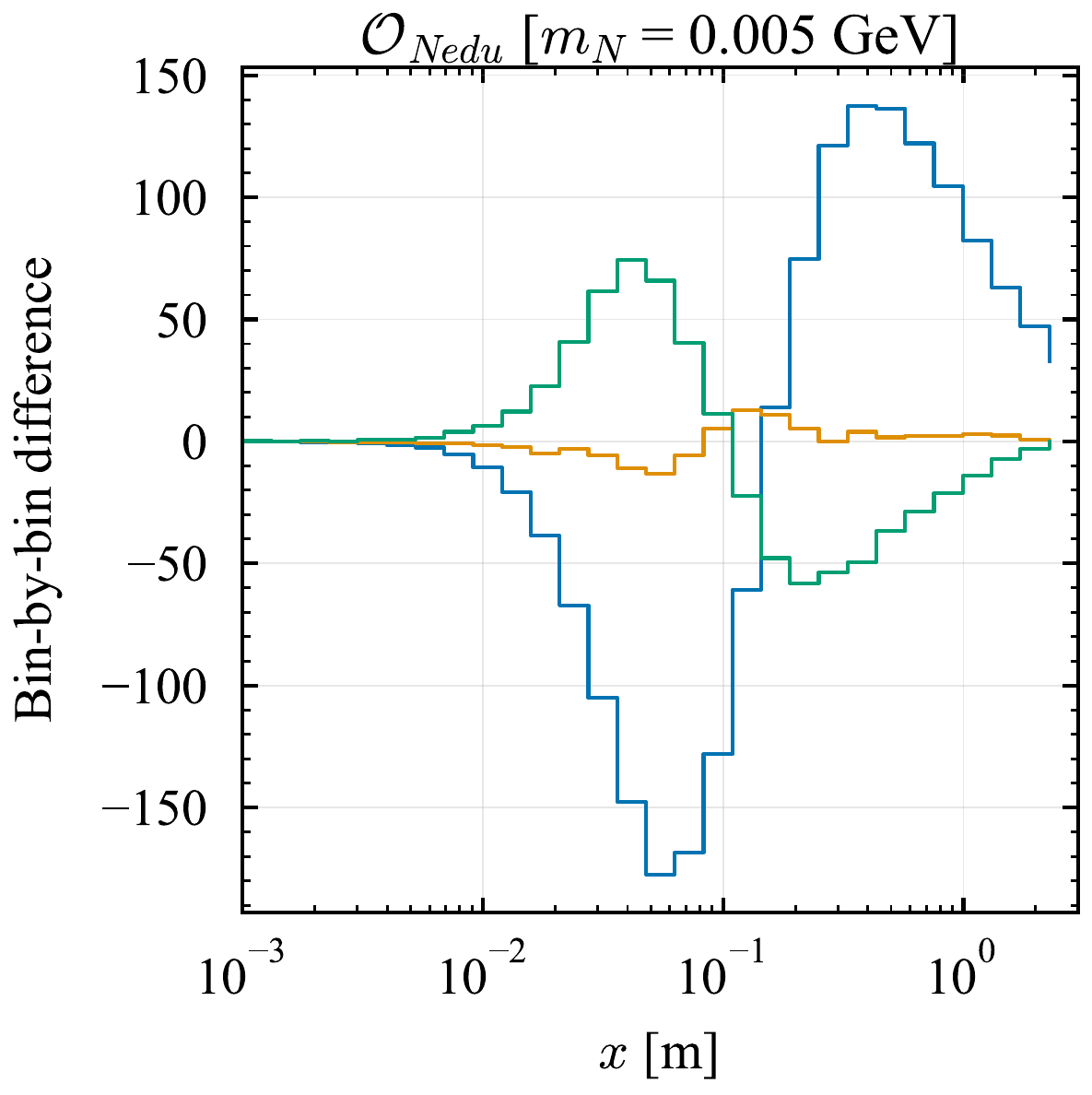}
    \includegraphics[scale=0.24]{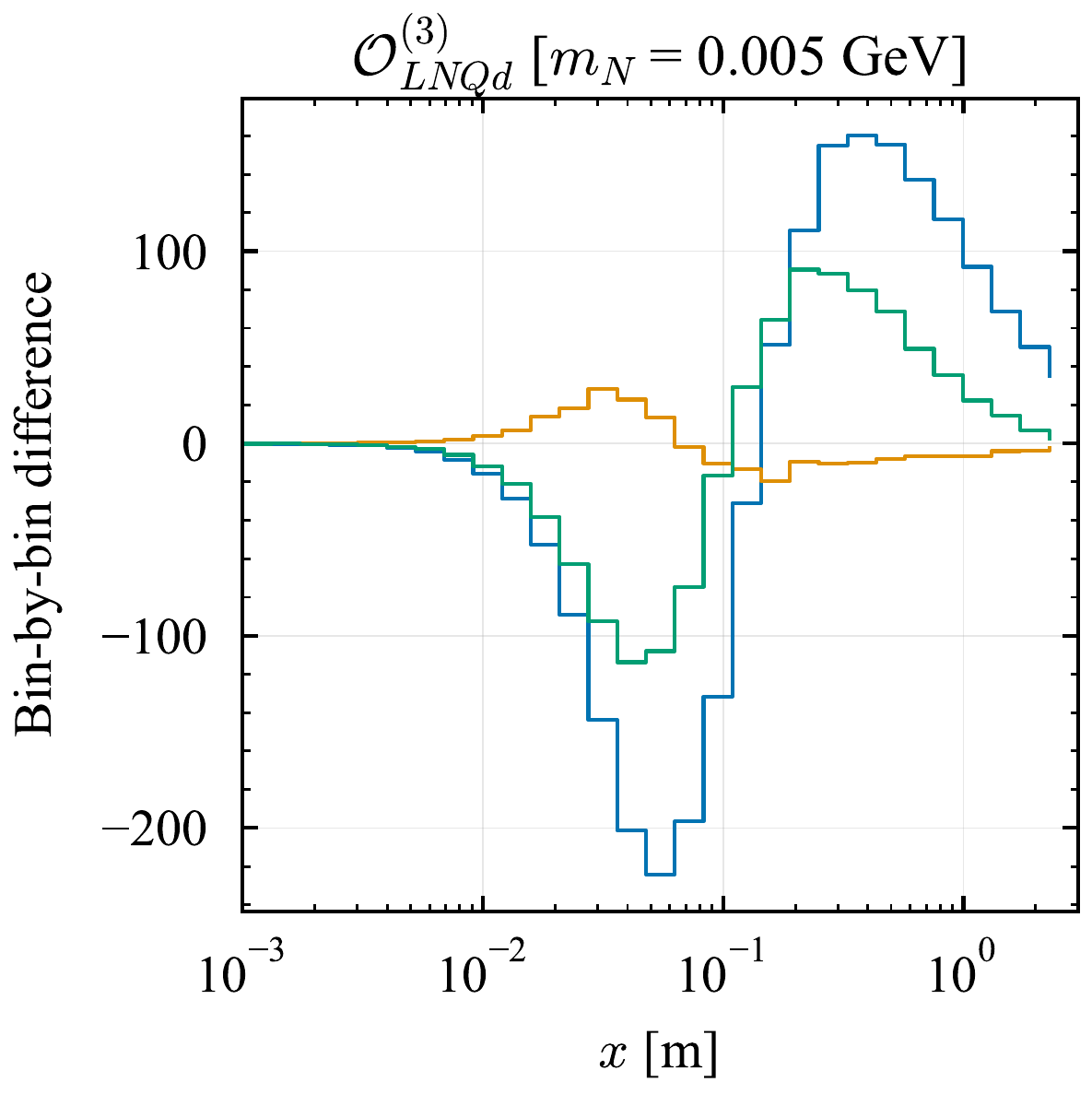}
    \caption{Similar to Fig.~\ref{fig:vertical_distance_positron_B} for RHNs produced in kaon decay. }
    \label{fig:vertical_distance_positron_K}
\end{figure}

\begin{figure}[!htbp]
    \centering
    \includegraphics[scale=0.25]{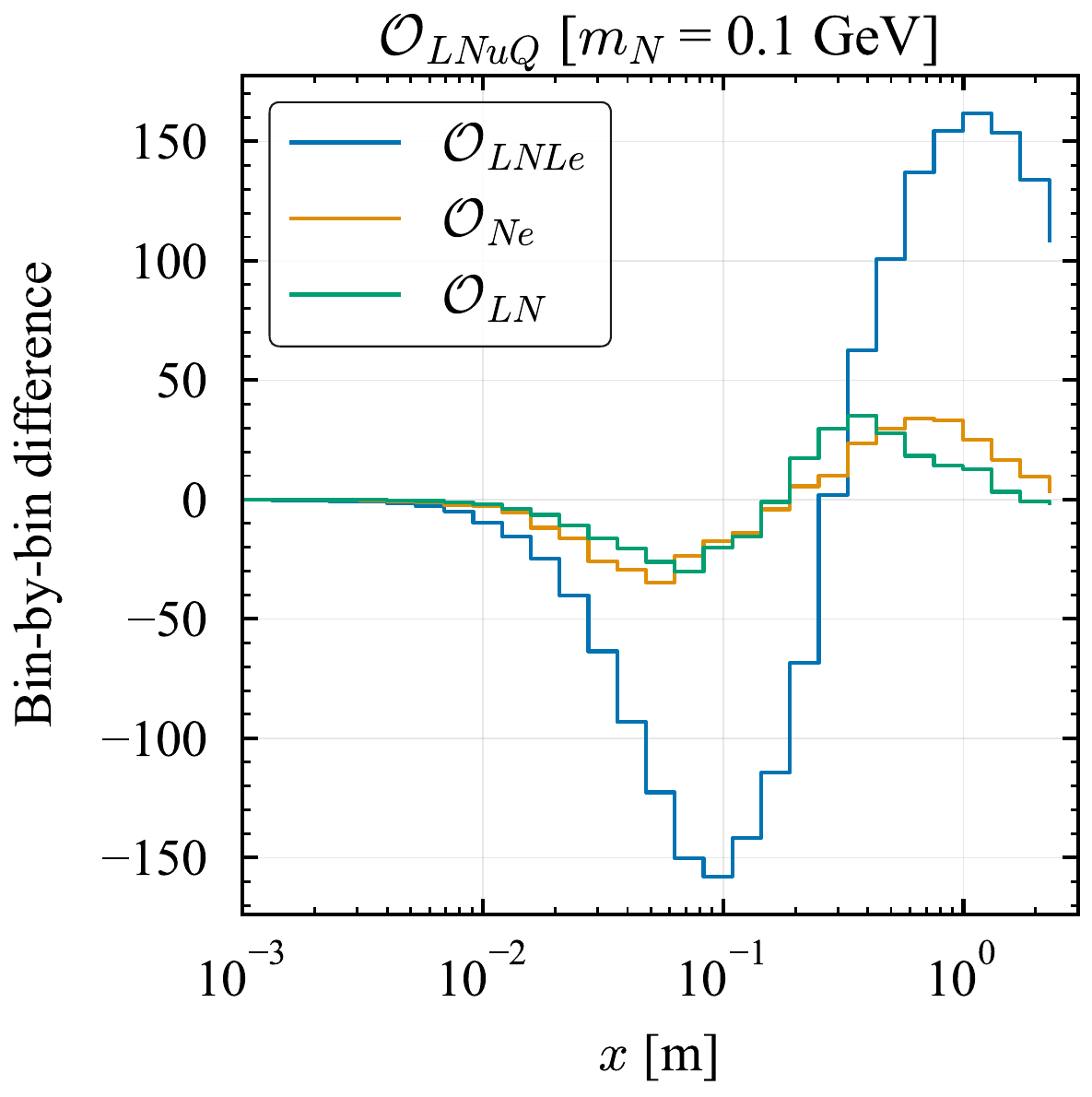}
    \includegraphics[scale=0.25]{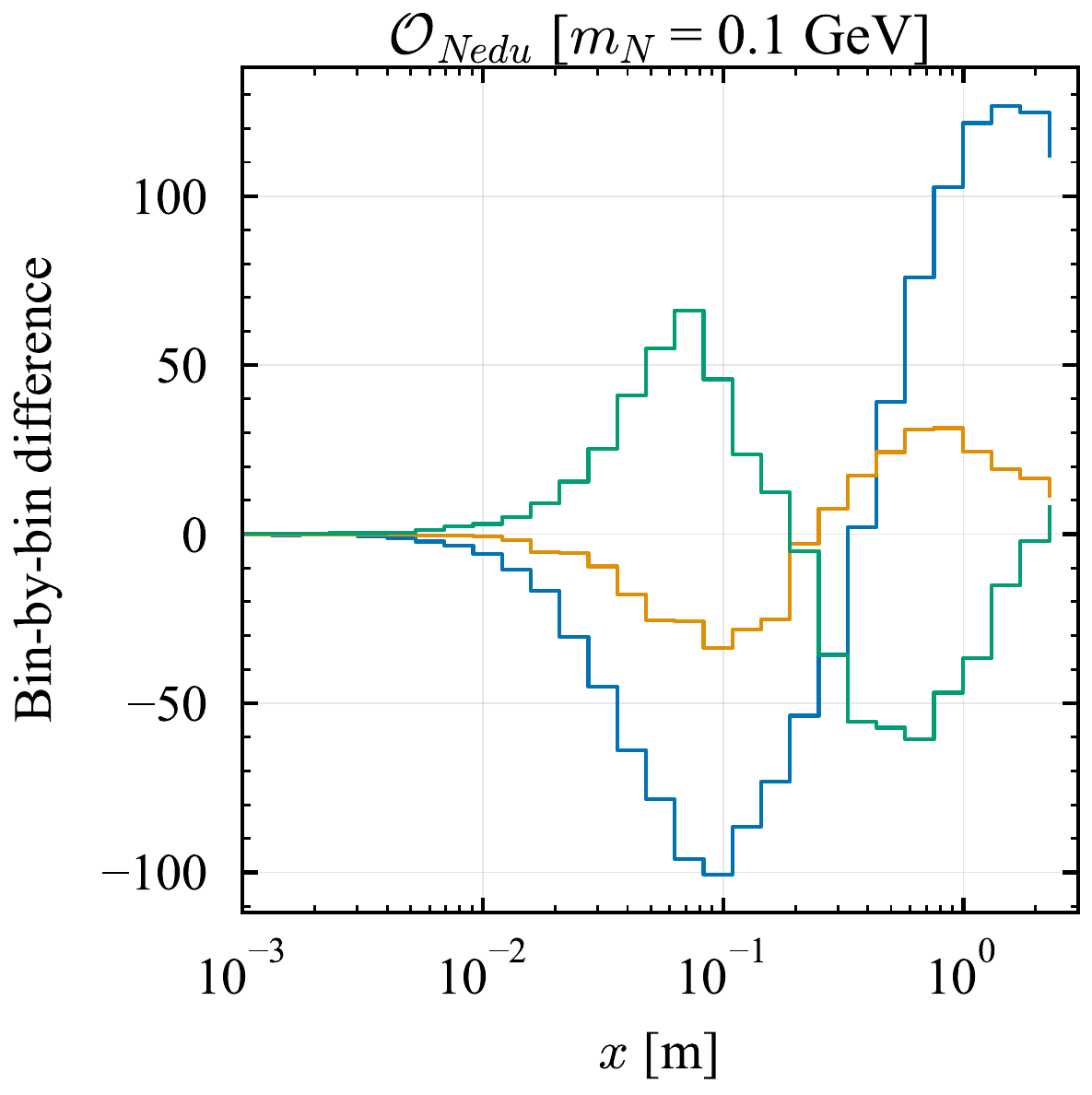}

    \includegraphics[scale=0.25]{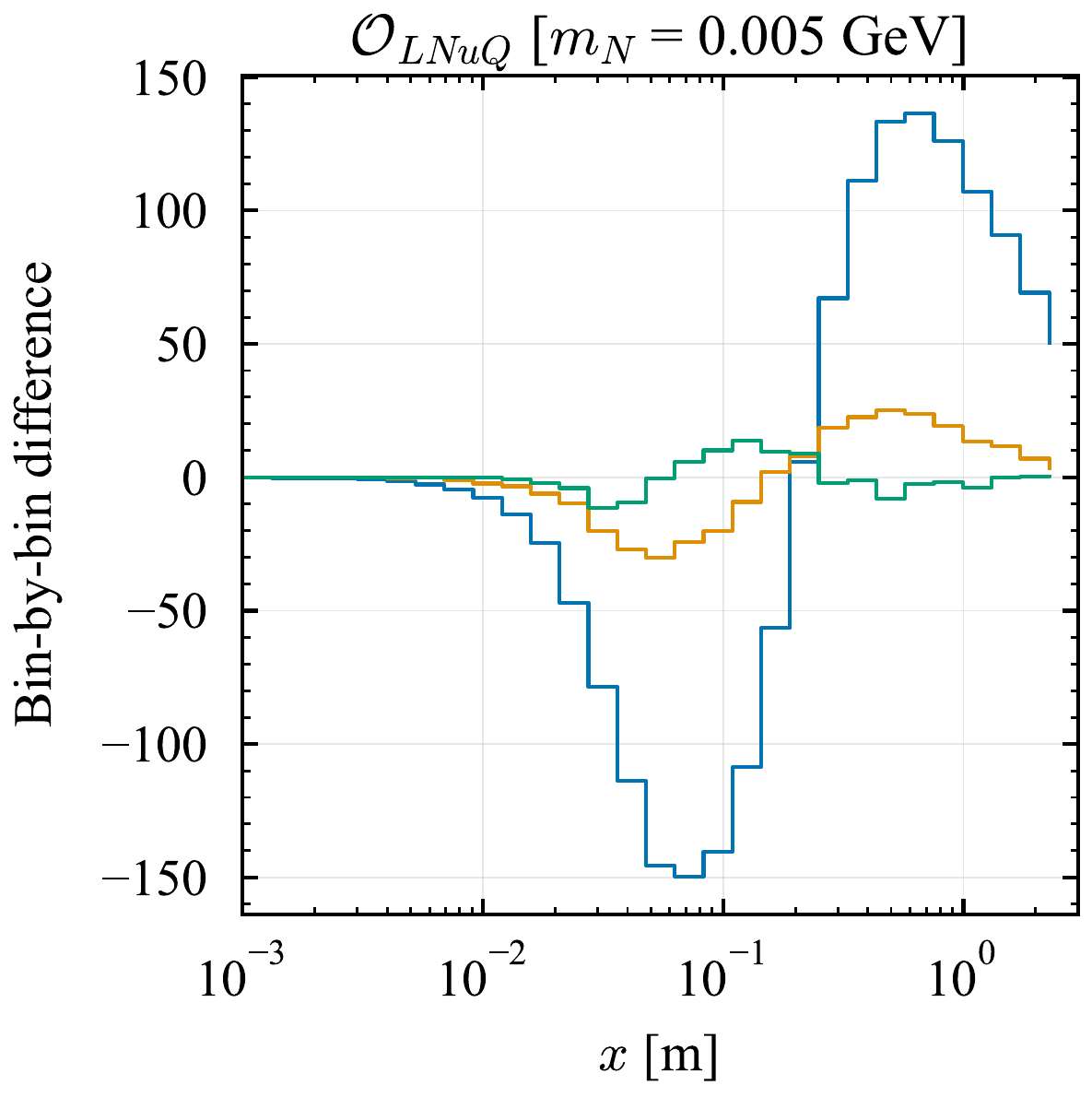}
    \includegraphics[scale=0.25]{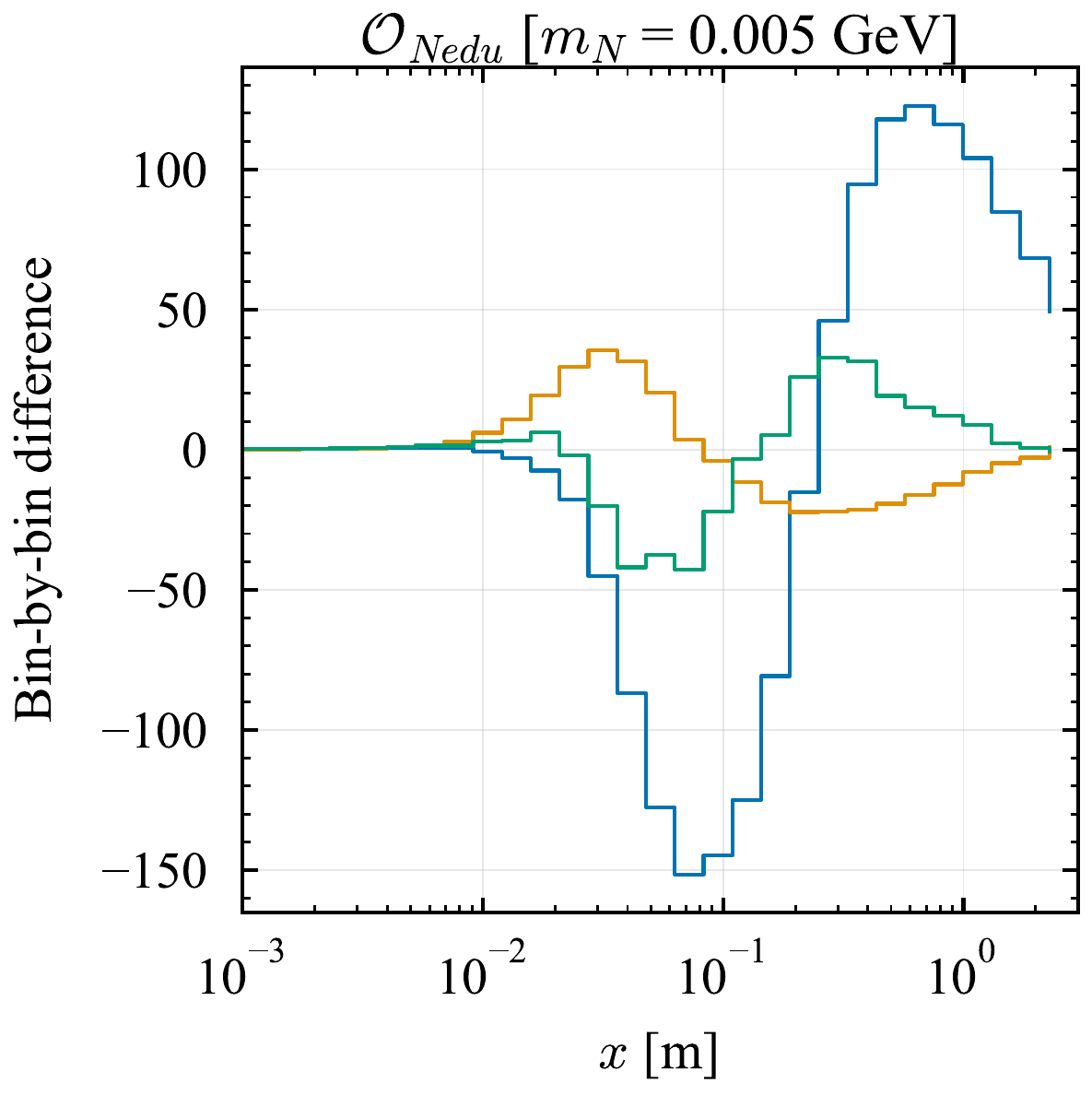}
    \caption{Similar to Fig.~\ref{fig:vertical_distance_positron_B} for RHNs produced in pion decay.}
    \label{fig:vertical_distance_positron_pi}
\end{figure}

In Figs.~\ref{fig:vertical_distance_positron_B}, \ref{fig:vertical_distance_positron_D}, \ref{fig:vertical_distance_positron_K}, and \ref{fig:vertical_distance_positron_pi} we present the difference in the positron spatial distributions for the decay of Dirac and Majorana RHNs produced via $B$, $D$, $K$ and $\pi$ decay, respectively, for three representative RHN masses: 1~GeV (near the B-meson mass threshold), 100~MeV (accessible to multiple meson decay channels), and 5~MeV (low-mass regime where polarization effects are maximized). All distributions are normalized to $10^4$ events to allow a direct comparison between different scenarios. However, in our statistical analysis, we properly account for the actual expected event yields based on the production cross sections in Fig.~\ref{fig:production_cross_sections}. 

Several important trends emerge from these figures. Examining the first row of Fig.~\ref{fig:vertical_distance_positron_B}, we observe that the difference in event counts for the Dirac and Majorana RHNs is largest for the $\mathcal{O}_{LNLe}$ decay operator in combination with the $\mathcal{O}_{LNuQ}$ production operator. This can be understood by considering the interplay between the RHN polarization and production cross sections for $B$ meson decays.
From Fig.~\ref{fig:production_cross_sections}, we observe that two-body $B$ meson decays dominate over three-body decays near $m_N = 1$~GeV for $\mathcal{O}_{LNuQ}$. Since RHNs produced from two-body decays have polarization $P = 1.0$, the angular distributions of their decay products are maximally different for Dirac and Majorana RHNs. In contrast, for $\mathcal{O}_{Nedu}$ and $\mathcal{O}^{(3)}_{LNQd}$, three-body decays dominate the production mechanism. The polarization of RHNs from three-body decays for these operators for $m_N = 1$~GeV is significantly smaller than unity (see Fig.~\ref{fig:pol_bmeson}), resulting in smaller amplitude differences for the Dirac and Majorana RHNs.

This can also be seen from the results for other values of $m_N$ for $\mathcal{O}_{Nedu}$ and $\mathcal{O}^{(3)}_{LNQd}$. Comparing 
$m_N  = 1$~GeV with $m_N  = 0.1$~GeV and 0.005~GeV, we observe that the amplitude of the  difference increases significantly as the RHN mass decreases. This is because the polarization of RHNs approaches unity for lower masses even for RHNs produced through three-body decays, thus enhancing the sensitivity of the angular distributions to the nature of the RHN.

Also, from the first row of Fig.~\ref{fig:vertical_distance_positron_B}, we observe that the amplitude of the difference between the Dirac and Majorana cases decreases in the order $\mathcal{O}_{LNuQ}$, $\mathcal{O}_{Nedu}$, and $\mathcal{O}^{(3)}_{LNQd}$, with $\mathcal{O}_{LNuQ}$ exhibiting the largest amplitude. At first glance, this may seem counterintuitive given that Fig.~\ref{fig:pol_bmeson} shows that $\mathcal{O}^{(3)}_{LNQd}$ produces more positively polarized $\bar{N}$ compared to $\mathcal{O}_{Nedu}$ for 
$m_N = 1$~GeV. However, this can be understood by noting that the $\mathcal{O}_{Nedu}$ induces two-body decays unlike $\mathcal{O}^{(3)}_{LNQd}$, and yields $\bar{N}$ with unit polarization. This leads to a more pronounced difference between the Dirac and Majorana RHNs, explaining the hierarchy in the amplitudes.

Another interesting feature is evident from the first row of Fig.~\ref{fig:vertical_distance_positron_D}, which shows the distribution corresponding to RHNs produced from $D$ meson decays for a representative mass of $m_N = 1$~GeV. Comparing the top-left panel ($\mathcal{O}_{LNuQ}$) with the top-center panel ($\mathcal{O}_{Nedu}$), we observe that the $\mathcal{O}_{LNLe}$ distribution exhibits an overall sign flip between the two cases. This behaviour can be directly traced to the polarization of RHNs produced in $D$ meson decays via these two operators, as shown in Fig.~\ref{fig:pol_bmeson}. For $m_N = 1$~GeV, the $\mathcal{O}_{LNuQ}$ operator generates a highly polarized RHN with $P \sim +1$, while the $\mathcal{O}_{Nedu}$ operator leads to $P \sim -0.75$. Since the slope of the 
$\cos\theta_{\ell\ell}$ distribution is governed by \(\sum_{i = 7}^{13} P(\bar{N}) \mathcal{C}_i K_i\), a change in the sign of the polarization reverses the slope of the $\cos\theta_{\ell\ell}$ distribution, leading to the relative sign flip for $\mathcal{O}_{LNuQ}$ and $\mathcal{O}_{Nedu}$. This demonstrates how the sign of the RHN polarization directly impacts the differences between the Dirac and Majorana scenarios and reinforces our earlier approximate calculations that showed the correlation in the positron spatial distribution in the detector and the $\cos\theta_{\ell\ell}$ distribution. We do not attempt a similar qualitative explanation for the $\mathcal{O}^{(3)}_{LNQd}$ operator because the RHN polarization is close to zero, leading to significant interference effects among the various angular contributions. Consequently, the resulting spatial distributions are more complex, smaller in amplitude and cannot be easily interpreted using the simple approximations that work well for the highly polarized cases discussed above. In contrast, for $\mathcal{O}_{LNLe}$ and $\mathcal{O}_{Nedu}$, the RHN polarization is close to the extremal values $\pm 1$, allowing for a more straightforward understanding of the observed distribution shapes in terms of the underlying angular correlations. This flipping effect also does not occur with the other meson decays considered since the corresponding polarizations are always positive for all the masses and operators.

\subsection{Statistical analysis}

To quantify the statistical significance with which a Majorana RHN hypothesis can be rejected given that the RHN is Dirac,  we perform a $\chi^2$ analysis using the test statistic,
\begin{equation}
    \chi^2 = \sum_{i=1}^{30} 2 \left[(1 + \alpha_i) N_{\text{Majorana}}^i - N_{\text{Dirac}}^i + N_{\text{Dirac}}^i \ln \frac{N_{\text{Dirac}}^i}{(1 + \alpha_i) N_{\text{Majorana}}^i}\right] + \frac{\alpha_i^2}{\sigma_i^2}\,,
\end{equation}
where $N_{\text{Majorana}}^i$ and $N_{\text{Dirac}}^i$ are the expected event yields in the $i$-th bin for each case, and $\alpha_i$ are bin-dependent nuisance parameters that account for systematic uncertainties. We adopt conservative systematic uncertainties: $\sigma_i = 0.74$ per bin for $B$ meson channels and $\sigma_i = 1.0$ per bin for $D$, $K$ and $\pi$ meson channels. These values account for uncertainties arising from the simulation of the forward region using different event generators. 
 We exclude bins with $N_{\rm {Majorana}}^i = 0$ from the $\chi^2$ calculation to avoid undefined contributions. For bins with $N_{\rm{Dirac}}^i = 0$, we compute the $\chi^2$ using the above formula, as it approaches a well-defined limit. 

For each RHN mass and operator combination, we compute the $\chi^2$ values for the three spatial observables: the total electron-positron separation $X(e^+ - e^-)$, the electron displacement $X(e^-)$, and the positron displacement $X(e^+)$. In this analysis, the Dirac and Majorana event distributions are normalized to the expected number of events in the FASER2 detector for the given mass and operator choice, accounting for the production cross sections, decay probabilities, and detector acceptance.

\begin{figure}[t]
    \centering
    \includegraphics[scale=0.29]{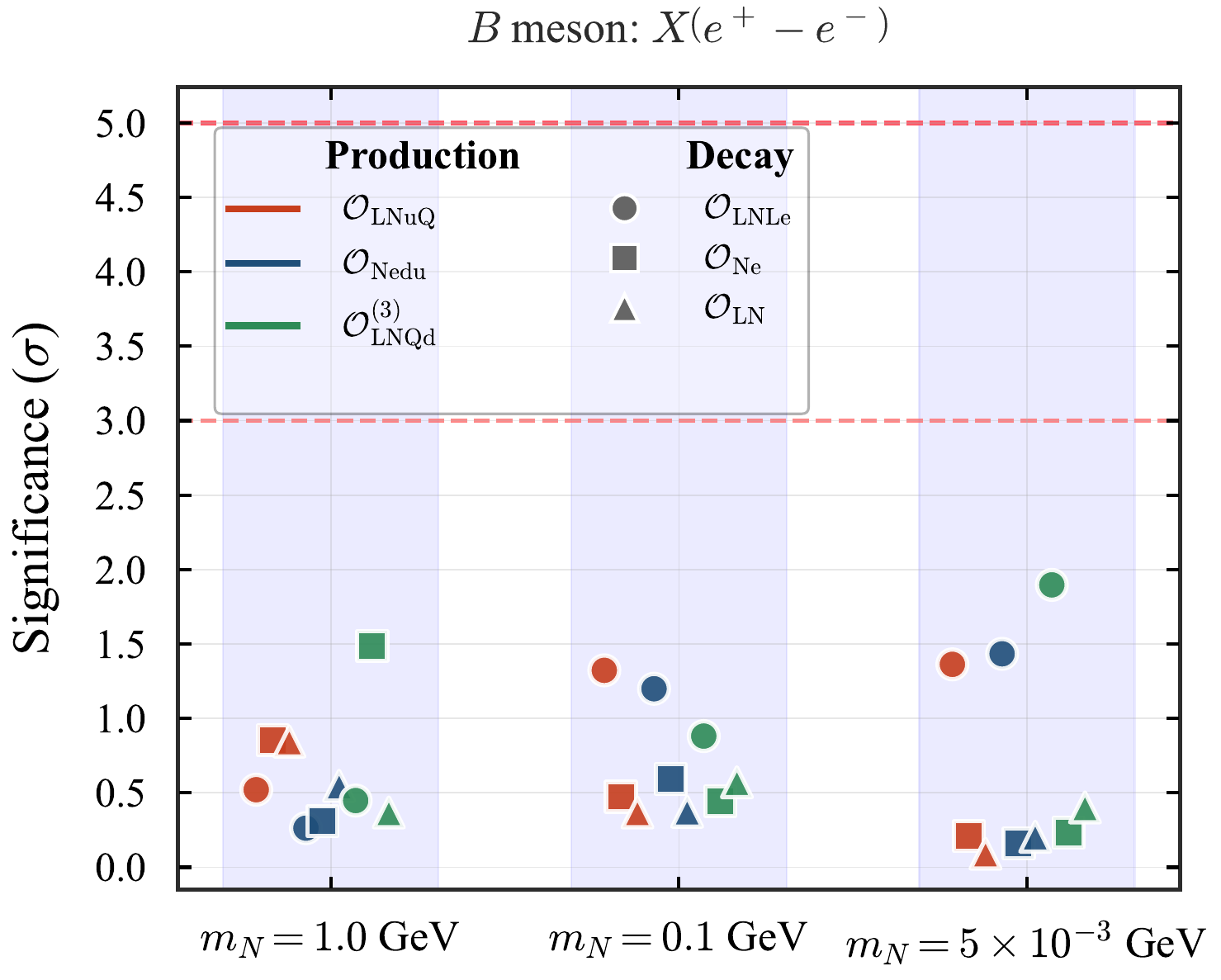}
    \includegraphics[scale=0.29]{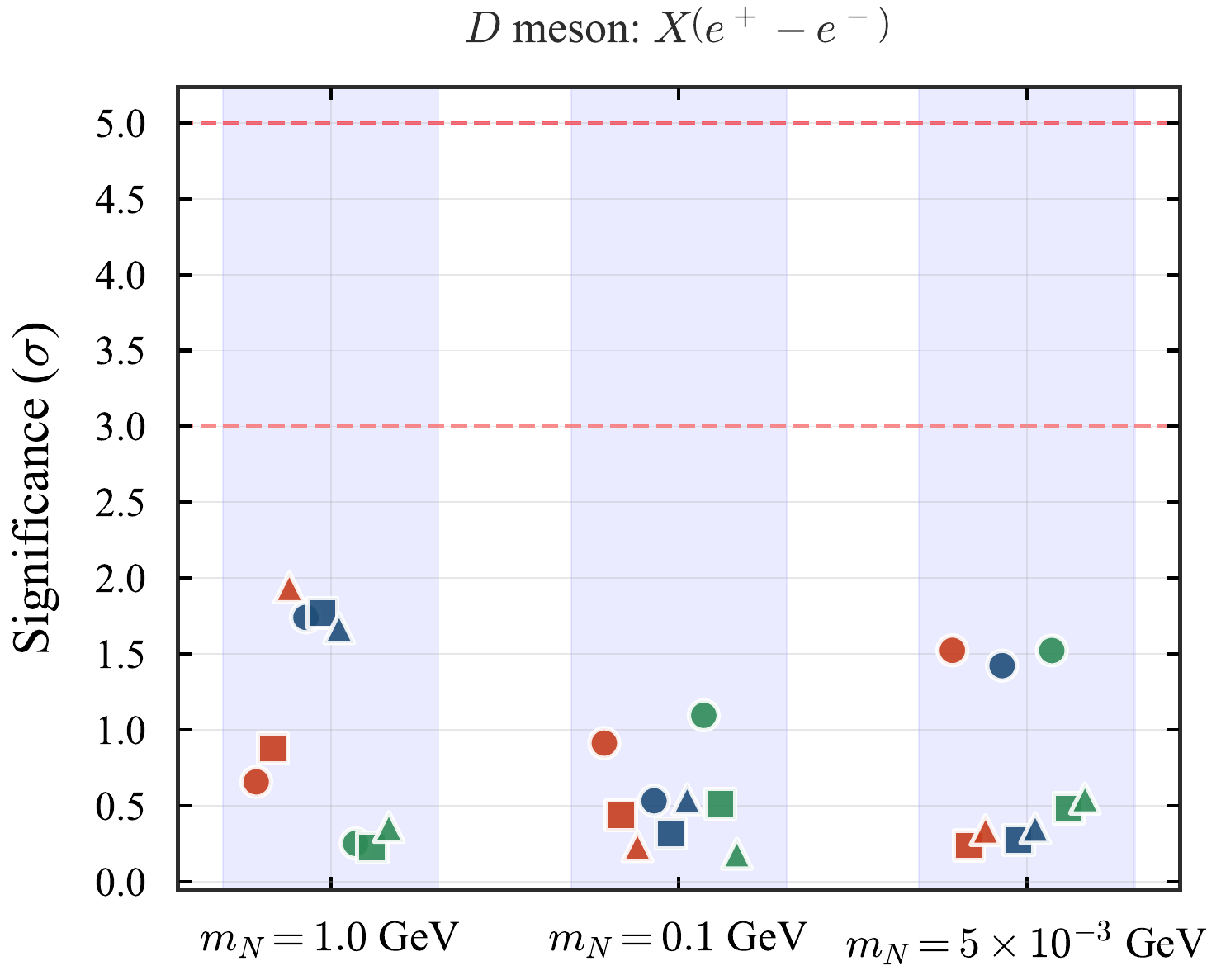}
    
    \includegraphics[scale=0.29]{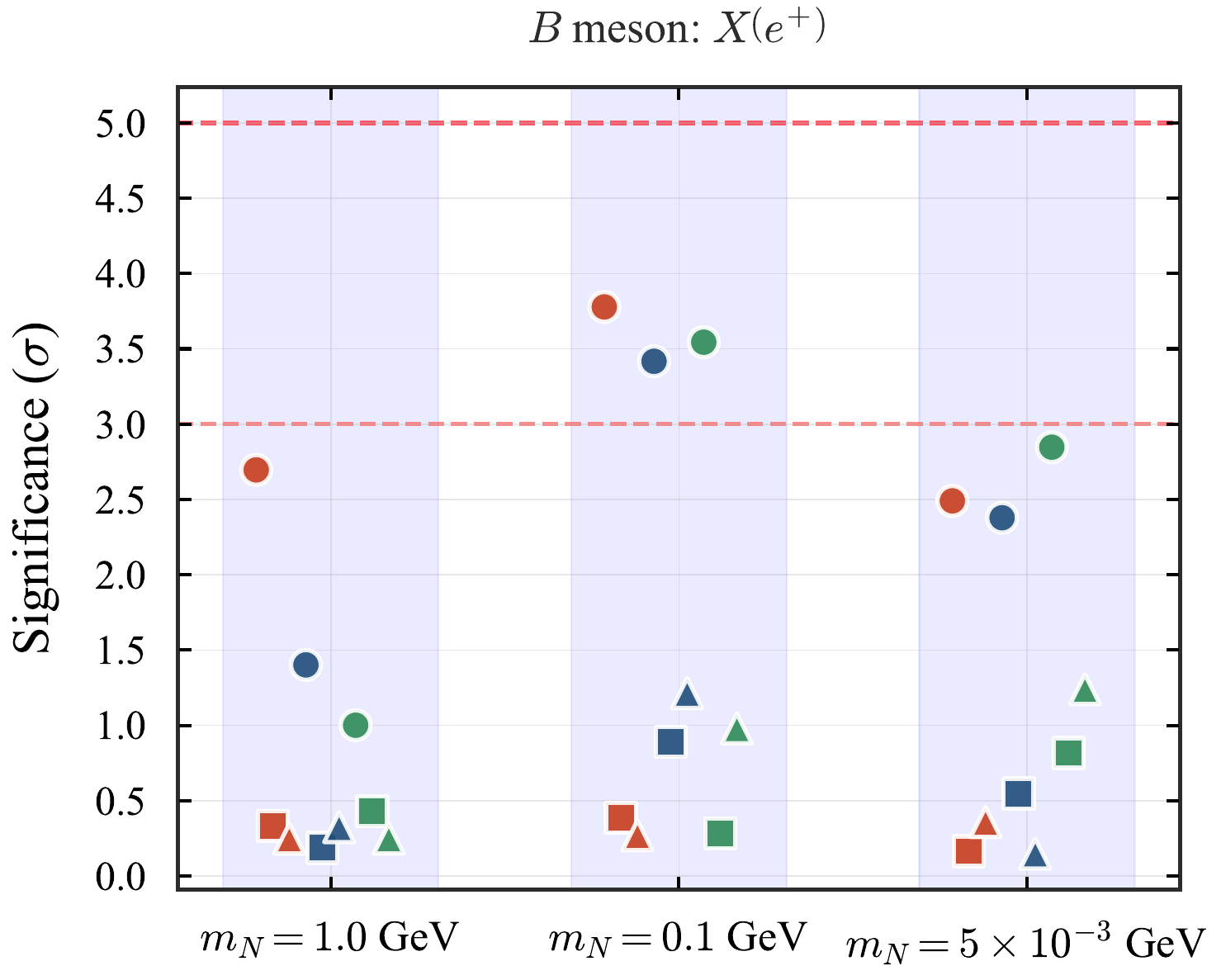}
    \includegraphics[scale=0.29]{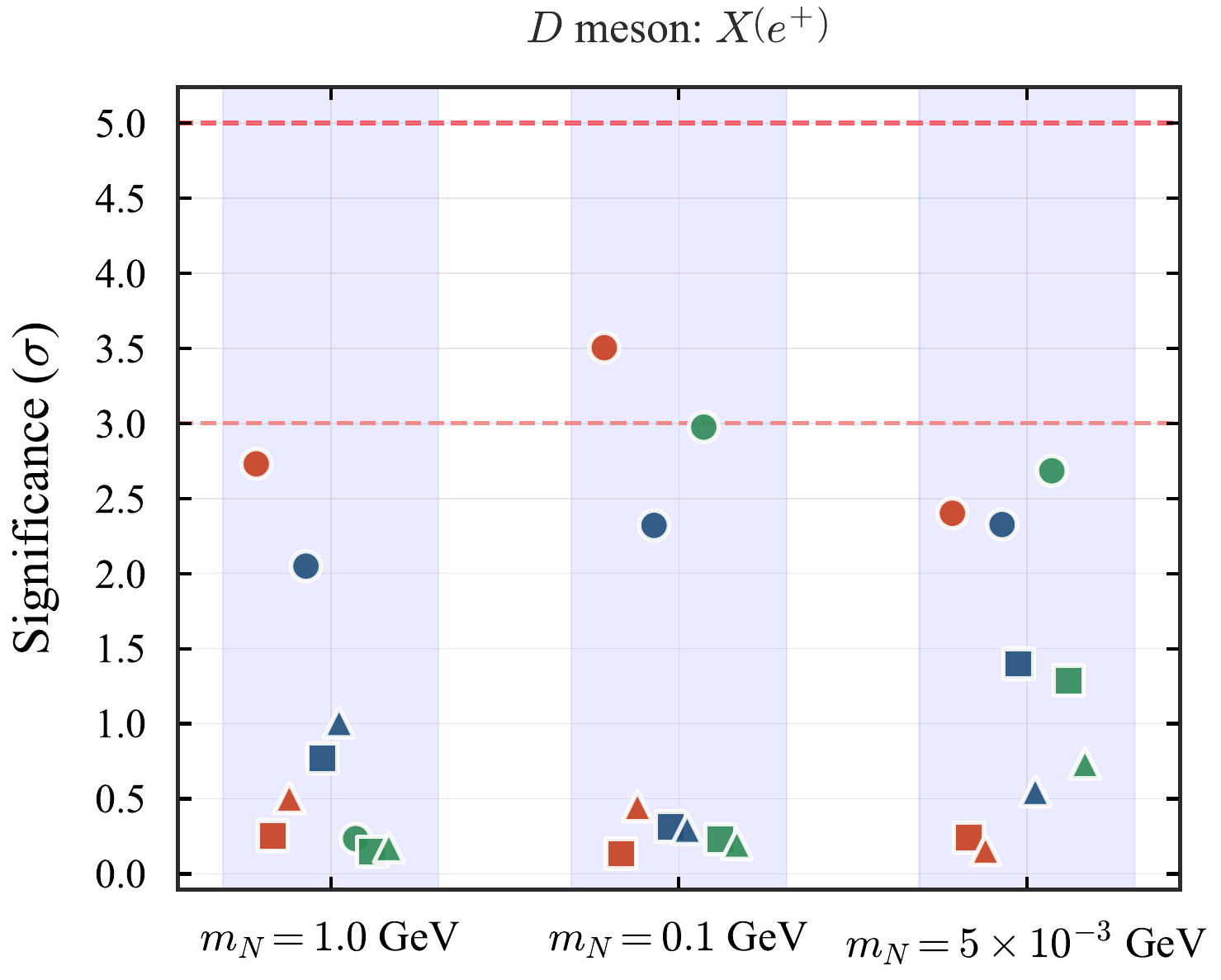}
    
    \includegraphics[scale=0.29]{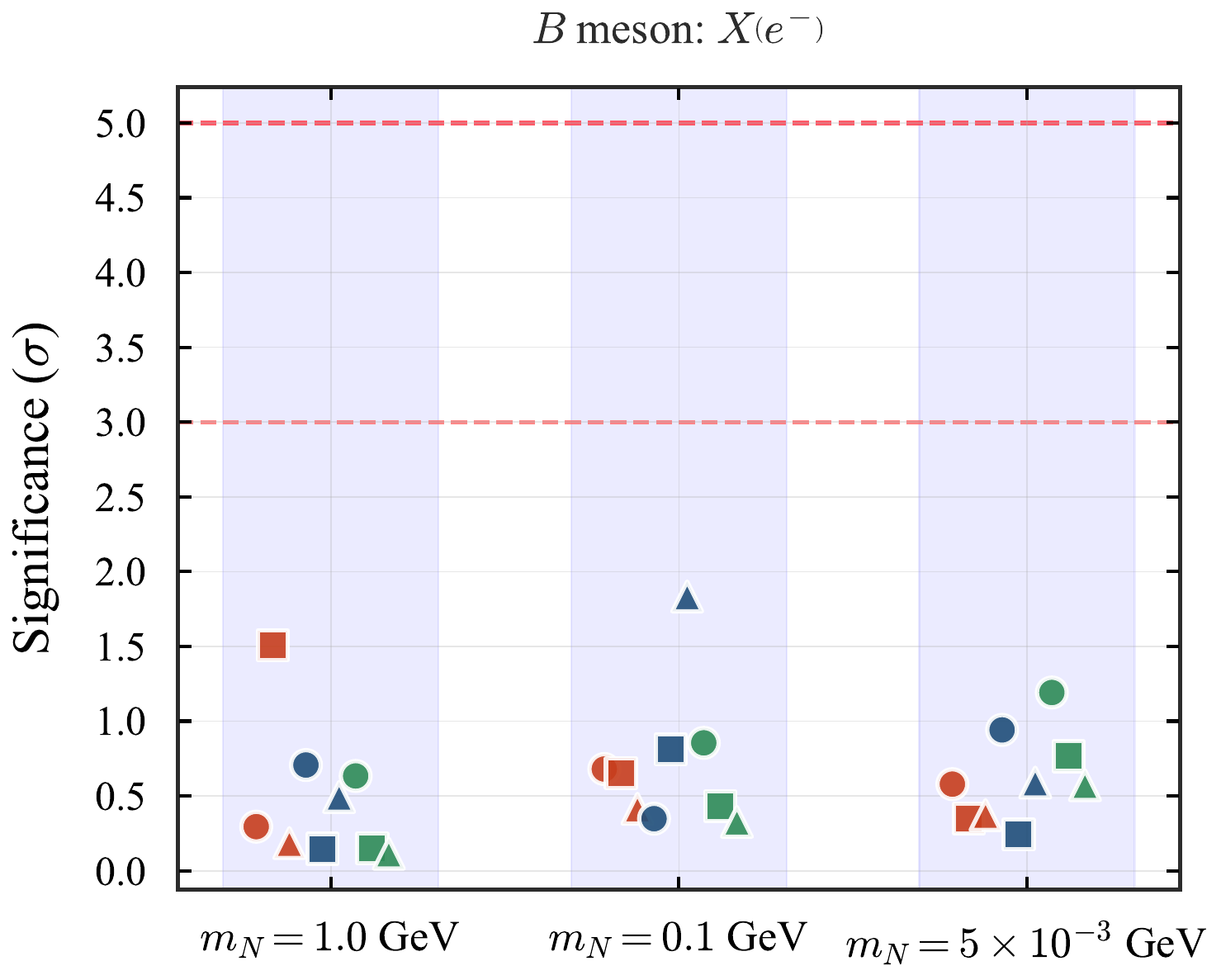}
    \includegraphics[scale=0.29]{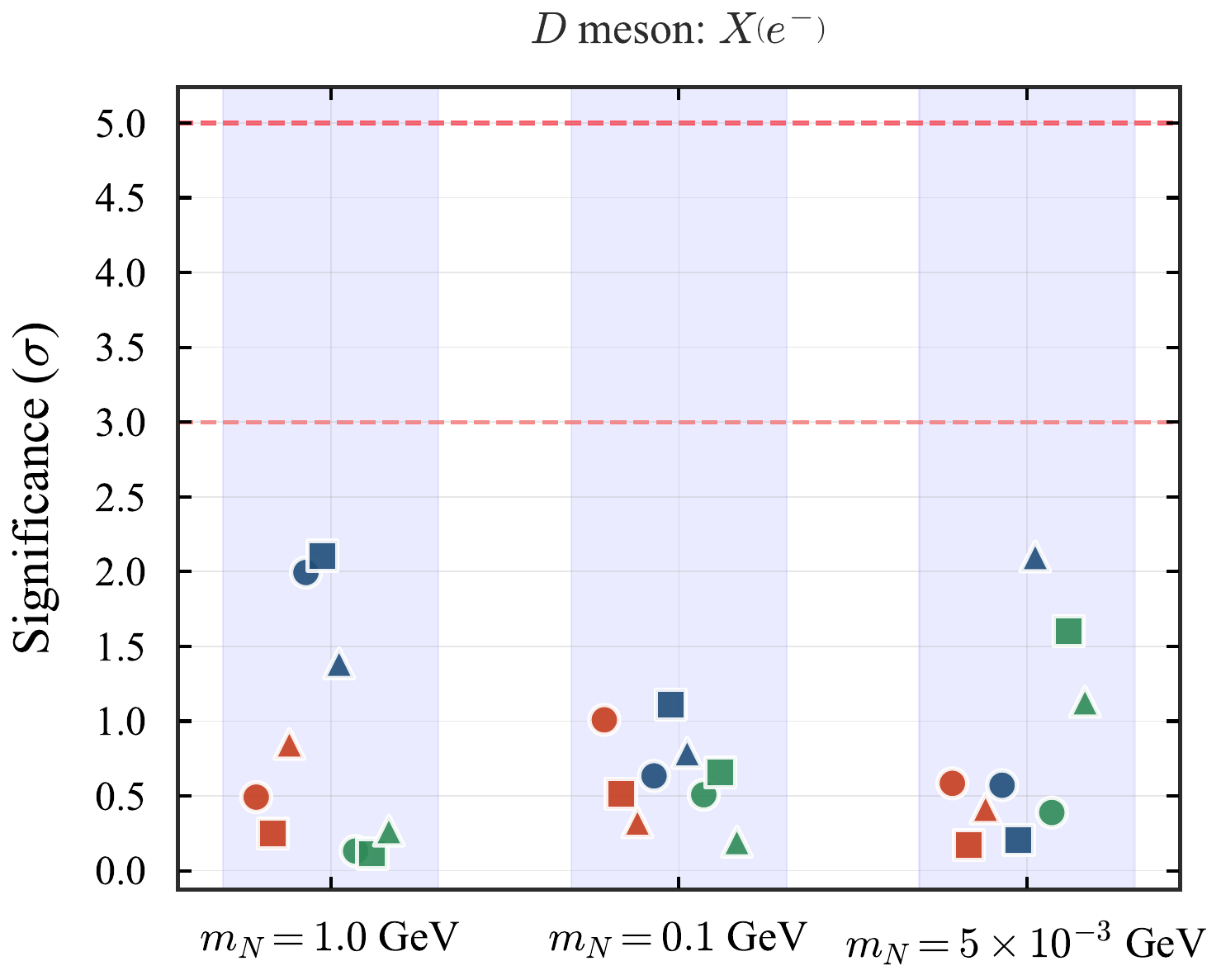}
    
    \caption{
    Statistical significance with which a Majorana RHN hypothesis can be rejected given that the RHN is Dirac, for three spatial observables: total $e^+e^-$ separation $X(e^+ - e^-)$ (top), positron displacement $X(e^+)$ (center), and electron displacement $X(e^-)$ (bottom). Results are shown for RHN masses of 1~GeV, 100~MeV, and 5~MeV, produced in $B$ meson decay (left panels) and $D$ meson decay (right panels) for the nine combinations of production and decay operators. }
    \label{fig:chi2_plot_all_masses_BD_mesons}
\end{figure}

\begin{figure}[t]
    \centering
    \includegraphics[scale=0.27]{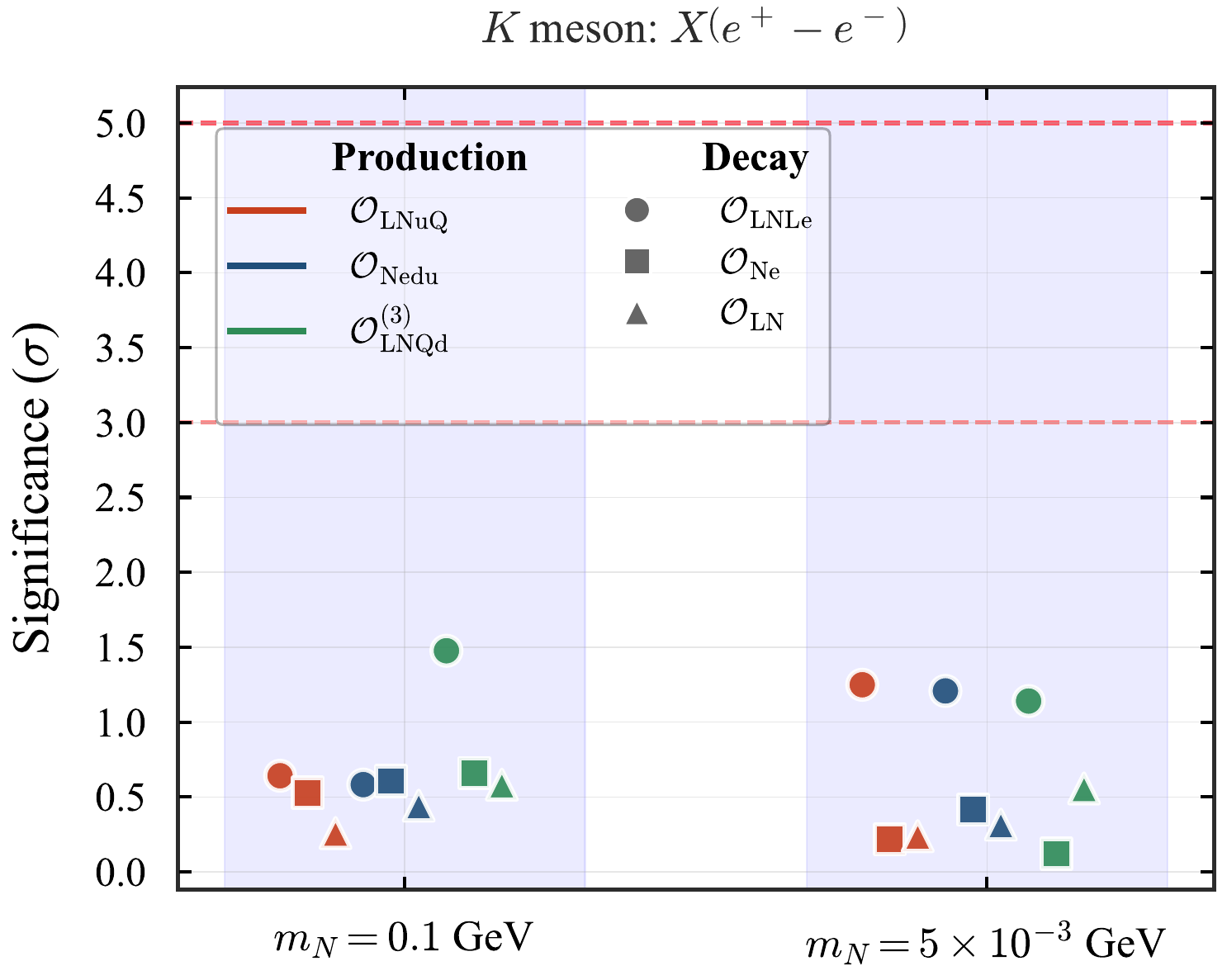}
    \includegraphics[scale=0.27]{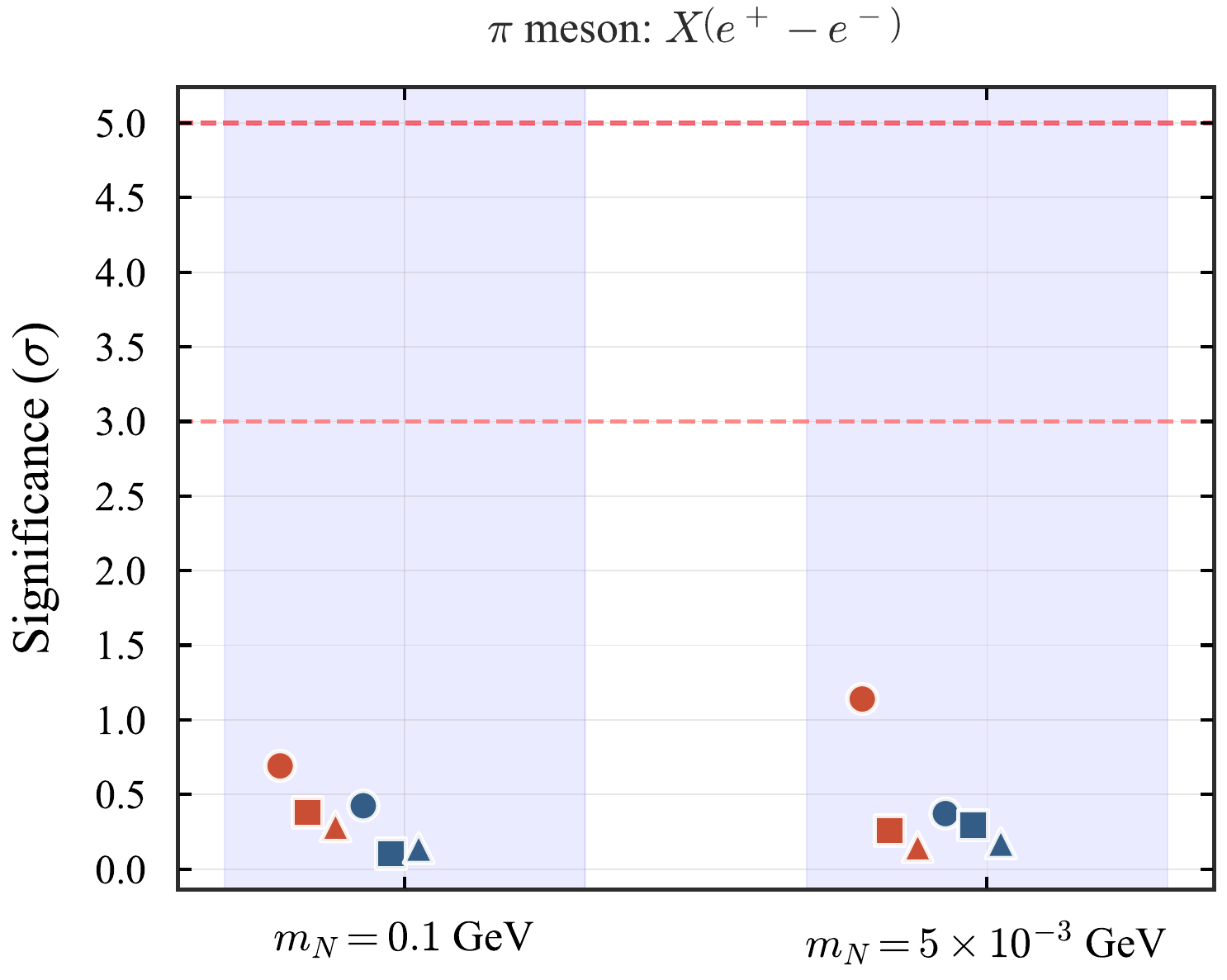}
    
    \includegraphics[scale=0.27]{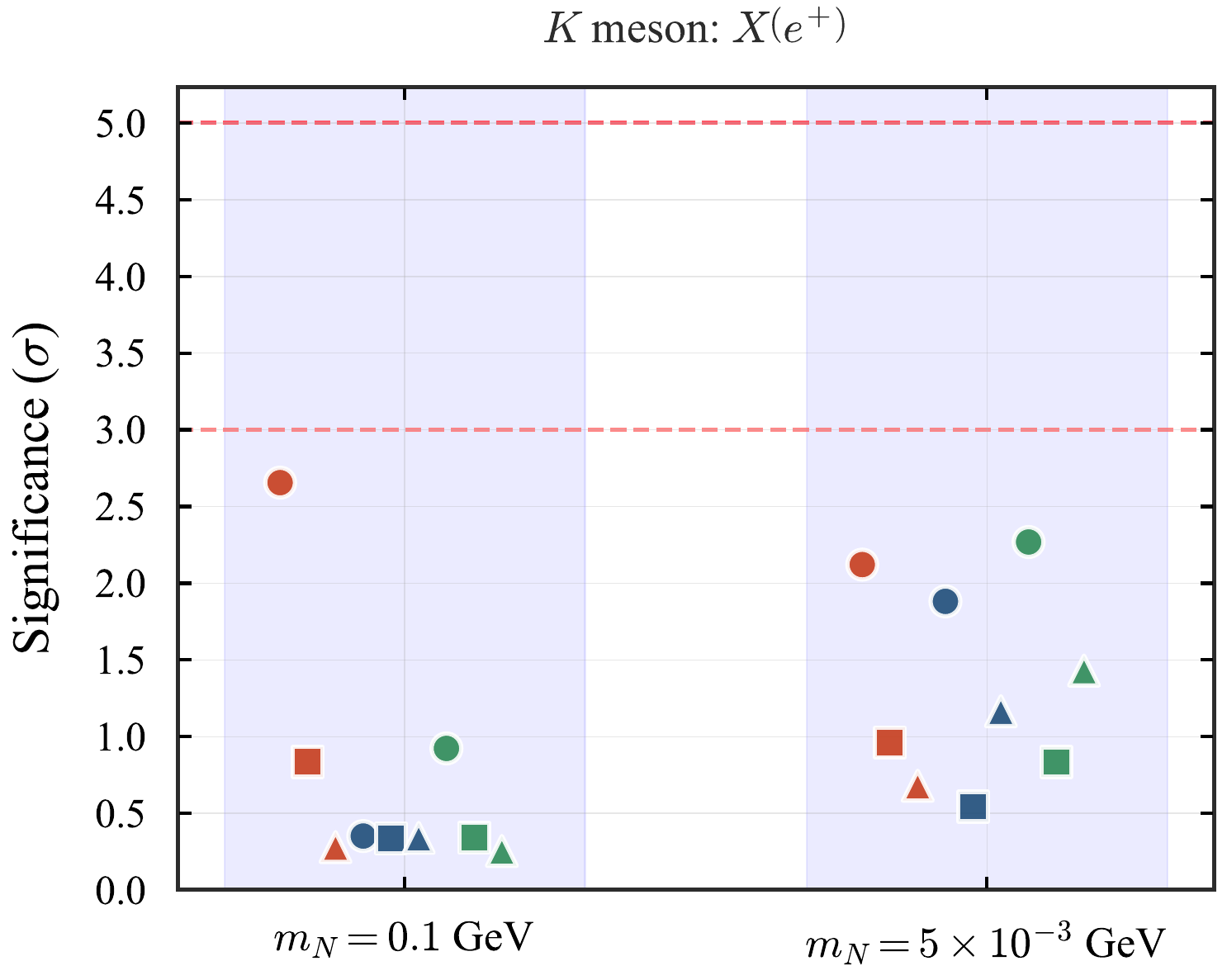}
    \includegraphics[scale=0.27]{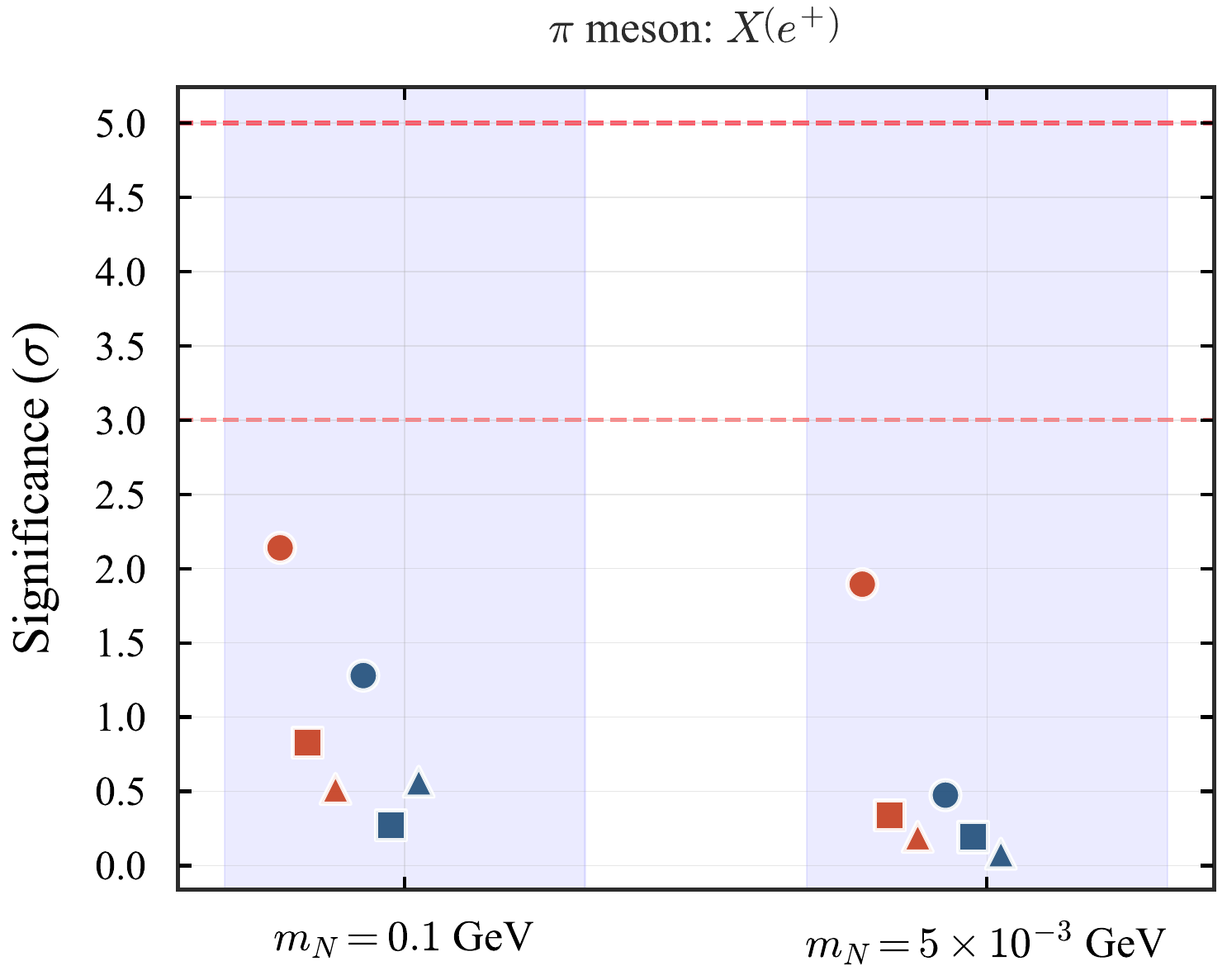}
    
    \includegraphics[scale=0.27]{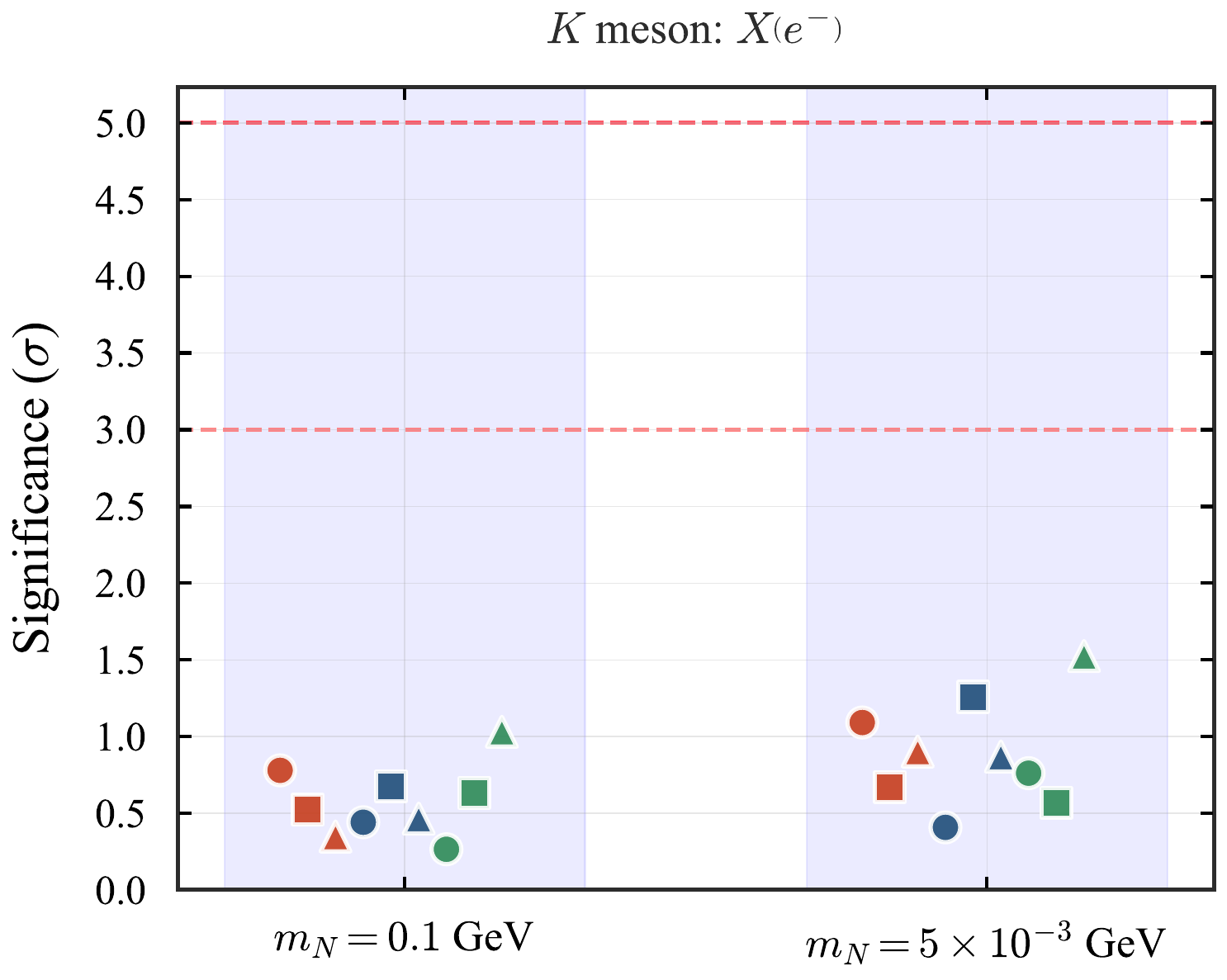}
    \includegraphics[scale=0.27]{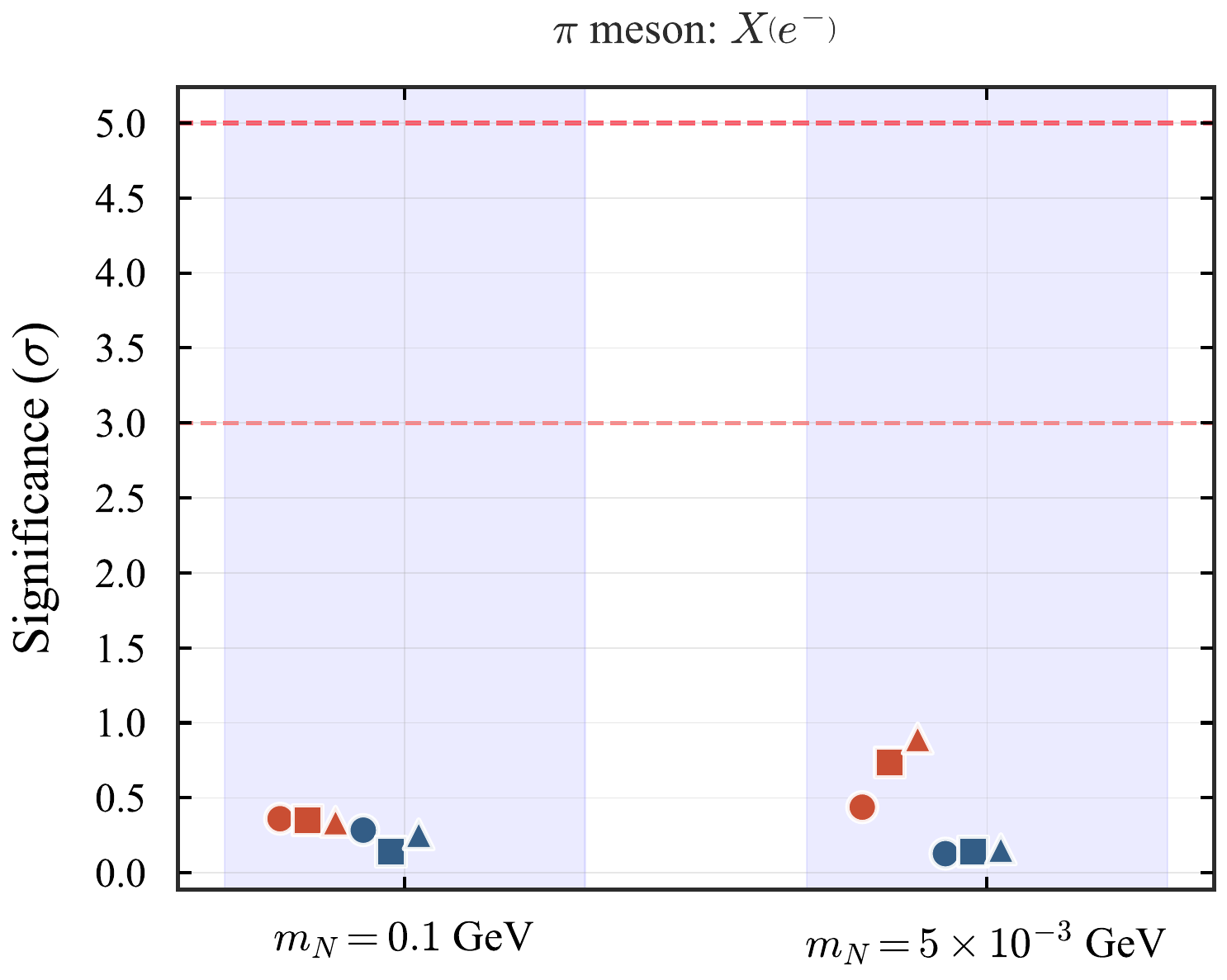}
    
    \caption{
    Similar to Fig.~\ref{fig:chi2_plot_all_masses_BD_mesons} for RHNs produced in kaon decay (left panels) and pion decay (right panels).}
    \label{fig:chi2_plot_all_masses_Kpimesons}
\end{figure}

The results of our statistical analysis are presented in Figs.~\ref{fig:chi2_plot_all_masses_BD_mesons} and~\ref{fig:chi2_plot_all_masses_Kpimesons}. Each panel shows the statistical significance $\sigma = \sqrt{\chi^2}$ for the three observables as a function of the RHN mass. Different colors and symbols indicate the nine combinations of SMNEFT production and decay operators. Figure~\ref{fig:chi2_plot_all_masses_BD_mesons} shows results for $B$ mesons (left panels) and $D$ mesons (right panels), while Fig.~\ref{fig:chi2_plot_all_masses_Kpimesons} shows results for $K$ (left panels) and $\pi$ mesons (right panels. The rows correspond to the  horizontal separation $X(e^+ - e^-)$ (top row), the positron distribution $X(e^+)$ (middle row), and the electron distribution $X(e^-)$ (bottom row).

Note that $X(e^+)$ consistently provides the greatest statistical power for distinguishing Dirac from Majorana RHNs.
$X(e^+)$ yields $\gtrsim 2.5\sigma$ sensitivity for RHNs in the entire mass range produced via $B$, $D$ and $K$ mesons and decaying through $\mathcal{O}_{LNLe}$.
In the $B$ and $D$ meson channels, the statistical significance reaches more than 3$\sigma$ for $m_N = 0.1$~GeV for all production operators in the $B$ meson channel, and for $\mathcal{O}_{LNuQ}$ and $\mathcal{O}^{(3)}_{LNQd}$ in the $D$ meson channel.  This enhanced sensitivity can be attributed to the highly polarized RHNs ($P \approx 1$) that maximizes the angular asymmetries in the decay products.
The sensitivity generally, but not always, decreases as the RHN mass approaches the kinematic threshold of the parent meson, where the available phase space becomes constrained and the polarization effects are reduced. It should also be noted that the total number of events also affects the magnitude of the $\chi^2$ values, as fewer events lead to smaller differences. 

\begin{figure}[t]
    \centering
    \includegraphics[scale=0.5]{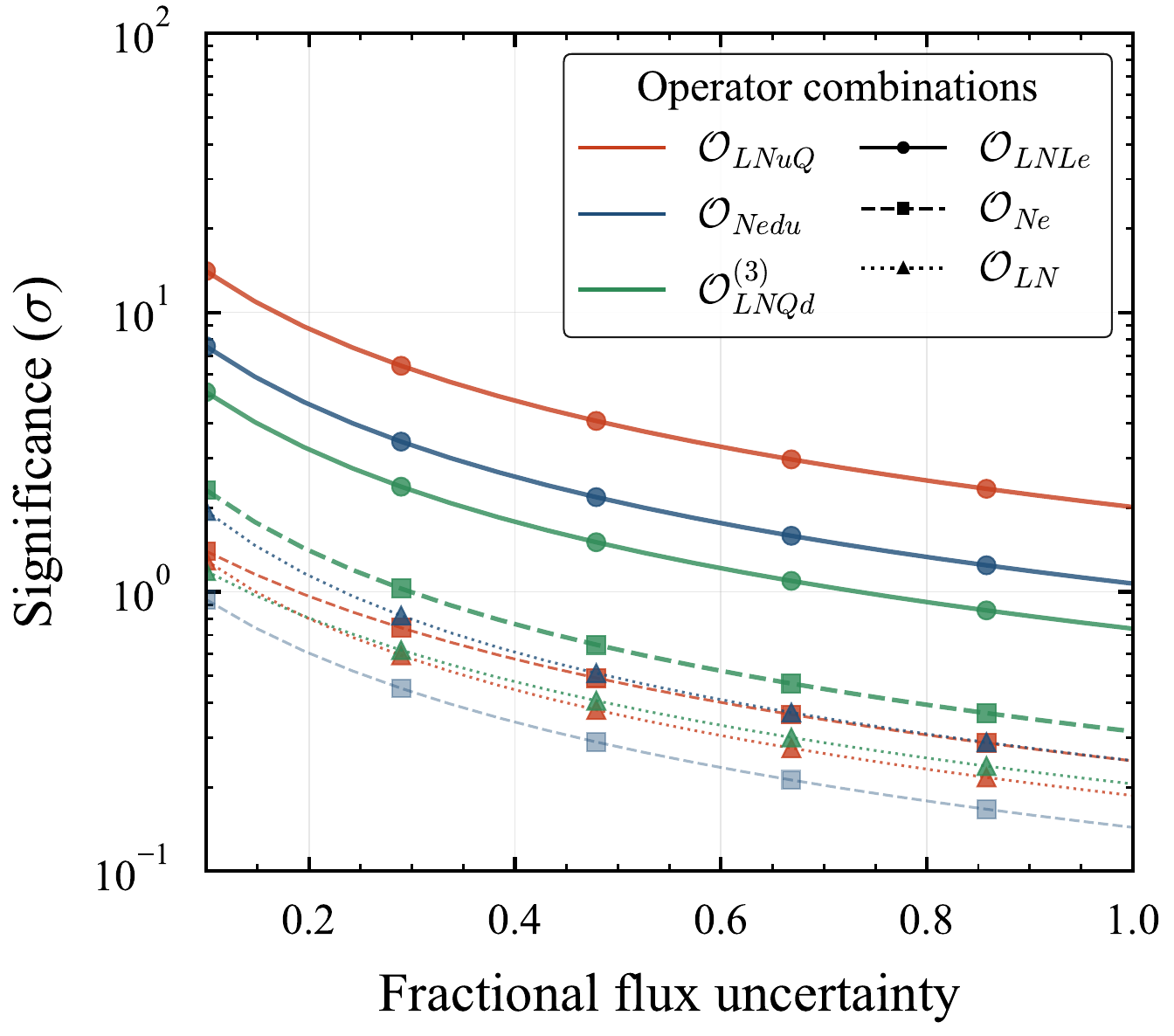}
    \caption{
    Statistical significance with which a Majorana RHN hypothesis can be rejected given that the RHN is Dirac,
    as a function of the fractional flux uncertainty for the positron distribution $X(e^+)$ arising from 1~GeV RHN decay produced via $B$ meson decay. Each curve corresponds to one of the nine operator combinations with the colors corresponding to the production operators and line-types corresponding to the decay operators.}
    \label{fig:sig_vs_unc}
\end{figure}

Although our results demonstrate promising sensitivity for distinguishing between Dirac and Majorana RHNs, it is important to acknowledge that experimental uncertainties, particularly those associated with forward meson flux modeling, represent a significant limiting factor in the statistical power of these measurements. The substantial systematic uncertainties we have incorporated (74\% for $B$ meson channels and 100\% for lighter meson channels) reflect the current theoretical understanding of forward hadron production at the LHC. Future improvements in these uncertainties through better theoretical modeling or dedicated calibration measurements could substantially enhance the discriminating power between Dirac and Majorana RHNs. Figure~\ref{fig:sig_vs_unc} illustrates the relationship between significance and fractional flux uncertainty for $X(e^+)$ originating from 1~GeV RHNs produced by $B$ mesons, for the nine operator combinations. We normalize the distributions to $10^4$ events. As expected, the statistical significance decreases sharply with increasing uncertainty. Notably, many additional operator combinations could become accessible if the forward flux uncertainties are reduced.

In the preceding analysis, we considered the decay of one meson species at a time. However, in the actual FASER2 experiment, it will not be possible to identify which specific meson decay produced the RHN. Consequently, it is necessary to consider the combined contribution from all meson decays simultaneously. However, as shown in Fig.~\ref{fig:production_cross_sections}, for each production operator and across the full range of RHN masses, there is typically one decay channel that dominates. For instance, in the case of $\mathcal{O}_{LNuQ}$, two-body decays of $D$ mesons dominate for all values of $m_N$. Therefore, the results based on $D$ meson decays should remain approximately valid, within the flux uncertainties. Similarly, for $\mathcal{O}_{Nedu}$, two-body $D$ meson decay dominates at $m_N = 1$~GeV, while two-body $K$ meson decay dominates for $m_N = 0.1$~GeV and $0.005$~GeV. Accordingly, the results from these dominant decay modes can be considered reliable within the associated uncertainties.

Despite these limitations, the clear trends and operator-specific features identified in our analysis suggest that targeted experimental searches focusing on low-mass RHNs and specific decay channels could provide a viable path toward experimentally determining the fundamental nature of RHNs. 

\section{Summary}
\label{Section5}

We investigated the potential to distinguish between Dirac and Majorana RHNs using spatial distributions of their decay products at the FASER detector. Our analysis, conducted within the SMNEFT framework, demonstrates that the angular correlations inherent in RHN decays provide a powerful experimental probe for determining the fundamental nature of these particles.
Our key findings are as follows.

\textbf{Kinematic signatures.} The angular distributions of the decay products in the RHN rest frame are fundamentally different depending on whether the RHN is a Dirac or Majorana particle. These differences, particularly manifested in the $\cos\theta_{\ell\ell}$ distribution, arise from the operator structure governing the decay and are maximized when RHNs are highly polarized. Upon boosting to the laboratory frame, these angular asymmetries translate into measurable spatial separations of charged tracks at the detector.

\textbf{Experimental sensitivity.} Our Monte Carlo simulations demonstrate that the \\ FASER2~(2025) detector configuration in Table~\ref{tab:detector_configurations} provides substantial sensitivity to probe the Dirac or Majorana nature of RHNs with masses between 5~MeV and 1~GeV. The positron spatial distribution is the most discriminating observable. FASER can achieve a sensitivity $\gtrsim 2.5\sigma$ for RHNs in the entire mass range produced via $B$, $D$ and $K$ mesons through any of the production operators ($\mathcal{O}_{LNuQ}$, $\mathcal{O}_{Nedu}$, $\mathcal{O}^{(3)}_{LNQd}$) and decaying through $\mathcal{O}_{LNLe}$. For RHNs with a mass of 0.1~GeV produced from $B$ meson decay via any production operator, and from $D$ meson decay via $\mathcal{O}_{LNuQ}$ and $\mathcal{O}^{(3)}_{LNQd}$, and decaying through $\mathcal{O}_{LNLe}$, FASER can achieve discrimination between Dirac and Majorana RHNs at greater than the $3\sigma$ level. The discrimination power is generally enhanced for lower RHN masses where polarization effects are maximized, and for operators that produce highly asymmetric angular distributions.

\textbf{Operator dependence.} The discriminating power depends strongly on the specific SMNEFT operators governing RHN production and decay. The $\mathcal{O}_{LNLe}$ decay operator consistently provides the strongest discrimination capability, while production through two-body meson decays generally yields higher sensitivity than three-body decays due to the enhanced RHN polarization. This operator dependence provides additional handles for the experimental validation of the underlying physics.

\textbf{Systematic uncertainties.} While our analysis allows for substantial systematic uncertainties (74\%-100\% per bin) associated with forward hadron production modeling, the robust nature of the kinematic signatures ensures that the discriminating between Dirac and Majorana RHNs remains experimentally accessible. The clear operator-specific patterns and mass-dependent trends identified in our study provide concrete targets for future experimental searches. Much higher sensitivities can be achieved with a concerted effort to reduce uncertainties in the forward meson flux.

Our results establish that kinematic measurements of RHN decay in forward LHC detectors offer a viable experimental strategy for addressing the fundamental question of the Dirac or Majorana nature of neutrinos.


\acknowledgments

We thank A.~Das for collaboration during the early stages of this work. 
S.K.A is supported by JST SPRING, Grant Number JPMJSP2119. A.D. is supported in part by the U.S. National Science Foundation under Grant No.~PHY-2309937. D.M. is supported in part by the U.S. Department of Energy under Grant No.~de-sc0010504. 
\newpage
\bibliographystyle{jhep}
\bibliography{bibliography}

\begin{thebibliography}{10}
\bibitem{hepjo}
{Shivasankar K.A}, ``Hepjo is a high-performance library for numerical
  computations in particle physics, written in mojo similar to scipy,
  scikit-hep in python.''
\end{thebibliography}
\end{document}